\documentclass[remotesensing,article,accept,moreauthors,pdftex]{Definitions/mdpi} 

\firstpage{1} 
\pubvolume{1}
\issuenum{1}
\articlenumber{0}
\pubyear{2023}
\copyrightyear{2023}
\datereceived{} 
\dateaccepted{} 
\datepublished{} 
\hreflink{https://doi.org/} % If needed use \linebreak
\extrafloats{100}
\Title{Mid-Infrared Observations of the Giant Planets} %MDPI: the article type in the document is different with that in Susy, please confirm
\TitleCitation{Mid-Infrared Observations of the Giant Planets}

\Author{Michael T. Roman $^{1,2}$\orcidA{}}% and Arrate Antu\~{n}ano $^{3}$}
\AuthorCitation{Roman, M.T.}
\address{%
$^{1}$ \quad School of Physics and Astronomy, University of Leicester, Leicester LE1 7RH, United Kingdom {m.t.roman@leicester.ac.uk}\\%MDPI: please add city and postal code. email is different with that in Susy, please confirm
$^{2}$ \quad Facultad de Ingeniera y Ciencias, Universidad Adolfo Ib\'{a}\~{n}ez, Av. Diagonal las Torres 2640, Pe\~{n}alol\'{e}n, Santiago, Chile
}

\abstract{The mid-infrared spectral region provides a unique window into the atmospheric temperature, chemistry, and dynamics of the giant planets. From more than a century of mid-infrared remote sensing, progressively clearer pictures of the composition and thermal structure of these atmospheres have emerged, along with a greater insight into the processes that shape them. Our knowledge of Jupiter and Saturn has benefitted from their proximity and relatively warm temperatures, while the details of colder and more distant Uranus and Neptune are limited as these planets remain challenging targets. As the timeline of observations continues to grow, an understanding of the temporal and seasonal variability of the giant planets is beginning to develop with promising new observations on the horizon.}

\keyword{giant planets; atmospheres; dynamics; atmospheres; chemistry}

\begin{document}
%%%%%%%%%%%%%%%%%%%%%%%%%%%%%%%%%%%%%%%%%%

\section{Introduction}

The mid-infrared region of the electromagnetic spectrum provides a unique and important window into the atmospheric physics and chemistry of the giant planets. Linking the near- and far-infrared, it spans a range of wavelengths (variously defined), over which the dominant source of planetary radiation transitions from scattered sunlight to intrinsic thermal emission. As the scattered solar component fades with increasing wavelength, the various features and colors that define the planets' appearances in visible and near-infrared images give way to distinct thermal structures shaped by the temperatures and chemistry of these atmospheres. Against this changing backdrop of scattered and emitted radiation, numerous molecules leave their distinct spectral signatures, indicative chemical abundances, kinetic temperatures, and ambient pressures. The observation and analysis of reflected and radiant energy can thus be used to reveal the composition, temperature, and structure of a planetary atmosphere from afar, providing remote measurements of fundamental properties largely inaccessible by other means. 

From more than a century of mid-IR observations, a rich picture of the four giant planets' atmospheres has emerged. Now, with the anticipated results from the new JWST promising to revise our knowledge of these planets in the years ahead  \citep{norwood2016giant}, we use this opportunity to look back and take stock of the field. In this review, we examine remote sensing of the Solar System's giant planets across the mid-infrared. We trace an observational history from its modest beginnings to present-day efforts, highlighting what we have learned along the way and what questions remain for future work.   %MDPI: Please confirm if the italics should be retained.  
%MDPI: please recheck the whole paper and remove all unnecessary italics in maintext
%MTR: I have removed them throughout. 

\subsection{The Mid-Infrared}
\begin{figure}[H]
\includegraphics [width=1.\linewidth, trim=0in 0.in .2in .0in, clip]{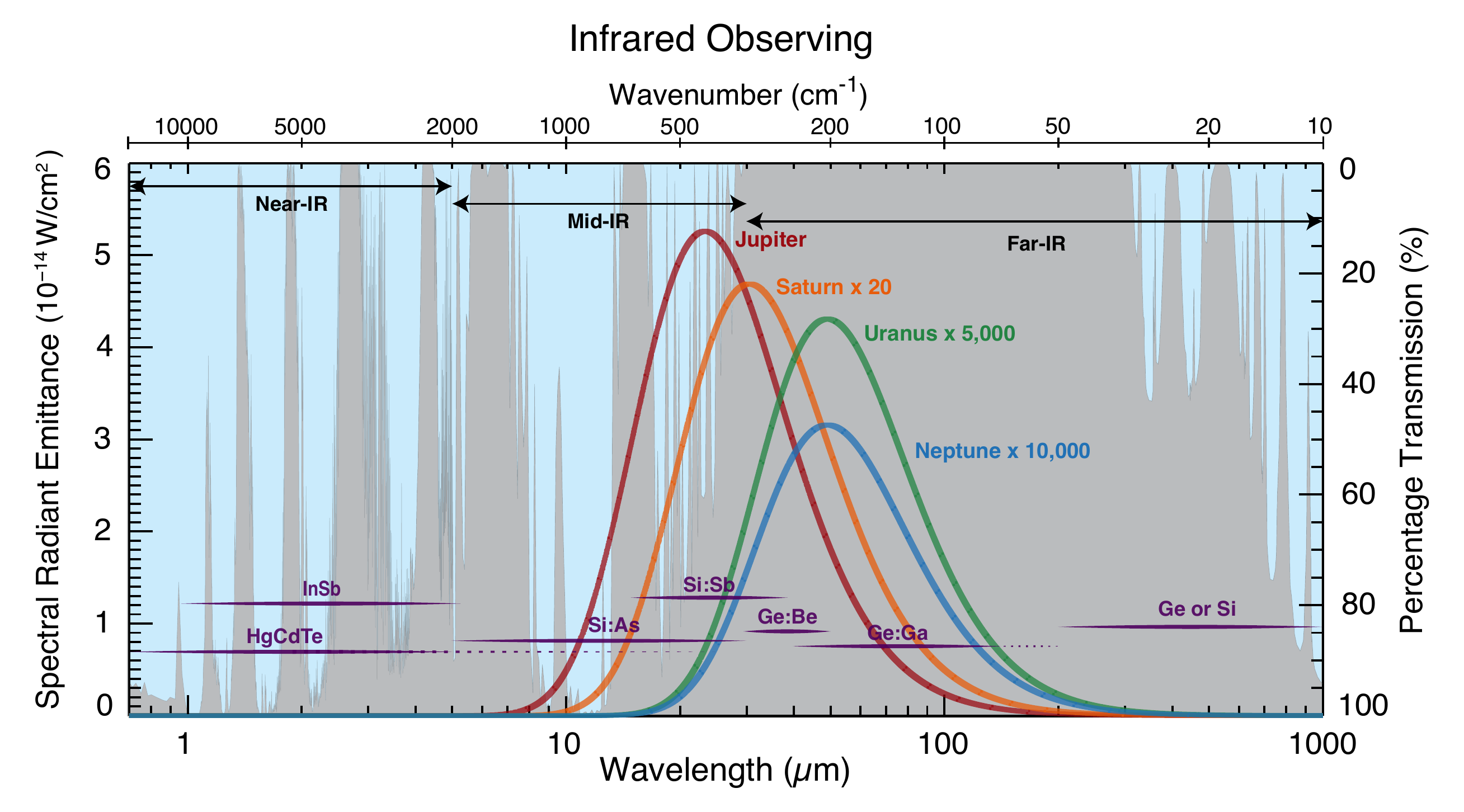}
\caption{Idealized blackbody emittance of the giant planets compared to the telluric atmospheric transmission and typical detector sensitivities across the infrared. The assumed boundaries of the near-, mid-, and far-infrared regions are indicated. Colored curves show the black body spectral radiant emittance for the effective temperatures of the giant planets, accounting for their distances, scaled and labeled for clarity. Jupiter and Saturn peak in the mid-infrared, while Uranus and Neptune peak in the far-infrared. The atmospheric transmission is indicated by the blue--gray interface varying between 100\% (full transmission) and 0\% (total attenuation) from the top of the atmosphere down to a surface altitude of 2,640 m (corresponding to the altitude of the Very Large Telescope (VLT) at Cerro Paranal, with a precipitable water vapor (PWV) of 1.66 mm at an air mass of 1.15 \citep{noll2012atmospheric,jones2013advanced}). Characteristic ranges for various thermal detectors are shown in purple  \citep{fazio1994infrared,haller1994advanced,glass1999handbook,tokunaga2002infrared,rieke2003detection,mclean2012infrared,ives2014aquarius}. \label{fig:ir_observing}}
\end{figure}

Infrared radiation (IR) occupies the region of the electromagnetic spectrum between visible light and radio waves (specifically, microwaves), corresponding to wavelengths from about 750\,nanometers to 1\,millimeter. In the modern literature, it is commonly divided into three subdivisions---near-, mid-, and far-infrared (see Figure \ref{fig:ir_observing}). The precise boundaries of these divisions are generally not agreed upon and differ widely across various disciplines and applications. The International Commission on Illumination (CIE)\footnote{\textit{International Standard CIE S 017:2020 ILV: International Lighting Vocabulary}, 2nd edition}, 
%MDPI: footer is not allowed in this journal. so all footers were moved into maintext like this, please conifrm %MTR:  I think footnotes are very much appropriate in this review article format and strongly preferred.  Regardless, Mr. Zhang said I may use them, so I have reinstated them. 
for example, defines the mid-IR as radiation with wavelengths of only 1.4 to 3\,microns, while the International Organization for Standardization (ISO)\footnote{\textit{ISO 20473:2007, Optics and Photonics---Spectral bands}} adopts a much broader range for the mid-IR, spanning from 3 to 50\,$\upmu$m. In some engineering literature, the infrared is divided into five regions, classified as short-wave (1--3\,$\upmu$m), mid-wave (3--5\,$\upmu$m), long-wave (8--12\,$\upmu$m), and very-long (12--30\,$\upmu$m) infrared (e.g., \citep{guan2021recent}), with significant variation in the defined demarcations. 
%MDPI: please confirm change in this citation. the following highlights are the same

\vspace{-18pt}
  %MDPI: Please use commas to separate thousands for numbers with five or more digits (not for four digits) in the picture, e.g., "10000" should be "10,000"
%MDPI: Please change the hyphen (-) into minus sign ($-$, "U+2212"), e.g., "-1" should be "$-$1".
%MDPI: please move Figure 1 after its first citation and revise ref citation or please add Figure 1's citation in previous content

In astronomy and planetary science literature, the mid-infrared typically refers to wavelengths between roughly 5 $\upmu$m and 20 to 30\,$\upmu$m \cite{larson1980infrared, glass1999handbook,tokunaga2002infrared,mampaso2004infrared}. These bounds are a natural consequence of practical constraints, namely astronomical detector technology and the transparency of Earth's atmosphere.  This adopted lower limit around 5\,$\upmu$m roughly coincides with the longest wavelengths detected by most common near-infrared detectors, typically composed of indium antimonide (InSb) or mercury-cadmium-telluride (HgCdTe) (see Figure \ref{fig:ir_observing}). At longer wavelengths, arsenic- or antimony doped silicon (Si:As and Si:Sb) Impurity Band Conduction (IBC) detectors are typically employed, sensitive to ranges of \mbox{$\sim$6--27\,$\upmu$m} and $\sim$14--38\,$\upmu$m, respectively, followed by germanium photoconductive detectors and bolometers in the far-infrared  \citep{fazio1994infrared,haller1994advanced,glass1999handbook,tokunaga2002infrared,rieke2003detection,mclean2012infrared,ives2014aquarius}. Case in point, the JWST Near Infrared Camera (NIRCam) instrument uses HgCdTe detectors for the 0.6--5\,$\upmu$m region, while the JWST Mid-Infrared Instrument (MIRI) uses a Si:As detectors to measure radiation from 5 to \mbox{28\,$\upmu$m  \citep{rieke2005overview,wells2006miri,rieke2015mid}}. As discussed in the next section, the so-called atmospheric window of infrared transparency provides a natural upper boundary to the mid-infrared around 30\,$\upmu$m, beyond which little infrared radiation is transmitted through the atmosphere.  

For the purpose of this review, we will adopt the definition of the mid-infrared as radiation between 5 and 30\,$\upmu$m (or 2,000--333 cm$^{-1}$, in terms of wavenumber) and limit our scope to remote sensing within this wavelength range. We will also restrict ourselves to the giant planets without our Solar System, leaving the growing number of extrasolar planet infrared observations to other reviews \cite{encrenaz2014infrared,pluriel2023hot}.
%MTR: added the above, including new references

\subsection{Atmospheric Transmission, Emission, and Mid-Infrared Sub-Bands}
Gaseous absorption, primarily by telluric water vapor, renders the Earth's atmosphere largely opaque to extraterrestrial infrared radiation as seen from the ground at various wavelengths. Between about 30 $\upmu$m and several hundred microns, the atmosphere is nearly continuously opaque\footnote{The atmospheric transparency begins to increase once again approaching the millimeter region, which has been used to sense deeper into the atmospheres of the giant planets than that which can be accessed by visible and infrared \mbox{observations  \citep{depatermicrowaves,encrenaz2002microwave,depaterh2spossible}}.}, marking the adopted cutoff between the mid- and far-infrared (see Figure \ref{fig:ir_observing}). Owing to this absorption, the far-infrared (or sub-millimeter) spectral region is only accessible from extremely high-altitude, airborne, and space observatories \cite{naylor1991atmospheric}. 

\begin{figure}[H]
\includegraphics [width=1.0\linewidth, trim=0in 0.in 0in .1in, clip]{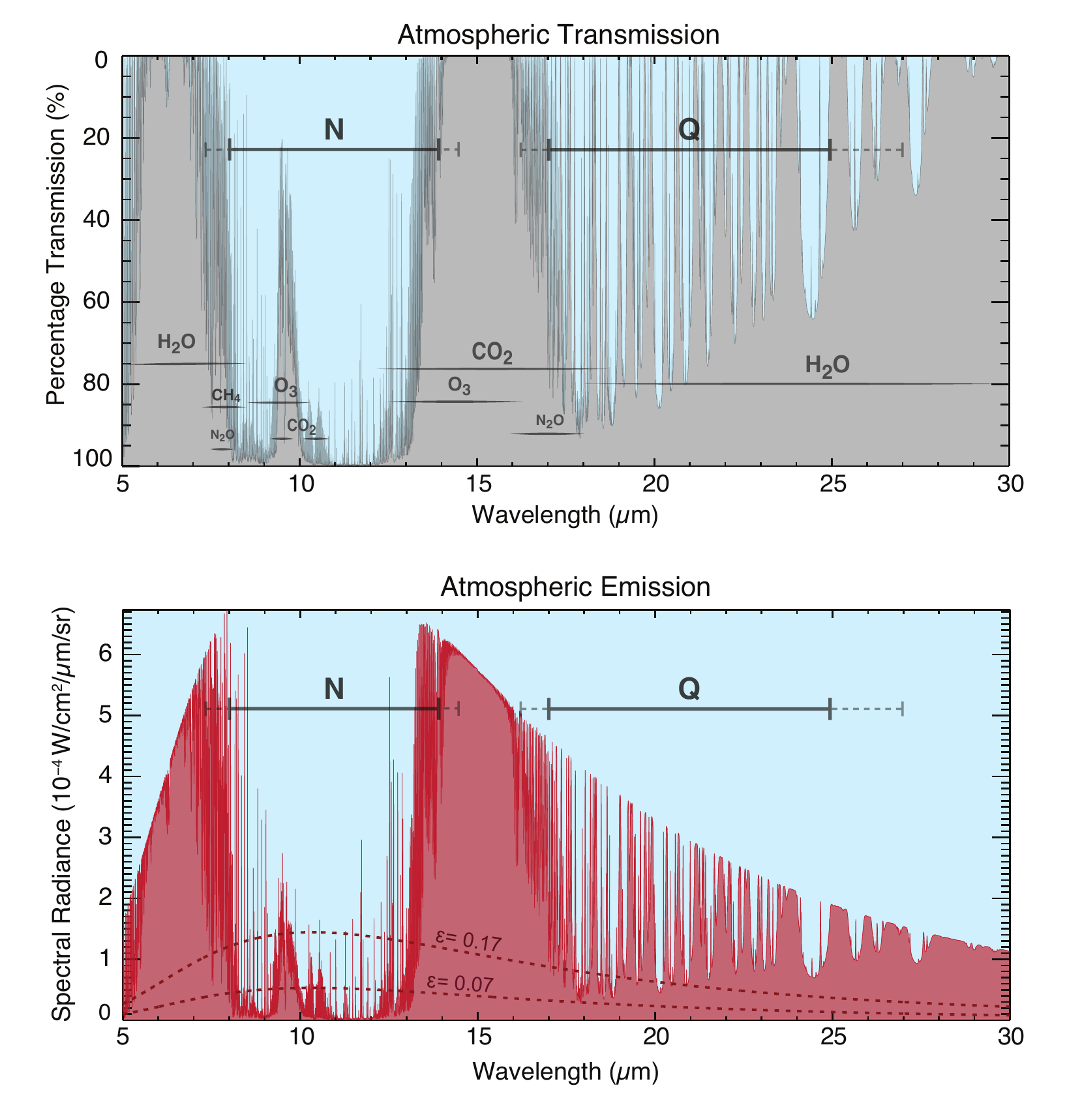}
\caption{The atmospheric transmission (top) and emission (bottom) in the mid-IR, for conditions at Cerro Paranal, as described in Figure \ref{fig:ir_observing}. Transmission is indicated by the blue--gray interface varying between 100\% (full transmission) and 0\% (total attenuation) from the top of the atmosphere down to the surface \citep{noll2012atmospheric,jones2013advanced}. Emission from the atmosphere is indicated by the red shaded curve, assuming annual average temperatures, 1.5 mm PWV, and an airmass of 1.16. Additional emission from the telescope is shown for two different assumed values of emissivity, spanning typical values found in the literature ($\epsilon$= 0.07--0.17) \citep{wiedemann1996science,kaspar2017near,holzlohner2021bolometric}, and typical ambient temperature of 280 K. The total telluric thermal radiance is orders of magnitude greater than that received from the giant planets.\label{fig:midirtrans}}
\end{figure}

Between 5 and 30\,$\upmu$m, the atmospheric transmission is more variable and frequency dependent, with H$_2$O, CO$_2$, O$_3$, CH$_4$, and nitrous oxides contributing to the opacity \citep{gebbie1951atmospheric,taylor1957atmospherictrans,smette2015molecfit} (see Figure \ref{fig:midirtrans}). Strong absorption by CO$_2$ between 14 and 17\,$\upmu$m effectively blocks the atmospheric window near its center, and thus mid-infrared is typically divided into two sub-regions known as the N and Q bands in photometric systems. The precise ranges of these bands are not universally standardized, but the N band is typically recognized as ranging from roughly 8 to 14\,$\upmu$m, while the Q band extends between 17 and 25--27\,$\upmu$m  \citep{glass1999handbook, tokunaga2002infrared}. These bands are often divided further into various sub-bands for filtered imaging (e.g., Q1, Q2, Q3, etc.), naturally demarcated by the numerous absorption lines  \citep{thomas1976intermediate, glass1999handbook}. 

Additionally, corresponding to a narrow window of atmospheric transparency between 4.6 and 5.0\,$\upmu$m, the M band straddles the rough boundary between the near- and mid-infrared.  It has been grouped with the mid-infrared in at least some literature ({e.g.,}  \citep{kendrew2010mid}), although it is more commonly considered as a near-infrared band \citep{glass1999handbook}.

The gases in Earth’s atmosphere do not only absorb radiation---they also emit, with an emission spectrum characteristic of the atmospheric temperature and composition. Given the Earth’s effective temperature of 255 K, the atmosphere’s black body thermal emission peaks near 12~$\upmu$m (see Figure \ref{fig:midirtrans}). Likewise, the telescope itself inescapably emits thermal radiation corresponding to the observatory’s ambient temperature (typically 280--290 K at the VLT, for example \citep{noll2013cerro,holzlohner2021bolometric}) leading to an additional source of thermal radiation that also peaks in the N-band \citep{holzlohner2021bolometric}.  This combined telluric emission easily overwhelms the faint celestial emission from the colder, distant atmospheres of the outer planets.  One solution to this problem is to actively cool the instrument and to place the telescope above as much of the Earth atmosphere as possible, ideally well into space. However, when space is out of reach, observations from the ground are still possible over much of the mid-IR owing to specialized techniques developed by observers over the past century. 

The standard approach is to attempt to remove the thermal contribution of the sky and telescope by a process known as chopping and nodding \citep{papoularchopnod1983}. Chopping entails oscillating the telescope's secondary mirror at a frequency of several hertz, cycling on and off target in order to isolate and subtract the sky's thermal contribution from the total signal.  Likewise, nodding attempts to remove the residual, non-uniform emission from the telescope by alternating the telescope's pointing every few minutes. By this approach, measurements of Uranus' 13-$\upmu$m emission, for example, can be made from the ground despite being roughly 100,000 times fainter than the combined sky and telescope emission \citep{roman2020uranus}.  However, even this approach cannot overcome the atmosphere's considerable infrared opacity beyond the atmospheric window, and significant portions of the infrared spectrum (e.g., $\sim$5.5--8\,$\upmu$m, 13.5--17\,$\upmu$m, and 25--30\,$\upmu$m) remain inaccessible from the ground. 

\vspace{-18pt}

%This approach is  from dry mountain tops (e.g., Cerro Paranl and Maunakea) 

\subsection{Why We Observe in the Mid-Infrared}
\begin{figure}[H]
\includegraphics[width=1.0\linewidth,trim=.0in 0in 0.2in 0in]{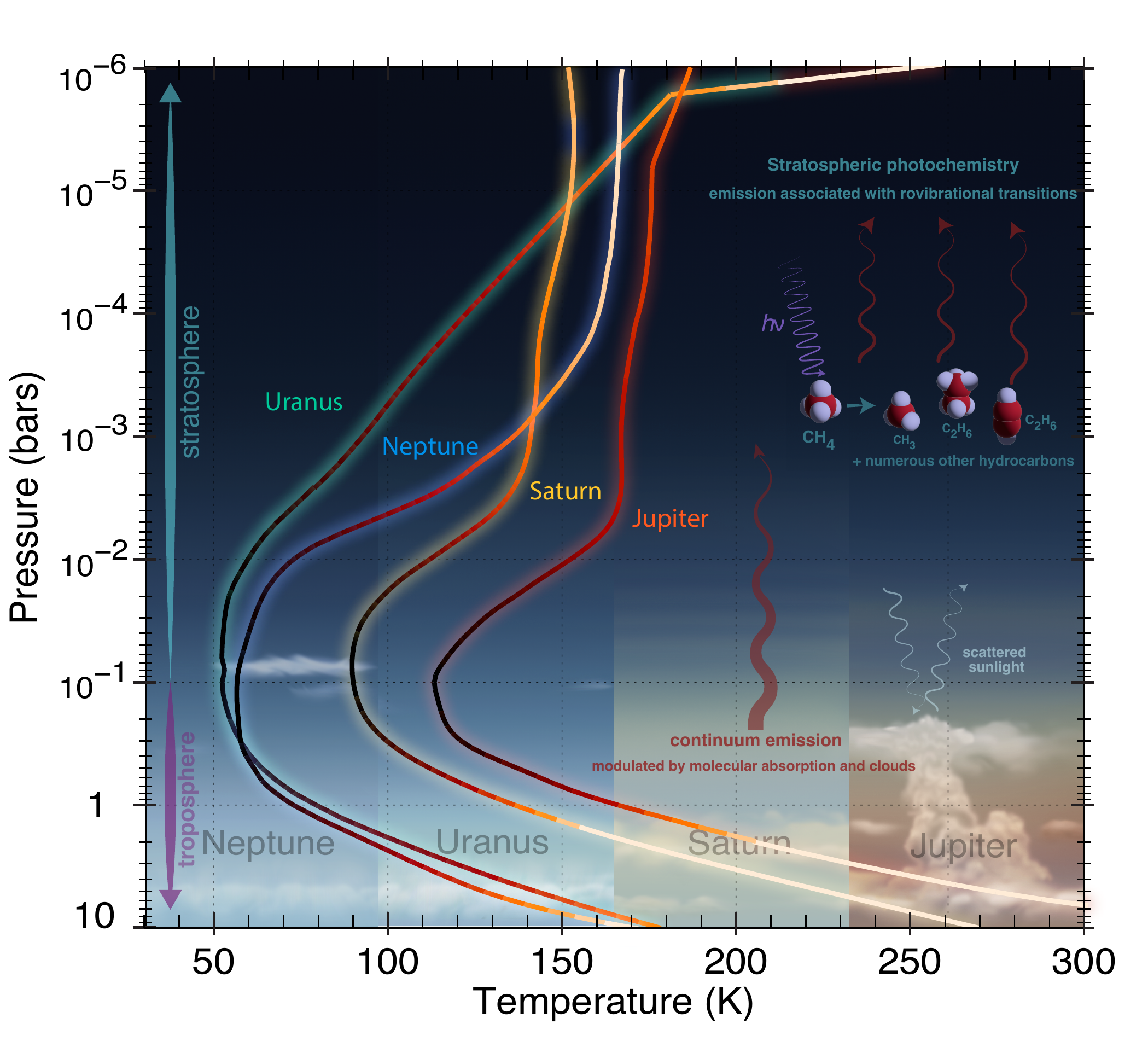}
\caption{Contributions to observed mid-infrared emission from the giant planet atmospheres. Vertical temperature profiles for Jupiter, Saturn, Uranus, and Neptune are shown for pressures ranging from 10 bar to 1 microbar. While scattered sunlight from aerosols weakly contributes at shorter wavelengths, the mid-IR is dominated by intrinsic thermal emission from the atmosphere.  The mid-IR emission originates from heights above the cloud layers, within the upper troposphere and lower stratosphere. Stratospheric emission is primarily associated with various stratospheric hydrocarbons that result from methane photochemistry.\label{fig:temprofs}}   %MDPI: Please change the hyphen (-) into minus sign ($-$, "U+2212"), e.g., "-1" should be "$-$1".
%MTR: On the y-axis?  Changed. 
\end{figure}

While observations of scattered sunlight at visible wavelengths define our most familiar views of the giant planets, they do not reveal a complete picture of the important processes that shape these atmospheres.  A complementary understanding of the atmospheric environment, within and above the clouds, can be achieved with infrared observations. Although scattered sunlight from aerosols can contribute to mid-IR radiances (particularly at shorter wavelengths), mid-IR is dominated by intrinsic emission from the atmosphere, indicative of temperature and composition (see Figure \ref{fig:temprofs}). 

With effective temperatures\footnote{In astronomy, the effective temperature relates the observed emission from an astronomical source to that of a perfect black body of known temperature. The term and application date back to at least the 19th century, where, for example, it was applied to estimate the surface temperature of the Sun  \citep{violle1874temperaturesun,langley1878temperaturesun}.  For giant planets with thick atmospheres, it generally corresponds to emission from the altitudes at which the atmospheric gases become opaque to infrared radiation---typically around 100 mbar.} of less than 125 K, the idealized black body emission of the Solar System's giant primarily emit energy in the infrared region of the electromagnetic spectrum.  The spectral radiant black body emission of Jupiter’s and Saturn’s peak in the mid-infrared, while emission from the colder atmospheres of Uranus’ and Neptune’s peak at longer wavelengths of the far-infrared (see Figure \ref{fig:midirtrans}).  In either case, considerable energy is radiated in the mid-infrared, and this thermal emission is relatively more accessible to observers on Earth's surface than that radiating in the far-infrared.  Understanding the temperature structure and energy budget of these giant planets, therefore, requires measurements of mid-infrared radiances. This idealized picture of the mid-infrared emission is, however, complicated---and greatly enriched---by the presence of radiatively active molecules, which profoundly alter the emission spectrum. 

The mid-infrared is home to rotational-vibrational transitions of numerous molecules found in the giant planet atmospheres, including CH$_4$, C$_2$H$_6$ C$_2$H$_2$, NH$_3$, PH$_3$, H$_2$O, C$_2$H$_4$, CH$_3$, GeH$_4$, AsH$_3$, C$_6$H$_6$, CO$_2$, and more  \citep{moses2004stratospherejupiter,moses2005photochemistry,fouchet2009saturn,moses2020icegiantchem}.  In spectroscopic observations, these state transitions show up as emission or absorption features, depending on the vertical temperature and chemical structure within the atmosphere. The intensity of spectral lines is dependent on both the abundance of the emitting or absorbing molecule and its ambient temperature\footnote{This is under the assumption of local thermodynamic equilibrium, which may be assumed generally valid at pressures corresponding to the tropospheres and lower-to-mid stratospheres. The ambient pressure of the molecule's environment also shapes the spectroscopic signature through pressure broadening}. If the ambient temperature is known, the molecular abundance can be inferred, typically by comparison of the observations with simulations from theoretical radiative transfer models (e.g., \citep{irwin2008nemesis}). Alternatively, if the molecular abundances are known, the observed spectrum can be used to constrain the temperature. The greater the spectroscopic resolution of the observations, the better the vertical resolution of the inferred temperatures or abundances. Imaging essentially provides an integrated radiance over a finite passband and, therefore, yields poorer vertical resolution (effectively vertically averaging), but it typically has the advantage of greater angular (spatial) resolution and radiometric sensitivity.

The detection and measurement of specific molecules can provide unique insight into processes active in giant planets’ atmospheres.  Some species are expected as a result of a solar composition atmosphere in thermodynamic chemical equilibrium (CH$_4$, for example  \citep{kuiper1949new,weidenschilling1973atmospheric}), but are nonetheless useful as indicators of temperatures, vertical structure, and circulation  \citep{orton1980saturn,conrath1983thermalSaturn,fletcher2009methane,fletcher2017jupiter}.   Others are unexpected given the ambient temperatures and bulk chemistry, and they require specific mechanisms to explain their abundances (for example, CO$_2$  \citep{burgdorf2006detection} and H$_2$O  \citep{feuchtgruber1997external} in Uranus atmosphere, implying external, meteoric sources).  

N-band (8--14 $\upmu$m) spectroscopy and imaging have been used in numerous investigations to infer temperatures, chemistry, and aerosol abundances in the troposphere and stratosphere of Jupiter (e.g., \citep{orton1975thermal,orton1991thermal,fletcher2010thermal,fletcher2016mid,fletcher2017moist, antunano2020characterizing}) and Saturn (e.g.,\citep{courtin1984compositionsaturn,noll1986abundances,sada199613c,fletcher2007meridional,fletcher2008temperature,hesman2009saturn,fletcher2009retrievals,fletcher2012originsaturnvortex,fletcher2018saturn,blake2021refining,blake2023saturn}). For Uranus and Neptune, the N band has been used to measure stratospheric emission associated with hydrocarbons (e.g., \citep{tokunaga1983new,orton1983observational,orton1992thermalneptune, fletcher2010neptune,greathouse2011spatially,orton2014mid,fletcher2014neptune,dePater2014neptune,roman2020uranus,roman2022subseasonal}), but interpretations have been limited by larger uncertainties in both temperatures and chemical abundances.

The dominant components of giant planets’ atmospheres are hydrogen and helium, the collision of which produces continuum absorption dependent on the pressure and temperature. Since the abundance of hydrogen and helium are homogeneous and relatively well constrained by the overall atmospheric density, this collision-induced absorption (CIA) provides a powerful, unambiguous indicator of the atmospheric thermal structure. Q-band observations (17--25\,$\upmu$m) are dominated by this absorption, and have thus successfully been used to infer atmospheric temperatures in the upper tropospheres of all the Solar System giant planets (e.g.,  \citep{orton1991thermal, fletcher2009retrievals, fletcher2010thermal, depater2010multijupiter,fletcher2014neptune,orton2014a,orton2015thermal,fletcher2018hexagon,fletcher2020jupiter,roman2020uranus,roman2022subseasonal,blake2023saturn}). Hydrogen \emph{emission} can also be found in the Q band, and this can additionally serve as a remote sensing thermometer of the stratosphere, as discussed in Section \ref{sec:chemtempprofs}.

\begin{figure}[H]
\includegraphics [width=.99\linewidth, trim=.0in .0in .0in .0in, clip]{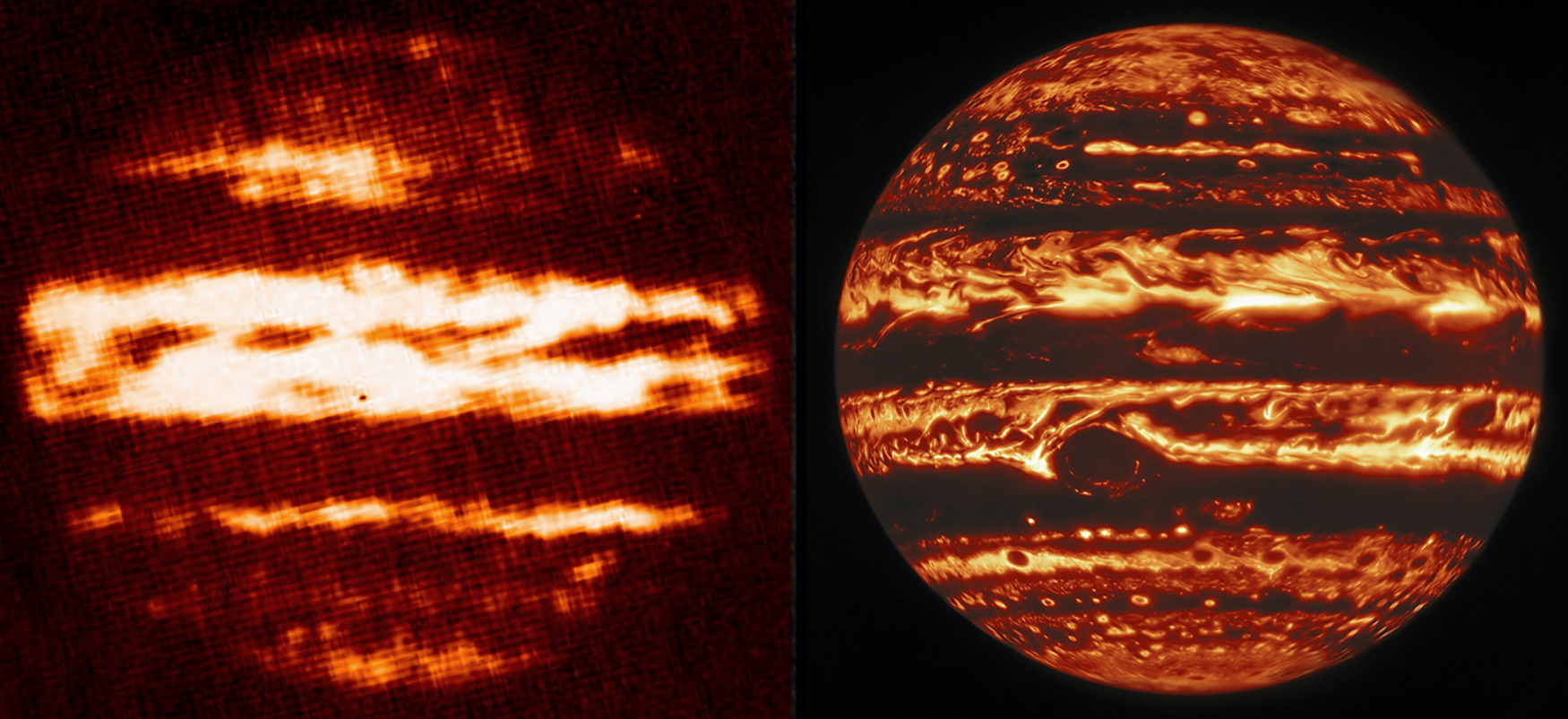}
\caption{Images of Jupiter at roughly 5\,$\upmu$m. Brighter regions indicate strong thermal radiation emerging from the atmosphere below the clouds, while darker patches reveal opaque clouds, silhouetted by underlying thermal emission.  The image on the left is one of the earliest examples of 5-$\upmu$m imaging, made with the Hale 200-in (5-m) telescope at Palomar Observatory in 1973  \citep{westphal1974five}.  The 4.7-$\upmu$m image on the right is from the Near-InfraRed Imager (NIRI) instrument \citep{hodapp2003gemini} at Gemini North in Hawai'i, composed of multiple images captured in 2017  \citep{gemini5micron}. The images reveal the dramatic improvement in the imaging quality, as well as changes in cloud structure over the past half-century.\label{fig:5micron}}
\end{figure}

Observations at roughly 5\,$\upmu$m have notably been used to study Jupiter's deep atmosphere  \citep{gillett1969jupspec,westphal1969observations,keay1973high,westphal1974five,fink1978germane,bjoraker1986gas,bjoraker2015jupiter,bjoraker2018gas, bjoraker2020jupiter}, producing striking,  high-contrast images of Jupiter's clouds silhouetted against the underlying thermal emission as shown in Figure \ref{fig:5micron}. (See  \citep{bjoraker2022spatial} for a review of 5-$\upmu$m imaging of Jupiter, and see \citep{wong2023deep} for a review of Jupiter's deep clouds). Saturn show less contrast at 5\,$\upmu$m owing to thicker, scattering hazes, but such observations have helped constrained vertical cloud structure and deeper chemistry  \citep{momary2006zoology,bjoraker2006ammonia,bjoraker2007saturn,fletcher2011saturn,barstow2016probing,yanamandra2015probing}.  As a consequence of their colder temperatures and weakly scattered sunlight, the Ice Giants have smaller radiances at 5\,$\upmu$m, and as a result, have largely been unexplored at this \mbox{wavelength  \citep{encrenaz2004first, fletcher2010neptune, orton2014a}}. JWST promises to provide the first detailed observations of the Ice Giants in this spectral region in the years ahead. It will be one of many observational breakthroughs that JWST promises in the mid-infrared, as it observes the giant planets with unprecedented precision and sensitivity. 

\section{A Historical Overview: Observing The Giant Planets in the Mid-Infrared }
Infrared characterization of the giant planets developed in parallel with the evolution of broader infrared astronomy in general.  It is marked by advances in theory, technology, and techniques that sparked new discoveries, inevitably prompting new theories, technologies, and techniques.  Unlike galactic and most stellar astronomy, however, planetary astronomy has the advantage of being able to apply remote sensing with relatively generous proximity, including very close encounters via robotic spacecraft. Robotic missions to the giant planets have afforded leaps in knowledge in recent decades, building upon and complementing a long history of ground-based observing. Beginning with basic measurements of effective planetary temperatures, mid-infrared investigations grew to provide critical insight into the chemistry, structure, and dynamics of the giant planets. 

Repeatedly over this observational history, Jupiter, by virtue of its superior size, proximity, and brightness, was naturally investigated first and most thoroughly. Successful investigations were then typically extended to Saturn shortly thereafter. Finally, Uranus and Neptune, owing to their great distances and cold temperatures, were investigated if and when even viable, consistently lagging many years behind the Gas Giants in thermal and chemical characterization.

\subsection{Beyond the Visible: Measuring Heat from the Giant Planets}
In the closing year of the 18th century, William Herschel demonstrated that radiant heating from the Sun extended beyond the red light of the visible spectrum \cite{herschel1800xiv}, arguably marking the birth of infrared astronomy.  Over the following century, quantitative investigations of this “invisible thermometrical spectrum'' and the infrared properties of materials developed alongside innovations in optics and instrumentation (e.g.,  see early reviews by~\cite{coblentz_review}).  As theory and tools developed, astronomers pushed their observations further into the uncharted infrared spectrum while aiming their instruments at increasingly fainter celestial targets. 

Beginning in the late 1850s, the first successful measurements of the “non-luminous” radiation from the Moon's surface were made using telescopes equipped with sensitive early thermopiles, which converted observed radiative energy into electrical energy that was then read as needle deflection on a galvanometer. Increasingly sensitive radiometers were subsequently developed, including the Langley bolometer\footnote{
Giving rise to an enduring limerick, dubiously attributed to Langley's student:\begin{center}“Prof. Langley devised a Bolometer.\\
It's really a sort of Thermometer.\\
It'll detect the heat\\
Of a Polar Bear's feet\\
At a distance of Half-a Kilometer.''  \citep{pasachoff1985contemporary,walker2000briefhistory}\end{center}} in 1878  \citep{langley1880bolometer}, which has found continued use in modern submillimeter instruments  \citep{holland2002bolometers} (e.g., Herschel-PACS  \citep{balog2014herschel}).  Lacking spectrometers with dispersive prisms and gratings tuned to the infrared  \citep{coblentz_review,stroke1967diffraction,palik1977history}, early observers simply used glass filters, transparent to visible light but opaque to thermal radiation, to remove and isolate the thermal component from the total observed radiation.  This filtering approach provided ratios of relative band radiances,  leading to the first (somewhat disputed) estimates for the extreme diurnal range of lunar surface temperatures  \citep{rosse1870radiation,langley1889temperature,boys1890iii, very1898probable1,very1898probable2}. 

\begin{figure}[H]
\includegraphics[width=.99\linewidth]{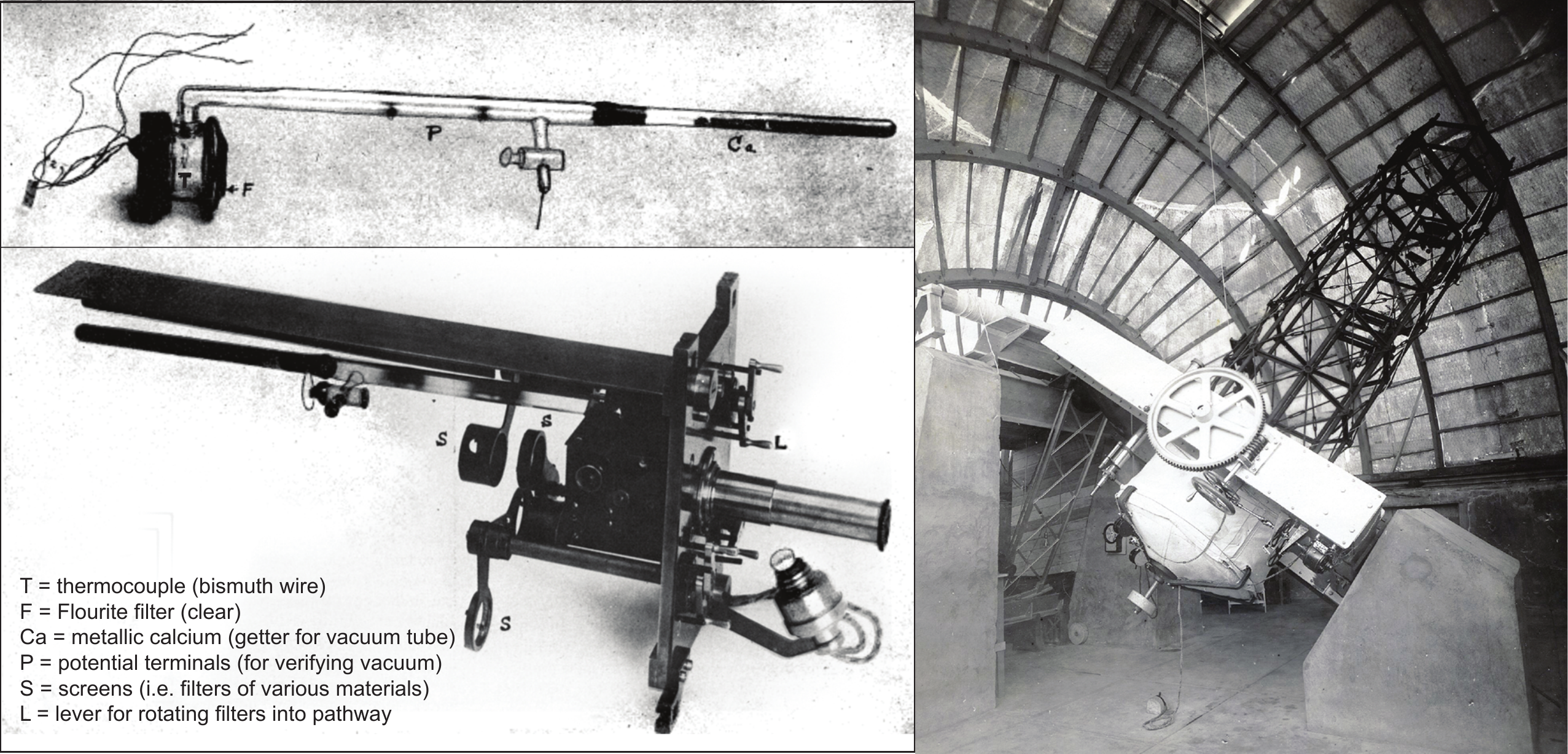}
\caption{Photographs of the Coblentz-Lampland radiometer, used to make groundbreaking measurements of thermal from stars and the giant planets  \citep{coblentz1923measurments}. Top left: Evacuated glass tube containing the thermocouple---a wire of bismuth that converts temperature differences to electric voltage via the thermoelectric effect. The thermocouple is kept in an evacuated tube, with the vacuum maintained by the presence of reactive calcium metal (serving as a getter). Bottom left: The thermocouple is placed into the radiometer, which was fastened to the photographic plate holder of the telescope. The thermocouple was placed in the optical path between the target and eyepiece, while filters of different passbands were selectively rotated in and out of view. By these means, coarse spectra could be inferred.  Right: The Lampland 40-inch telescope of the Lowell Observatory, on which the radiometer was mounted for much of Coblentz's planetary work, inside its dome, ca. 1909 (\textit{Image credit: Slipher, E.C. , “The 42-inch Lampland Telescope inside of its dome,” Lowell Observatory Archives, \url{https://collectionslowellobservatory.omeka.net/items/show/1047}}, accessed on 22 December 2022). \label{fig:radiometer}} %MDPI: please add access date (day month year); MTR: Added.
\end{figure}

By the early 20th century, improved radiometers and observing techniques were combined with larger reflecting telescopes to provide the first quantitative estimates of thermal emission from the planets. In pioneering work by Coblentz and \mbox{Lampland  \citep{coblentz1922further,coblentz1923measurments,coblentz1924new,coblentz1925some,coblentz1927temperatures}} and Pettit and Nicholson  \citep{pettit1923measurements,pettit1924radiation}, observations of Venus, Mars, Jupiter, Saturn, and Uranus were made between 1914 and 1924 using sensitive new radiometers and a series of filters \footnote{According to \cite{coblentz1925some,menzel1926planetary}, a water cell 1 cm in thickness transmits radiation that lies
between 0.3 and 1.4 $\upmu$m, while the filters of quartz, glass, and fluorite are transparent up to 4 $\upmu$m, 8 $\upmu$m, and 12.5 $\upmu$m, respectively.} in order to separate observed radiances into five discrete spectral bands ranging between 0.3 $\upmu$m to 15 $\upmu$m\footnote{Although, in reality, atmospheric transmission drops significantly beyond 13 $\upmu$m, with very little transmission between 14~$\upmu$m and 15 $\upmu$m; see Figure \ref{fig:midirtrans}.}, extending progressively further into the mid-infrared (see Figure \ref{fig:radiometer}). The combination of filtered observations thus provided the first rough spectra of the giant planets. Analysis of these spectral data revealed Jupiter and Saturn to have temperatures (at the effective emission layers) of 120--140 k and 125--130 K, respectively, while Uranus was colder yet, with an upper limit of 100 K  \citep{menzel1923water,menzel1926planetary}---not far from modern estimates\footnote{Modern estimates of the effective temperatures: 124.4  $\pm$ 0.3, 95.0  $\pm$ 0.4, and 59.1 $\pm$ 0.3 for Jupiter, Saturn, and Uranus, respectively \citep{Astrophysical_Quantities2002}.} of the planetary effective temperatures. These measured temperatures indicated the giant planets were cold---not much warmer than expected for equilibrium with the solar heating---and therefore contributed to evidence that the low density of the outer planets could only be explained by a bulk composition rich in hydrogen  \citep{menzel1930hydrogen}.

Over the following decades, improvements in technology and technique continued.  Advances in mid-infrared bandpass filters and gratings (i.e., those that transmit only in the mid-infrared) enabled improved calibration of stars and planets by allowing for the direct comparison with known blackbody cavities at the telescope  \citep{sinton1960radiometric}. Errors due to the drift in detector response and changing sky radiance were minimized by shifting the sensor on and off target at high frequency  \citep{pfund1929resonance,kivenson1948infra,sinton1960radiometric}---an approach that evolved into the chopping and nodding technique still used today to remove the thermal signal of the sky and telescope  \citep{westphal_ir_200in_1963,papoularchopnod1983}. By the 1960s, photometric systems utilizing mercury-doped germanium detectors cooled by liquid hydrogen allowed for increased sensitivity in the mid-infrared spectral region  \citep{westphal_ir_200in_1963}. 

\begin{figure}[H]
\includegraphics[width=.99\linewidth]{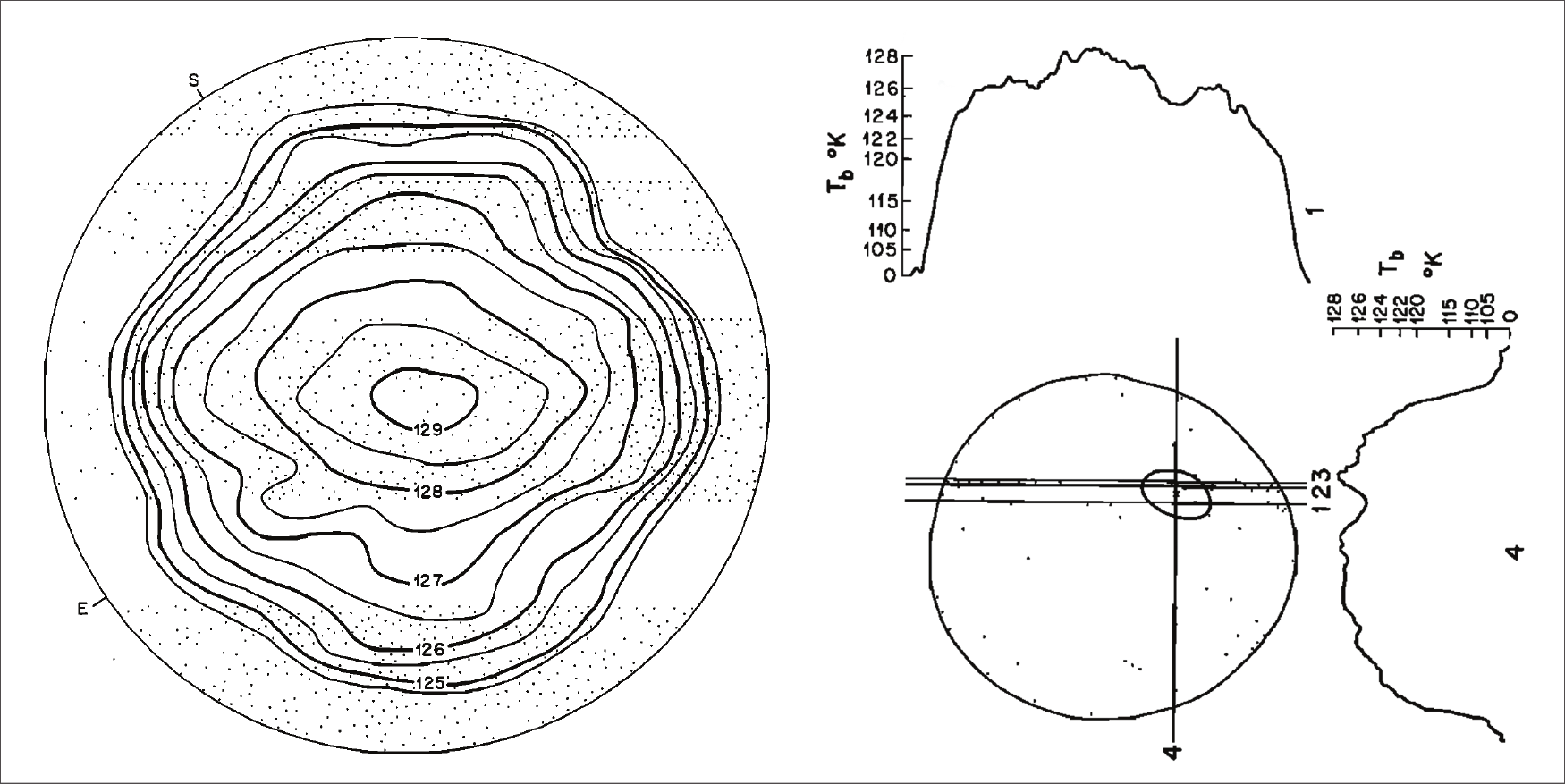}
	\caption{Among the earliest maps of Jupiter's mid-infrared radiances, reproduced from Murray et~al.   \citep{murray1964observations}. Left: Contours show brightness temperatures derived from 8 to 14 $\upmu$m observations, made over five nights in mid-December 1963 using the Hale 200-inch (5.1 m) Telescope of the Palomar Observatory. The contours indicate modest limb-darkening and early hints of possible zonal thermal structure, with the visible belt-zone structure superimposed (oriented south pole upwards). Right: Perpendicular lines represent scans passing through the Great Red Spot (GRS). Corresponding brightness temperature curves (as numbered just to the right and above) show a depression in temperature at the location of the GRS.\label{fig:jupiter1964}}
\end{figure}

Utilizing the new detectors and techniques, observations in the early 1960s provided the first truly spatially resolved photometry of the thermal radiation emitted by a giant planet. Beginning in 1962, radiances across the disk of Jupiter were measured at 8--14 $\upmu$m using the Palomar Observatory Hale 200-inch telescope---the world's largest telescope at the time.  These spatially resolved data revealed thermal limb-darkening indicative of temperatures increasing with depth; temperature contrast ($\sim$0.5\,K) between the warmer darker belts and the cooler brighter zones \cite{murray1964observations}; and that the Great Red Spot (GRS) was 1.5--2.0 K cooler than the surrounding disk (see Figure \ref{fig:jupiter1964}). 

Soon after, the first observations of Saturn and Uranus at 10 $\upmu$m  \citep{low1964infraredsaturn} and \mbox{17--25~$\upmu$m}  \citep{low1966observations,low1966infrareduranus} were made and flux calibrated by comparison to recently defined photometric standard stars. These observations yielded a 20-$\upmu$m brightness temperature and 95~$\pm$~3~K for Saturn, roughly consistent with modern values, and 55 $\pm$ 3~K for Uranus, which established “the current lower limit to the brightness temperature of a celestial object which can be measured in the infrared”  \citep{low1966infrareduranus}.  The opacity of Earth's atmosphere limited the infrared spectrum that can be obtained from the ground, leading observers to seek greater heights.

Observations from airborne observatories began in the 1960s, with rockets  \citep{harwit1966results, houck1975jupiter}, balloons  \citep{danielson1966stratoscope,strong1965infraredballoon}, and jets  \citep{aumann1969internaljet,armstrong1972far} rising above Earth's moist lower atmosphere. A 12-inch telescope flown on a modified Lear jet (NASA 701) in 1968 captured thermal radiances from Jupiter and Saturn using a series of broad filters with bandpasses sampling the spectrum between 1.5 $\upmu$m to 350 $\upmu$m.  Analysis of these brightness temperatures showed that both Jupiter and Saturn radiated roughly twice as much energy as they receive  \citep{aumann1969internaljet, armstrong1972far}\footnote{The energy balances---ratios of emitted to received radiation---for Jupiter and Saturn have been revised down to 1.7 and 1.8, respectively, following Voyager measurements  \citep{pearl1991albedo}; however, following Cassini, the balance for Jupiter was raised to 2.1  \citep{li2018jupiterenergy}, while Saturn was shown to be seasonally variable  \citep{li2010saturnpower}}.

Neptune's thermal emission was finally measured years later, when, in 1972, observations were made from the newly constructed high-altitude observatory on Maunakea. At over 4,200-m above sea level, the observatory's altitude sits above a majority of the Earth's attenuating water vapor, permitting observations further into the mid-infrared.  Both Uranus and Neptune were observed between 17 and 28 $\upmu$m using a liquid-helium-cooled bolometer mounted on a 2.24-m  telescope  \citep{simon1972lowbolometer,morrison1973temperaturesUandN}. Surprisingly, it was discovered that Neptune had a brightness temperature of 57.2  $\pm$  1.6 K at 24-$\upmu$m ---warmer than that of Uranus (54.7  $\pm$  1.6 K) despite Neptune's greater distance from the Sun  \citep{morrison1973temperaturesUandN,rieke1974infrared,macy1977detection}. Combined with observations in the visible, far-infrared, and millimeter wavelengths, this led to the conclusion that Neptune radiates excess heat---$2.4^{+1.3}_{-0.9}$ times as much power as it absorbs  \citep{murphy1974evidence}---similar to Jupiter and Saturn.  Uranus was evidently the outlier, as the only giant planet apparently lacking an internal heat source. 

Observations continued to improve over the following decades, refining these initial temperature measurements with effective temperatures constrained at longer and longer wavelengths, including far-infrared  \citep{wright1976recalibration,loewenstein1977effective}, sub-millimeter  \citep{whitcomb1979submillimeter}, millimeter  \citep{epstein1970mars,ulich1973planetary,werner1978one}, and microwave  \citep{kellermann1970thermal,mayer1971microwave} wavelengths.  Spatially resolving the temperature structure on the Ice Giants had to wait for spacecraft encounters and larger telescopes in the following decades. Meanwhile, the spectral resolution was quickly improving from the ground, allowing for the detection of discrete spectral signatures  \citep{connes1966near, larson1980infrared}. 

\subsection{A New Window into the Giant Planets' Atmospheric Composition}

In the 1970s, the focus of mid-infrared planetary studies arguably shifted from temperatures to chemistry. Until then, the detection and measurement of the atmospheric composition had been investigated primarily in the visible and near-infrared for the better part of a century  \citep{huggins1890vii,draper1879photographing,slipher1909spectra,adel1934constitution,wildt1932absorptionsspektren,dunham1933note,kuiper1949new,kiess1960hydrogen}, but such observations had only succeeded in spectroscopically identifying molecular hydrogen (H$_2$) and methane (CH$_4$) in the giant planets, plus ammonia (NH$_3$) in Jupiter and Saturn.  Based on assumptions of solar-composition in chemical and adiabatic equilibrium, the newly-constrained atmospheric temperatures of the planets were used to predict theoretical abundances of several hundred volatile compounds throughout Jupiter's atmosphere  \citep{lewis1969observability,weidenschilling1973atmospheric}, while photochemical models were predicting disequilibrium of stratospheric hydrocarbons such as ethane (C$_2$H$_6$) and acetylene (C$_2$H$_2$) due to photolytic destruction of CH$_4$  \citep{wildt1937photochemistry, strobel1969photochemistry}. The mid-infrared provided a promising new window to potentially detect these molecular signatures via fundamental rovibrational and pure rotational transitions. 

\textls[-15]{In 1973, excess radiance at 11--14-$\upmu$m---as seen in moderate resolution (R $\sim$ 50--66) spectra of Jupiter \cite{gillett1969jupspec} from the ground at $\sim$2,500-meter (8,200-ft) altitude---was correctly identified as the first evidence of stratospheric C$_2$H$_6$ and C$_2$H$_2$ on Jupiter, enhanced by an atmospheric temperature inversion \citep{ridgway1974jupiter, tokunaga1976ethane}, confirming photochemical model \mbox{predictions  \citep{wildt1937photochemistry, strobel1969photochemistry}}}. This \mbox{12-$\upmu$m} C$_2$H$_6$ enhancement was also seen in the spectrum from Saturn the following year~\citep{gillett1974saturn}. In light of these discoveries, previous observations of  Uranus' exceptionally weak 12--13-$\upmu$m emission were insightfully reinterpreted as evidence that Uranus was not necessarily colder, but potentially contained less stratospheric ethane than Neptune  \citep{rieke1974infrared}. We now know colder temperatures and lower ethane abundances \emph{both} contribute to Uranus' relatively weak 12--13-$\upmu$m emission compared to Neptune \citep{moses2020icegiantchem}. 

%Following these discoveries, the relatively weaker 12--13-$\upmu$m emission previously observed on Uranus relative to Neptune was insightfully interpreted as evidence that Uranus was either colder \emph{or} contained less stratospheric ethane than Neptune  \citep{rieke1974infrared}---we now know that \emph{both} factors are responsible  \citep{moses2020icegiantchem}. 

Following from theory and techniques applied in the analysis of terrestrial satellite \mbox{data  \citep{wark1969atmospheric,ohring1973temperature}}, spectral inversion techniques were at this time being developed for the giant planets in order to infer vertical temperature profiles and chemical \mbox{abundances  \citep{taylor1972temperature, rodgers1976retrieval}}. In particular, Taylor  \citep{taylor1972temperature} showed that measurements of the $\nu_4$ branch of CH$_4$ (at $\sim$7.74\,$\upmu$m) could be inverted to provide temperature profiles for relatively warm Jupiter and possibly Saturn.  For colder Uranus and Neptune, collision-induced rotational S(0) absorption by hydrogen at 25--40 $\upmu$m could be used to infer temperature profiles. Indeed, measurements of S(0) and S(1) collision-induced H$_2$ absorption were successfully used to retrieve upper tropospheric temperature structure in the giant planets from Voyager-IRIS spectra decades later  \citep{conrath1998thermal}. 

An instrumental leap forward came with the advent of high-resolution Fourier Transform Spectrometers (FTS) in the late 1960s  \citep{connes1966near,connes1968carbon,connes1969atlas}, which allowed for greater spectral resolution (R>>500) at longer wavelengths. With the promise of further discoveries already evident in modest-resolution high-altitude observations  \citep{gillett1969jupspec}, high-resolution mid- and far-infrared spectroscopy rapidly emerged, opening a window to new molecules and greater constraints on atmospheric composition and vertical temperature structure. 

Spectroscopy in the decade that followed yielded the first detections of CH$_3$D  \citep{beer1973abundance}, 13-NH$_3$  \citep{encrenaz1978spectrumnh3}, H$_2$O  \citep{larson1975detectionh2o}, PH$_3$  \citep{ridgway1976ph3,larson1977phosphine}, GeH$_4$  \citep{fink1978germane}, and CO  \citep{beer1975detectionco,larson1978evidenceco} in the atmosphere of Jupiter. Given that PH$_3$, GeH$_4$, and CO are not thermodynamically stable at the low temperatures and pressures at which they were detected, their presence suggested strong vertical mixing from below producing tropospheric disequilibrium chemistry  \citep{prinn1975phosphine,prinn1977carbon}. Similarly, CH$_3$D  \citep{fink1978deuteratedsaturn} and PH$_3$  \citep{bregman1975observationph3saturn, larson1980middle} were detected in Saturn's atmosphere, along with conclusive evidence of stratospheric C$_2$H$_6$  \citep{tokunaga1975c2h6detection} and tentative detection of C$_2$H$_4$  \citep{encrenaz1975tentativec2h4}. NH$_3$ was also found  \citep{encrenaz1974abundancenh3}, but at a factor of at least 20 less than on Jupiter, consistent with Saturn's colder temperature and deeper condensation levels.  Likewise, disequilibrium GeH$_4$ on Saturn was not detected until a decade later  \citep{noll1988evidencesaturngeh4}. With even greater distances and colder temperatures, the chemistry (and temperature structure) of Uranus and Neptune remained almost unconstrained in the mid-infrared until their encounter with Voyager 2. 

\subsection{Remote Sensing Up Close: Missions to the Giant Planets}
Beginning in the 1970s, robotic spacecraft missions to the giant planets permitted infrared remote sensing of the giant planets at relatively close proximity without attenuation from the Earth's atmosphere. Infrared radiometers on Pioneer\,10 and Pioneer\,11 flew by Jupiter in 1973 and 1974, respectively,  \citep{chase1974pioneer}, equipped with broadband filters (11- and 26-$\upmu$m-wide) centered at roughly 20\,$\upmu$m and 40\,$\upmu$m, respectively.  Though broadly filtered in wavelength, the spatially resolved measurements provided new, stronger constraints on the energy balance  \citep{ingersoll1975pioneer,ingersoll1976results} and thermal structure of Jupiter  \citep{orton1975thermal}.  Similarly, Pioneer\,11 observed Saturn in 1979, providing similar refinements of Saturn's thermal structure and energy balance  \citep{orton1980saturn,ingersoll1980pioneersaturn}, before continuing out towards interstellar space. The first to encounter Jupiter and Saturn, the Pioneer missions were envisioned as precursors to a more ambitious Mariner program mission to the giant planets, later renamed as the Voyager Program. 

Launched in 1977, Voyager\,1 and Voyager\,2 carried the Infrared Interferometer Spectrometer and Radiometer (IRIS) experiment---arguably the most fruitful infrared instrumentation in the history of solar system exploration. A combination of three instruments, IRIS included a Michelson interferometer that operated in the infrared from 2.5\,$\upmu$m to 55\,$\upmu$m (180 and 2400\,cm$^{-1}$) with a spectral resolution of R$\sim$42--558 (4.3\,cm$^{-1}$), in contrast to the Pioneer radiometer's filters. 

\textls[-15]{Voyager\,1 reached Jupiter in March 1979, followed four months later by Voyager\,2. Initial findings from these observations included refined estimates of the effective temperature and energy balance  \citep{hanel1981albedojupiter}; improved measurements of meridional thermal structure and cold anomaly of the Great Red Spot (GRS)  \citep{hanel1979infraredjupitervoy1}; confirmation of excess thermal emission near Jupiter's north magnetic pole  \citep{kim1985infrared} new constraints on the ammonia cloud density and particle sizes  \citep{marten1981studyammonia}; new constraints on the chemical abundances  \citep{hanel1979infraredjupitervoy1}, including that of helium  \citep{gautier1981heliumjupiter}; and the first detection of several new hydrocarbons, including C$_2$H$_4$, C$_3$H$_4$, and C$_6$H$_6$  \citep{kim1985infrared}.  Similarly, IRIS placed new constraints on the temperature structure and chemistry of Saturn during the Voyager 1 and Voyager 2 encounters in 1980 and 1981, respectively,  \citep{pirraglia1981thermal,conrath1983thermalSaturn,conrath1984helium,courtin1984compositionsaturn,hanel1983albedosaturn,bezard1984seasonal,conrath2000saturn}. Following the Saturn encounters, Voyager 1 began its extended mission on course to depart the solar system, while Voyager 2 continued onward towards the Ice Giants. }

The subsequent Voyager\,2 flybys of Uranus in 1986 and Neptune in 1989 marked watershed moments in the exploration of the outer planets. With unprecedented spatial resolution and phase-angle coverage, Voyager substantially improved constraints on the Bond albedos, effective temperatures, thermal structure, and energy balances  \citep{pearl1990albedo,pearl1991albedo, conrath1987helium,conrath1991helium} of both planets, confirming that Uranus was indeed anomalous in its lack of interior heat. In particular, the spatial resolution allowed the upper-tropospheric temperature structure of both planets to be mapped for the first time, revealing relative cold anomalies (2--4 K) at mid-latitudes compared to the warmer low and high latitudes  \citep{smith1989voyager}. This latitudinal structure was interpreted as evidence of mid-latitude upwelling and resulting adiabatic cooling as part of a meridional circulation cell, compensated by downwelling at the equator and poles  \citep{smith1989voyager,conrath1989neptune,conrath1990temperature}.  New constraints were also placed on the helium abundances of both planets  \citep{conrath1987helium, conrath1991helium}, although stratospheric hydrocarbons remained poorly constrained due to insufficient instrument sensitivity at wavelengths less than 25 $\upmu$m. Nonetheless, the Voyager 2 flybys of the Ice Giants remain to be the only close encounter with these distant worlds and the definitive account of their temperature structure. 

Notably, the IRIS observations also allowed for the first measurement of ortho-para hydrogen disequilibrium in the outer planets. Pressure-induced H$_2$ absorption at $\sim$17 $\upmu$m and $\sim$27 $\upmu$m result from transitions in ortho-H$_2$ and para-H$_2$ energy levels, respectively, and the ratio of these absorption features are theoretically dependent on \mbox{temperature  \citep{smith1978ortho,fletcher2018hydrogen}}. Conrath and Gierasch found ortho-para fractions were not in equilibrium with the retrieved temperatures on Jupiter, particularly at the equator, implying upwelling from warmer depths  \citep{conrath1983evidence}. Combined with implied zonal wind shear inferred from thermal wind relations, these observations provided powerful new insight into the atmospheric circulation on the giant planets  \citep{pirraglia1981thermal,gierasch1986zonal,conrath1998thermal, fletcher2014neptune, orton2015thermal}. 

Following the success of Voyager, the Galileo orbiter examined the Jupiter system over the course of 35 orbits between 1995 and 2003  \citep{bagenal2007jupiter, fischer2001mission,hunten1986atmospheric,russell2012galileo}. On-board instruments included the Near-Infrared Mapping Spectrometer (NIMS)  \citep{carlson1992NIMS}, which observed from 0.7 to \mbox{5.2 $\upmu$m}, and the Photopolarimeter-Radiometer (PPR) experiment  \citep{russell1992galileoPPR}, which observed in five mid-infrared spectral bands between 15 and 100 $\upmu$m.  The NIMS spectra, combined with contemporaneous visible imaging, found evidence of deep water clouds  \citep{nixon2001NIMSjupiter} and showed that most, \textit{but notably not all}, bright clouds blocking thermal emission extended vertically to the upper troposphere  \citep{irwin1997radiative,dyudina2001interpretation,irwin2002retrieval}.  The PPR was used to derive the \mbox{200--700-mbar} temperature field of the Great Red Spot (GRS) using four discrete mid-infrared filters centered on 15, 22, 25, and 37 $\upmu$m. These filtered data showed that the GRS was roughly 3 K colder than regions to its east and west, consistent with Voyager and previous investigations  \citep{orton1996galileo}. 

While Galileo was still in orbit around Jupiter, the next great flagship mission to the outer planets was already en route to Saturn. The Cassini-Huygens spacecraft launched in 1997, beginning its two-decade-long journey of exploration  \citep{matson2002cassini, dougherty2009saturn}. It observed Jupiter over a period of roughly six months, reaching its closest approach in December 2000 at just under 10 million kilometers  \citep{hansen2004cassini}, before entering orbit around Saturn in July 2004. 

\begin{figure}[H]
	\includegraphics [width=.9\linewidth, trim=.0in .0in .0in .0in, clip]{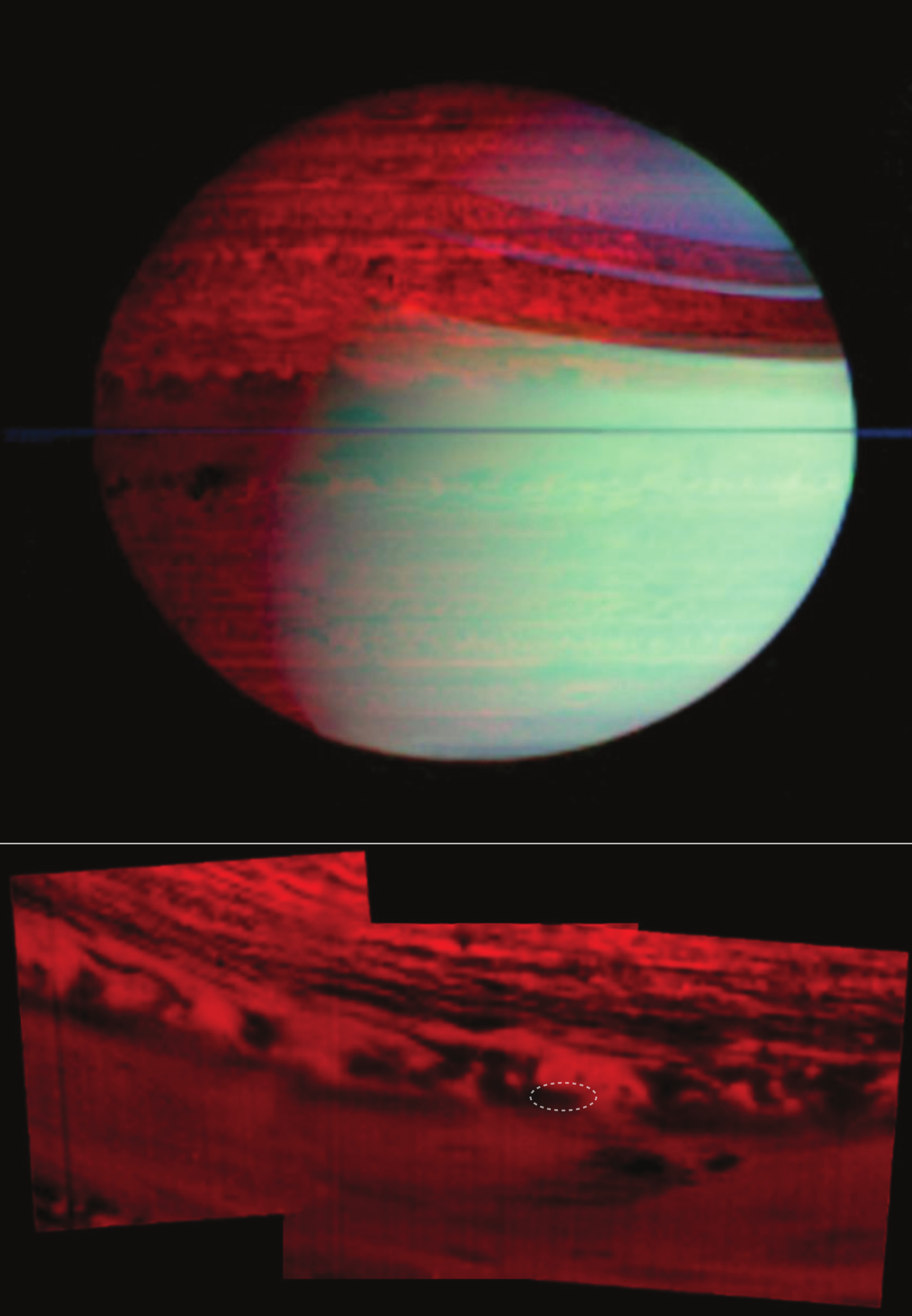}
	\caption{Images of Saturn from Cassini-VIMS. Top: False-color mosaic of Saturn from February 2006 showing thermal infrared radiation at 5.02-$\upmu$m (in red) and scattered sunlight at 1.07\,$\upmu$m and 2.71\,$\upmu$m (in blue and green, respectively). Discrete clouds appear silhouetted against the glow of Saturn's thermal emission at 5-$\upmu$m, while the rings cast a shadow upon Saturn's northern hemisphere. Bottom: The last images from VIMS, captured on 14 September, 2017, as the spacecraft made its final descent towards Saturn.  Thermal emission at 5 $\upmu$m appears brighter where the cloud opacity is less. The dotted ellipse marks the approximate location where the Cassini spacecraft soon thereafter entered into the atmosphere, concluding the mission. \textit{Image credits: NASA/JPL-Caltech/University of Arizona} \label{fig:saturn_vims}}
\end{figure}

The Cassini spacecraft was equipped with two state-of-the-art instruments sensitive to the infrared: the Cassini Visual and Infrared Mapping Spectrometer (VIMS)  \citep{brown2004cassini} and the Composite Infrared Spectrometer (CIRS)  \citep{flasar2004exploring,jennings2017composite}. VIMS was an improved successor to Galileo-NIMS  \citep{carlson1992NIMS} (even inheriting some of its mechanical and optical parts from the original NIMS engineering model  \citep{brown2004cassini}). As an imaging spectrometer, it produced spectra for each pixel (or \emph{spaxel}) in an image.  It was composed of a visible and infrared channel, allowing for measurements from the ultraviolet to the edge of the mid-infrared (0.3--5.1\,$\upmu$m). By simultaneously sensing both near-infrared scattering and thermal emission, VIMS allowed for new constraints on Saturn's cloud opacity and composition  \citep{sromovsky2010source,fletcher2011saturn,sromovsky2018vims,sromovsky2021evolution} (see Figure \ref{fig:saturn_vims}). For observations at longer wavelengths, Cassini's CIRS instrument was \mbox{used  \citep{flasar2004exploring,jennings2017composite}}.
Unlike Galileo's infrared instrument (PPR), Cassini's CIRS was a proper spectrometer. Following the FTS principles used since the 1960s, CIRS was composed of a mid-infrared Michelson interferometer and a far-infrared polarizing interferometer that could together provide spectra from 7.1 to 1000 $\upmu$m at a spectral resolution that could be set between 0.5 and 15.5 cm$^{-1}$.  %---7.1 to 16.7 $\upmu$m (600--1400 cm$^{-1}$) and 16.7–1000 $\upmu$m (10 and 600 cm$^{-1}$), respectively---at a spectral resolution that could be set between 0.5 and 15.5 cm$\^{-1}$. 

With its unprecedented spectral coverage, CIRS observations of Jupiter provided new constraints on temperature structure  \citep{simon2006jupiter}, energy balance  \citep{li2012emittedcirsjupiter}, cloud structure and composition  \citep{matcheva2005cloud,wong2004identification,bjoraker2018gas}, and chemical abundances, including that of NH$_3$  \citep{achterberg2006cassini}, PH$_3$  \citep{irwin2004retrievals, fletcher2009phosphine}, C2H2  \citep{nixon2007meridional} and C2H6  \citep{nixon2007meridional}, the D$/$H ratio  \citep{pierel2017d2hratios}, halides  \citep{fouchet2004upper}, and trace hydrocarbons  \citep{nixon2010abundances, sinclair2019jupiter}. Then, from its unrivaled vantage point in orbit around Saturn for more than 13 years, CIRS revolutionized our understanding of Saturn's seasonally variant chemistry and thermal structure  \citep{flasar2004exploring,fletcher2007characterising,fletcher2010seasonal,guerlet2011evolution,sinclair2013seasonalchem, sylvestre2015seasonal, fletcher2016seasonal,fletcher2018hexagon}. It placed the new and improved constraints on numerous molecular and isotopic abundances  \citep{hesman2009saturn,guerlet2010meridional,fletcher2009methane,guerlet2009vertical,teanby2006new,howett2007meridional,fletcher2009phosphine,hesman2012elusive,hurley2012latitudinal,abbas2013co2distribution,koskinen2016detection,pierel2017d2hratios,koskinen2018atmospheric}. The CIRS observations of Saturn remain the definitive measurements of the planet at mid-infrared wavelengths, and largely define our current knowledge of Saturn's temperature and chemistry (see  \citep{ingersoll2020cassini} for a comprehensive review).

\subsection{From High Above the Atmosphere: Observations from Space Telescopes}
While robotic spacecraft missions were venturing far into the outer Solar System, new discoveries were being made relatively closer to home with a series of space-borne telescopes. Though modest in size compared to ever-larger ground-based telescopes, these versatile observatories were unencumbered by telluric absorption, possessing a sensitivity only possible in the coldness of space.  

The Infrared Space Observatory (ISO) was the first such space observatory to make great contributions in mid-infrared (and far-infrared) observations of the giant planets.  Operated from 1995 to 1998, it was equipped with the Short Wave Spectrometer (SWS)---a scanning spectrometer sensing from 2.35 to 45.4 $\upmu$m with grating resolutions between 930 and 2450 ($\lambda$/$\Delta\lambda$) and a higher resolution Fabry–Pérot mode (20,600--31,000)~\mbox{ \citep{leech2003iso,sloan2003uniform}}. The combination of high spectral resolution and coverage led to the new detection of several molecules on all four giant planets  \citep{encrenaz1996first,bezard1998detection}, although Uranus and Neptune proved too faint to be observed below 7 $\upmu$m. Discoveries included the detection of water vapor in Saturn's troposphere at 5 $\upmu$m \citep{encrenaz1996first}; detection of new hydrocarbons (e.g.,  CH$_3$C$_2$H and C$_4$H$_2$) in Saturn's stratosphere  \citep{de1997first}; detection of stratospheric CO$_2$ $\nu_2$ bands on \mbox{Saturn  \citep{de1997first}}, Jupiter \citep{lellouch1997h_2o} and Neptune  \citep{feuchtgruber1997external}; and the first detection of methyl (CH$_3$)---a molecule diagnostic of the height to which methane is mixed---in the stratospheres of Saturn  \citep{bezard1998detection} and Neptune  \citep{bezard1999detection}. Numerous discoveries were also made at longer wavelengths with the Long Wavelength Spectrometer (LWS). See Encrenaz et al.,  \citep{encrenaz1999atmospheric} for an excellent summary. 

ISO was followed by the Spitzer Space Telescope, launched in 2003  \citep{werner2004spitzer}.  Sensing from 5.2 to 38 $\upmu$m with low (R$\sim$60--130) and moderate (R$\sim$600) resolution spectroscopy, the Spitzer-Infrared Spectrograph (IRS)  \citep{houck2004infrared} observed Neptune on four occasions between 2004 and 2006  \citep{meadows2008first,rowe2021neptune}, and Uranus in 2004  \citep{burgdorf2006detection} and 2007  \citep{orton2014a,orton2014mid}, near the time of the planet's equinox.  With a primary mirror of 0.85 m, Spitzer, like ISO, was not able to spatially resolve the Ice Giants' disks, but the observations nonetheless led to strong new constraints on the planets' disk-averaged temperature structure  \citep{orton2014a,rowe2021longitudinal,rowe2021neptune} and chemistry  \citep{orton2014mid}.  The observations yielded the first detections of C$_2$H$_6$ and possibly CH$_3$ on Uranus and the first detections of methylacetylene (C$_3$H$_4$) and diacetylene (C$_4$H$_2$) in both Ice Giants  \citep{burgdorf2006detection, meadows2008first}. 

Finally, it is worth noting that the new JWST promises to far surpass these previous mid-infrared space observatories and provide the definitive mid-infrared spectra of the giant planets. The Mid-Infrared Instrument (MIRI)  \citep{rieke2015mid} is capable of providing spatially resolved (integral field unit) spectra from 5 to 28 $\upmu$m with resolving powers from 1300 to 3700 ($\lambda$/$\Delta\lambda$). The telescope successfully launched on  25 December 2021 and is expected to be operational for 20 years.  All four giant planets will be observed in the first two years following launch.  With superior sensitivity and spatial resolution, the results are anticipated to greatly advance our understanding of Uranus and Neptune, in particular. 

\subsection{Matured Mid-Infrared Observing from the Ground}
\begin{figure}[H]
	\includegraphics [width=1.0\linewidth, trim=.0in .0in .0in .0in, clip]{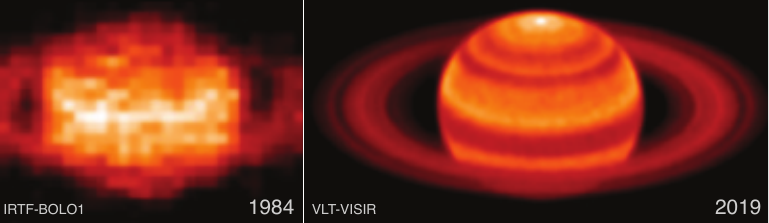}
	\caption{Improvements in mid-IR imaging, as illustrated by an early image of Saturn acquired with the IRTF-BOLO1 instrument in 1984 (left) compared to a recent image from the VLT-VISIR instrument in 2019  \citep{blake2023saturn}. \label{fig:saturn_comparison}}
\end{figure}

Back on the ground, improvements in detectors, telescopes, and observing techniques advanced ground-based observations to a quality rivaling spacecraft observations (e.g., see Figures \ref{fig:5micron} and \ref{fig:saturn_comparison}). The early contour maps of Jovian brightness temperatures from \mbox{Palomar  \citep{murray1964observations}} gave way to raster-scanned maps from the NASA Infrared Telescope Facility (IRTF) in the 1980s and 1990s \cite{orton1991thermal,Gezari1989Nature}, followed by the first modern 2-D array detectors in the 1990s\footnote{A notable example of this early 2-D array was the Mid-Infrared Array Camera (MIRAC), a 20 $\times $64-pixel Si:As IBC detector sensitive to radiation from 2 to 26\,$\upmu$m, made by a collaboration of the University of Arizona,
Smithsonian Astrophysical Observatory, and Naval Research Laboratory  \citep{hoffmann1993mirac,Hoffman1994In})}. By the mid-2000s, numerous mid-infrared instruments were in operation on 8-meter class telescopes, including:
Long Wavelength Spectrometer (LWS)  \citep{jones1993keck} at Keck; Michelle  \citep{glasse1997michelle} at Gemini North; The VLT Imager and Spectrometer for Mid-Infrared (VISIR)  \citep{lagage2004visir} at the Very Large Telescope (VLT); the Thermal-Region Camera Spectrograph (T-ReCS) at Gemini South,  \citep{deBuiz2005Trecs}; and the Cooled Mid-Infrared Camera and Spectrometer (COMICS)  \citep{kataza2000comics} at Subaru. Typically, planetary observations with these instruments have applied narrow-band filters covering spectral ranges between 8 and 13 $\upmu$m (the N-band) and 17 to 25 $\upmu$m (the Q-band), from which chemistry and/or temperatures were retrieved  \citep{fletcher2010thermal, orton2015thermal,fletcher2017moist,roman2020uranus,roman2022subseasonal,blake2023saturn}. Additionally, such observations were frequently used to complement contemporaneous spacecraft observation, providing greater spatial or temporal coverage than possible from orbit  \citep{fletcher2011thermalsaturn,fletcher2012originsaturnvortex,fletcher2017jupiter,fletcher2018hexagon}. % An example of the improvement in image quality over time can be seen in the montage of observations in Figure X.  Examples of the mid-infrared imaging of the planets in different spectral bands is shown in Figure X. 

In terms of spectroscopy, a notable workhorse of ground-based remote sensing at mid-infrared wavelengths over the past two decades is the Texas Echelon Cross Echelle Spectrograph (TEXES)  \citep{lacy2002texes}. Capable of the spectral resolving power of 15,000 to 100,000 ($\lambda$/$\Delta\lambda$) in windows between 5 and 25 $\upmu$m, TEXES has been used to great effect on IRTF and Gemini North to map chemistry and temperatures in Jupiter  \citep{fletcher2014origin,sinclair2018jupiter,fletcher2018jupiter,melin2018assessing,blain2018mapping,sinclair2020spatial}, \mbox{Saturn  \citep{greathouse2005meridional,moses2005latitudinal,greathouse2006first,fouchet2016stratospheric}}, and to a lesser extent Uranus  \citep{trafton2012mid,orton2019spatial} and Neptune  \citep{greathouse2011spatially,roman2022subseasonal}, with the exceptionally high spectral resolution needed to resolve fine lines. The resulting quality of retrieved maps of temperature, composition, and aerosols have been noted to even surpass previous spacecraft results for Jupiter  \citep{fletcher2016mid}.

Of the aforementioned mid-IR instruments, only VLT-VISIR and TEXES remain in operation as of 2023. Given the significant and unique information provided by mid-infrared ground-based observations, it can only be hoped that these continue to serve the community until the next generation of instruments is developed, at least.  

Looking ahead, promising future mid-infrared instruments to include a mid-infrared imager and spectrometer called MIMIZUKU (Infrared Multi-field Imager for Gazing at the UnKnown Universe)  \citep{kamizuka2012development}, developed for the planned 6.5-m telescope of the University of Tokyo Atacama Observatory (TAO), currently under construction in the Chilean Atacama at a remarkable 5640-m altitude  \citep{miyata2022university}. MIMIZUKU will cover a wavelength range of 2 to 38\,$\upmu$m with a spectral resolution of $\lambda$/$\Delta\lambda$ $\sim$60--230 and diffraction-limited (wavelength-dependent) angular resolution of 0.077--1.47 arcseconds.  This spatial resolution is exceptional by current far-infrared standards, although it will not surpass the current leading resolution of the larger VLT across much of the mid-IR (e.g.,\, TAO-MIMIZUKU's 0.7$''$ diffraction-limited resolution versus VLT-VISIR's 0.55$''$ resolution at 18\,$\upmu$m). The larger disks of Jupiter and Saturn will, therefore, be particularly well suited for MIMIZUKU, but all the Solar System's giant planets will benefit from its exceptionally broad spectral range, innovative technical design  \citep{kamizuka2022challenging}, and long-term monitoring capabilities in the years ahead.  MIMIZUKU has already seen its first light, having been successfully tested on the Subaru Telescope in 2018  \citep{kamizuka2020university}.  

Looking even further ahead, the European Southern Observatory's planned 39.3-m Extremely Large Telescope (ELT) first-generation instruments will include the Mid-infrared ELT Imager and Spectrograph (METIS)  \citep{brandl2012metis}. METIS promises to provide diffraction-limited imaging and medium resolution slit-spectroscopy from 3 to 13\,$\upmu$m (covering the M and N bands), as well as high resolution (R$\sim$100,000) integral field spectroscopy (IFU) from 2.9 to 5.3\,$\upmu$m  \citep{brandl2022status}. N-band imaging will be capable of an amazing 6.8-mas (milli-arcsecond) angular resolution over a 13.5$''$ $\times$ 13.5$''$ field of view (FoV).  The high spatial resolution and narrow FoV will make the instrument ideally suited for observing the small disks of Uranus and Neptune, while mosaicking or regional targeting will be required for Jupiter and Saturn. Likewise, the even narrower FoV of the M-band IFU (0.58$''$ $\times$ 0.93$''$) will be optimal for analyzing small-scale, 5-$\upmu$m atmospheric features with unprecedented resolution from the ground.  With METIS' first light expected in 2028  \citep{brandl2022status}, the complementary capabilities of MIMIZUKU, VISIR, and METIS promise exciting advances in mid-infrared observations from the ground over the next decade.

\section{What We Have Learned}
From more than a century of mid-infrared remote sensing, a picture of the general atmospheric thermal structure and chemistry of the giant planets has emerged. For Jupiter and Saturn, the picture can appear quite intricate, with complex structure, unexplained variability, and puzzling correlations across different heights and hemispheres. By comparison, our pictures of Uranus and Neptune in 2023 are little more than rough sketches, lacking details but nonetheless challenging our understanding of temporal variation in the outer solar system. 

\begin{figure}[H]
	\includegraphics [width=0.9\linewidth, trim=.1in .1in .3in .0in, clip]{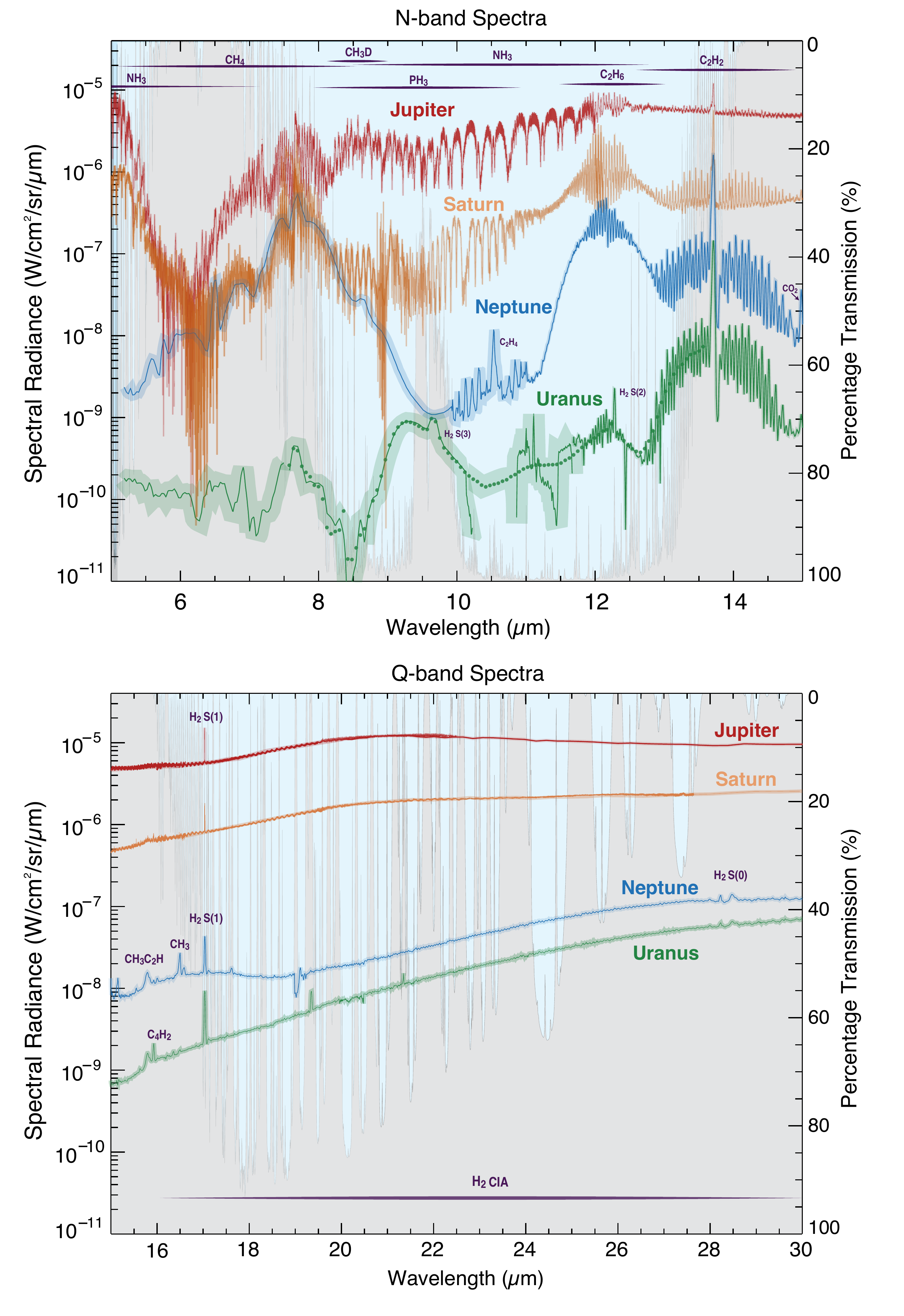}
	\caption{Observed mid-infrared spectra of giant planets in the N- and Q-bands (top and bottom panels, respectively).  The spectra of Jupiter (red) and Saturn (orange) are from ISO-SWS~\citep{encrenaz1996first} and Cassini-CIRS  \citep{li2012emittedcirsjupiter,li2010saturnpower}, while Uranus (green) and Neptune (blue) are disk-averaged radiances from Spitzer-IRS  \citep{orton2014a,rowe2021longitudinal,rowe2021neptune,roman2022subseasonal}. The rough uncertainty of the spectra (most evident for Uranus) is suggested by the faint transparent envelopes.  Select emission features are indicated, and the wavelengths at which different gases broadly contribute to spectra are indicated by the labeled horizontal lines (purple). The atmospheric transmission is indicated by the blue--gray interface varying between 100\% (full transmission) and 0\% (total attenuation) from the top of the atmosphere down to a surface, as in Figure \ref{fig:midirtrans}. \label{figspec}}
\end{figure}

\begin{figure}[H]
	\includegraphics [width=1.\linewidth, trim=.1in .0in .1in .0in, clip]{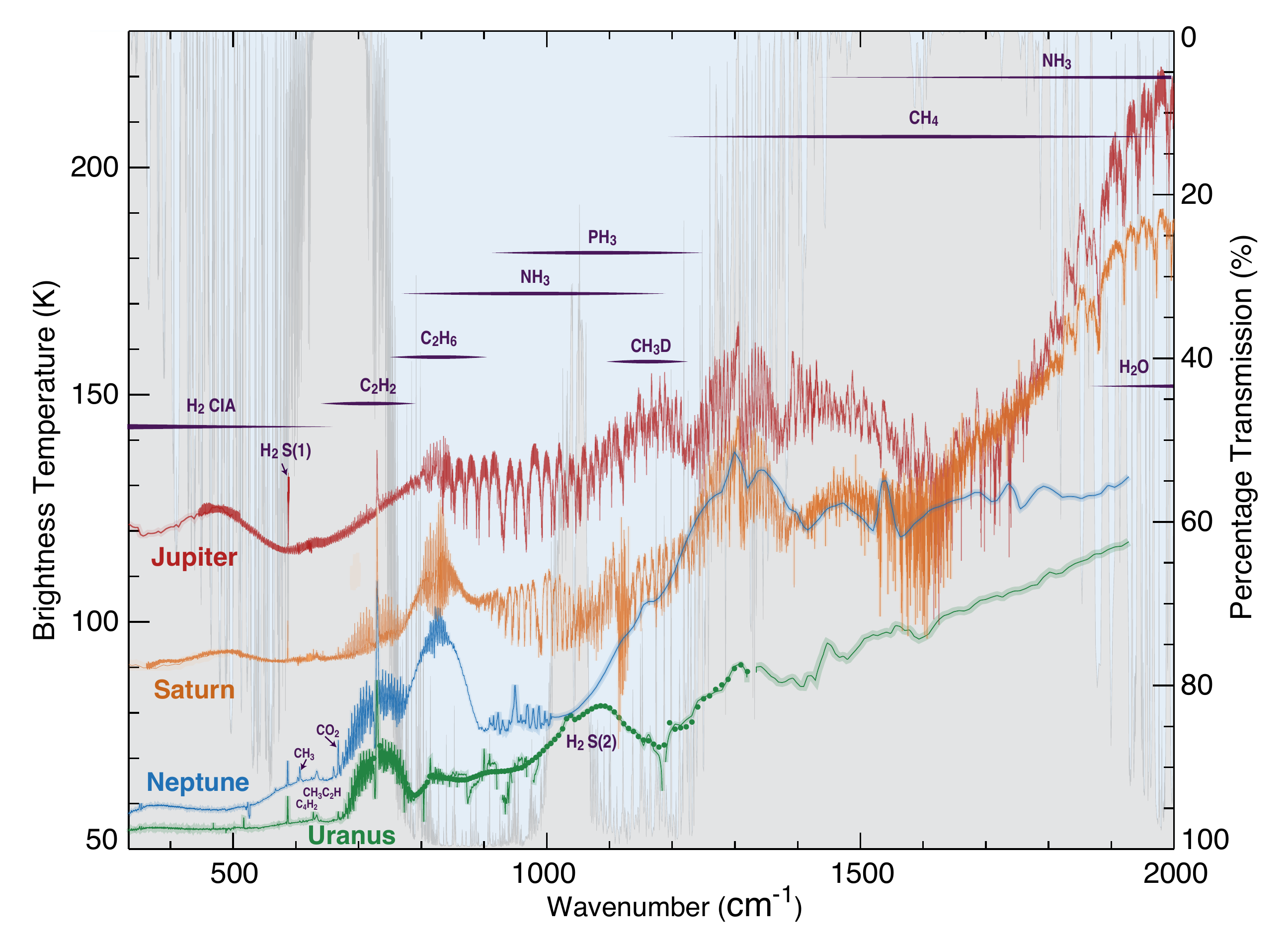}
	\caption{As in Figure \ref{figspec}, mid-infrared spectra of the giant planets, but now expressed in brightness temperature versus spectroscopic wavenumber. \label{fig:figspecwn}}
\end{figure}

Figures \ref{figspec} and \ref{fig:figspecwn} compare the observed mid-infrared spectra of the giant planets derived from ISO-SWS  \citep{encrenaz1996first} and Cassini-CIRS  \citep{li2012emittedcirsjupiter,li2010saturnpower} for Jupiter and Saturn, and Spitzer-IRS~\mbox{ \citep{orton2014a,rowe2021neptune,roman2022subseasonal}} for Uranus and Neptune. Figure \ref{fig:imgs} compares ground-based images in three key mid-infrared windows. 

\subsection{Chemistry and Temperature from Mid-IR Spectra}\label{sec:chemtempprofs}

\subsubsection{5--6 $\upmu$m}
From 5 to 6 $\upmu$m, scattered light and thermal emission contribute to the spectrum, modified by gaseous absorption. On Jupiter and Saturn, NH$_3$ and H$_2$O are the primary absorbers  \citep{encrenaz1996first,de1997first,drossart1998saturn}. Measurements of NH$_3$, have been used to provide insights into the accretion stage of the planets' formation histories. Analyses of the nitrogen ratios (at 5 to 6 $\upmu$m and $\sim$10--11 $\upmu$m) indicate identical values of these isotopic ratios for both Jupiter and Saturn, suggesting a similar history of primordial N$_2$ accretions during the formation of each planet  \citep{fouchet2004upper,fletcher2014origin}.  Likewise, the water abundance is important because oxygen is potentially telling of the carbon-to-oxygen (C/O) ratio, which is seen as diagnostic of the planet's formation history in the solar nebula  \citep{oberg2011effects,mousis2012nebular,lunine1989abundance}.  The quest for Jupiter’s and Saturn's deep water abundances has been a challenge since the mid-IR cannot sense well below the H$_2$O condensation level on either planet  \citep{taylor2004composition}. The Galileo probe ({in situ}) and Juno (microwave radiometer) have aimed to resolve this value for Jupiter, but uncertainties remain due to the inhomogeneous nature of Jupiter's atmosphere. A proper discussion is beyond the scope of this review, but see  \citep{bjoraker2020jupiter,bjoraker2022spatial}.

On Uranus and Neptune, this region of the spectrum was too weak to be observed by ISO-SWS, and even Spitzer-IRS spectra are in doubt  \citep{orton2014a,rowe2021longitudinal,rowe2021neptune}. The high opacity of Earth's atmosphere, particularly around 6 $\upmu$m, makes these observations impractical from the ground.  Observations with JWST-MIRI should provide the first comprehensive examination of this spectral region.

%\begin{figure}[H]
%\includegraphics [width=.97\linewidth, trim=.0in .0in .0in .0in, clip]{Figures/figure_mid_ir_spectra_15-30.pdf}
%\caption{Observed mid-infrared spectra of giant planets in the N- and Q-bands (top and bottom panels, respectively).  The spectra of Jupiter (red) and Saturn (orange) are from \textit{ISO-SWS} \citep{encrenaz1996first} and \textit{Cassini-CIRS}  \citep{li2012emittedcirsjupiter,li2010saturnpower}, while Uranus (green) and Neptune (blue) are disk-averages from \textit{Spitzer-IRS}  \citep{orton2014a,rowe2021neptune,roman2022subseasonal}. Telluric atmospheric transmission is faintly shown  (purple), scaled from opaque to transparent (bottom to top), for typical conditions on Maunakea. \label{figspec}}
%\end{figure} 

\vspace{-9pt}

\subsubsection{6--15 $\upmu$m}
From 6 to 15 $\upmu$m, the spectra are shaped by numerous strong emission and absorption features against a backdrop of the hydrogen-helium continuum emission from around the tropopause (roughly 100 mbar). On Jupiter and Saturn, absorption is produced by NH$_3$, PH$_3$, and H$_2$O, while CH$_3$D (at $\sim$9 $\upmu$m) and deeper CH$_4$ absorption is found in the spectra of all four giant planets. 

PH$_3$ is a disequilibrium species in the cold upper troposphere of Jupiter and Saturn, and its presence indicates vigorous vertical mixing on time scales less than that of chemical conversion  \citep{bregman1975observationph3saturn, larson1980middle,giles2017latitudinal}. It has yet to be detected on Uranus and Neptune  \citep{moreno2009search}. The spatial distribution reveals latitudinal variation in mixing, as discussed in Section \ref{sec2d}. 

\textls[10]{Combined measurements of CH$_4$ and CH$_3$D have been used to estimate the D/H ratio of the planets, providing powerful clues as to their formation history in the solar \mbox{nebula  \citep{lecluse1996deuterium}}. From theory, Jupiter and Saturn are expected to have D/H ratios consistent with the solar nebula, from which they derived most of their mass;  Uranus and Neptune, however, should have higher D/H ratios if they formed from proportionately larger, deuterium-rich icy cores.  Measurements have shown that D/H ratios on Uranus and Neptune are indeed a factor of a few larger than those of Jupiter and \mbox{Saturn  \citep{griffin1996first, feuchtgruber1999detection,lellouch2001deuterium, fletcher2010neptune, orton2014mid,pierel2017d2hratios,rowe2021longitudinal,rowe2021neptune}}. }

Nearly all the emission lines between 6 and 15 $\upmu$m are from stratospheric hydrocarbons, primarily CH$_4$ (peaking at 7.7 $\upmu$m), C$_2$H$_6$ ($\sim$ 12 $\upmu$m), C$_2$H$_2$ (13 to 15 $\upmu$m).  C$_2$H$_6$, C$_2$H$_2$, and other minor hydrocarbons (including methyl radicals (CH$_3$), ethylene (C$_2$H$_4$), methylacetylene (CH$_3$C$_2$H) and diacetylene (C$_4$H$_2$)),  are the result of photochemistry in the stratospheres of the giant planets  \citep{moses2020icegiantchem}. Methane from the troposphere is mixed up into the stratosphere, where it is then broken down by ultraviolet radiation, prompting a chain of chemical reactions that result in a mélange of new hydrocarbons  \citep{wildt1937photochemistry,strobel1969photochemistry,moses1992hydrocarbon,moses2005photochemistry, moses2018seasonal,moses2020icegiantchem}. Estimates of the abundances of these hydrocarbons have been used to infer vertical mixing within the atmospheres and constrain seasonal-chemical models of their formation and destruction, e.g.,  \citep{moses2018seasonal}. Emission from CH$_4$ has been used to infer stratospheric temperatures on Jupiter and Saturn since it is considered uniformly well mixed in the warm atmospheres of the Gas Giants  \citep{taylor1972temperature,greathouse2005meridional,fletcher2007characterising}, whereas it cannot necessarily be used as a thermometer on Uranus and Neptune given that colder temperatures are expected to condense methane and alter the distribution  \citep{greathouse2011spatially}. However, hydrogen is well mixed in all these atmospheres, and the H$_2$ S(0), S(1), S(2), S(3), and S(4) quadrupole emissions contribute at observed radiances roughly 28, 17, 12, 9.7, 8 $\upmu$m, respectively, to varying degrees. The S(2) and S(3) lines are weakly emitted from pressures near 1 $\upmu$bar, and though they are detected in the Spitzer-IRS observations of Uranus  \citep{rowe2021longitudinal}, they are generally lost in the forest of ethane lines on the other planets. The H$_2$ (S1) and H$_2$ (S0), observed at longer wavelengths, are most easily measured and have proven the most useful for evaluating temperatures and ortho-para fractions, as discussed below.

\textls[-15]{The relatively intense spectra of Jupiter and Saturn at wavelengths beyond $\sim$9 $\upmu$m is telling of their relatively warmer upper-tropospheric temperatures, as inferred from the earliest observations of these planets  \citep{coblentz1922further}. This can be seen in typical temperature profiles derived from spectra (see Figure \ref{fig:temprofs}). Neptune, however, appears relatively bright at \mbox{7--8 $\upmu$m}---comparable to Saturn and indicative of Neptune's surprisingly warm and methane-rich stratosphere. The large methane abundance is generally interpreted as evidence that Neptune has particularly strong vertical mixing, while Uranus is particularly \mbox{stagnant  \citep{rieke1974infrared,moses2018seasonal,moses2020icegiantchem}}. The stratospheric methane mole-fraction ((1.15 $\pm$ 0.10) $\times$ 10$^{-3}$ \mbox{ \citep{lellouch2015new,fletcher2010neptune}}) is greater than the expected value limited by the colder temperatures of the underlying tropopause (i.e., the cold-trapped minimum)   \citep{smith1989voyager,baines_UV_neptune_1990,baines1994clouds}. Moist convection has been discussed as a possible explanation for the stratospheric methane enhancement  \citep{stoker1986moist,lunine1989abundance, sinclair2020spatial}. Alternatively, another possible avenue for transferring methane from the troposphere to the stratosphere, despite the cold trap, was suggested following discoveries from thermal imaging.  Images from ground-based imaging show the south pole of Neptune to be warmer at the tropopause and lower stratosphere than elsewhere on the planet  \citep{orton2007evidencehotspot}.  Orton et al.  \citep{orton2007evidencehotspot} proposed that methane could potentially be seeping up from the troposphere at the warm pole before spreading to lower latitudes, avoiding cold-trapping. However, evidence of meridional transport or strong stratospheric methane gradients has yet to be found  \citep{greathouse2011spatially,roman2022subseasonal}.  Furthermore, the excess methane and potential associated hydrocarbon hazes are still not enough to explain the high stratospheric temperatures of Neptune, which exceed that expected from radiative heating models  \citep{appleby1986radiative,friedson1987seasonal,conrath1990temperature,marley1999thermal,li2018high}.  Additional modeling is necessary to explain these observations. }

The comparison of the planets' spectra at 12--14 $\upmu$m also reveals a striking difference between Uranus and the other giant planets. Uranus appears anomalously faint, with a conspicuous absence of C$_2$H$_6$ emission.  Modeling of the stratospheric photochemistry has suggested that this is a consequence of Uranus' apparently weak vertical mixing, which results in meager lower-stratospheric methane abundances (1.6 $\times$ 10$^{-5}$) and a lower-altitude homopause (7 $\times $ 10$^{-5}$ bars). This limits methane and hydrocarbon photochemistry to relatively higher pressures, where the dominant hydrocarbon reactions and loss rates differ.  With less CH$_4$ in the stratosphere, C$_2$H$_6$ is also less shielded and more easily photolyzed. This results in relatively lower ethane abundances (1.3  $\times$  10$^{-7}$ at 0.2 mbar)  \citep{orton2014mid} compared to Jupiter (2.08 $\times$ 10$^{-5}$  \citep{melin2020jupiter}, Saturn (9  $\times$ 10$^{-6}$  \citep{sinclair2013seasonalchem}), and Neptune (8.5 $\times$ 10$^{-7}$  \citep{fletcher2010neptune}).

%(1.3 $\times$ 10$^{-7}$) 1.3 $\times $ 10$^{-7}$ at 0.2 mbar

\subsubsection{15--30  {${\upmu}$}m}
Finally, from 15 to 30 $\upmu$m, the spectrum is dominated by the hydrogen-helium continuum emission from the upper troposphere and lower stratosphere.  At these wavelengths, the differences in radiances between the planets clearly express the relative temperatures around the tropopause ($\sim$40--200 mbar) (see Figure \ref{fig:temprofs}).  Uranus, with its apparent weak internal flux and vertical mixing of solar-absorbing methane, is overall coldest at these pressures, despite being nearer to the Sun than Neptune. Several small emission features can also be seen, including CO$_2$ on both Jupiter  \citep{feuchtgruber1997external,lellouch2002originh2oco2} and Saturn \cite{de1997first} at 14.98 $\upmu$m; CH$_3$C$_2$H (methylacetylene) and C$_4$H$_2$ (diacetylene) at 15.80 and 15.92 $\upmu$m, respectively, on all giant planets  \citep{de1997first,encrenaz1997giant,orton2014a,orton2014mid,rowe2021longitudinal,rowe2021neptune}; likewise CH$_3$ has been detected at 16.5 $\upmu$m, although only tentatively for Uranus  \citep{bezard1998detection,orton2014mid,rowe2021longitudinal}. Retrieved CH$_3$ on Jupiter and Saturn have been shown to be inconsistent with predicted values based on theoretical eddy diffusivity and CH$_3$ recombination rates  \citep{bezard1998detection}. Subsequent analysis of TEXES spectra also revealed a 3 $\times$  greater abundance of CH$_3$ in Jupiter's polar regions  \citep{sinclair2020spatial} than predicted by photochemical models  \citep{moses2017dust}. These inconsistencies suggest a need for additional sources of CH$_3$ production or uncertainties in chemical rates  \citep{sinclair2020spatial}, and the topic remains an area of active research. 

%\begin{figure}[H]
%\includegraphics[width=.97\linewidth,trim=.0in .5in .1in 0in]{Figures/tempprofs_giant_planets_rfs2.pdf}
%\caption{  \label{fig:temprofs}}
%\end{figure} 

Standing out among the emission features are the H$_2$ (S1) and H$_2$ (S0) hydrogen quadrupoles, observed at roughly 17 and 28 $\upmu$m.  These lines are unambiguously sensitive to the lower stratospheric temperatures within a larger continuum that is sensitive to the \textit{ortho} and \textit{para} fractions  \citep{smith1978ortho, conrath1990temperature,fletcher2018hydrogen}. Retrievals exploiting the H$_2$ S(1) quadrupole have been particularly important for the Ice Giants, where methane emission cannot be used as an unambiguous proxy for stratospheric temperature owing to its potentially variable distribution. Several studies have used the H$_2$ (S1) line to determine lower stratospheric temperatures and, combined with the H$_2$ and He continuum emission, derive vertical temperature profiles  \citep{greathouse2011spatially,orton2014a,rowe2021longitudinal,roman2022subseasonal}. Notably, the H$_2$ (S1) has also been used to confirm that Neptune's enhanced polar stratospheric emission and its changes in time are due primarily to variations in temperatures, as discussed in the next sections  \citep{roman2022subseasonal}. 

\subsection{Structure and Dynamics from Spatially Resolved Mid-IR Spectra and Imaging}\label{sec2d}

As current exoplanetary investigations demonstrate, a rich amount of atmospheric data can be inferred from an unresolved target  \citep{seager2010exoplanet,barstow2020outstanding}. However, constraining many of the processes shaping a three-dimensional atmosphere---often in unanticipated ways---requires observations to characterize the spatial structure.

\begin{figure}[H]
	\includegraphics[width=1.0\linewidth]{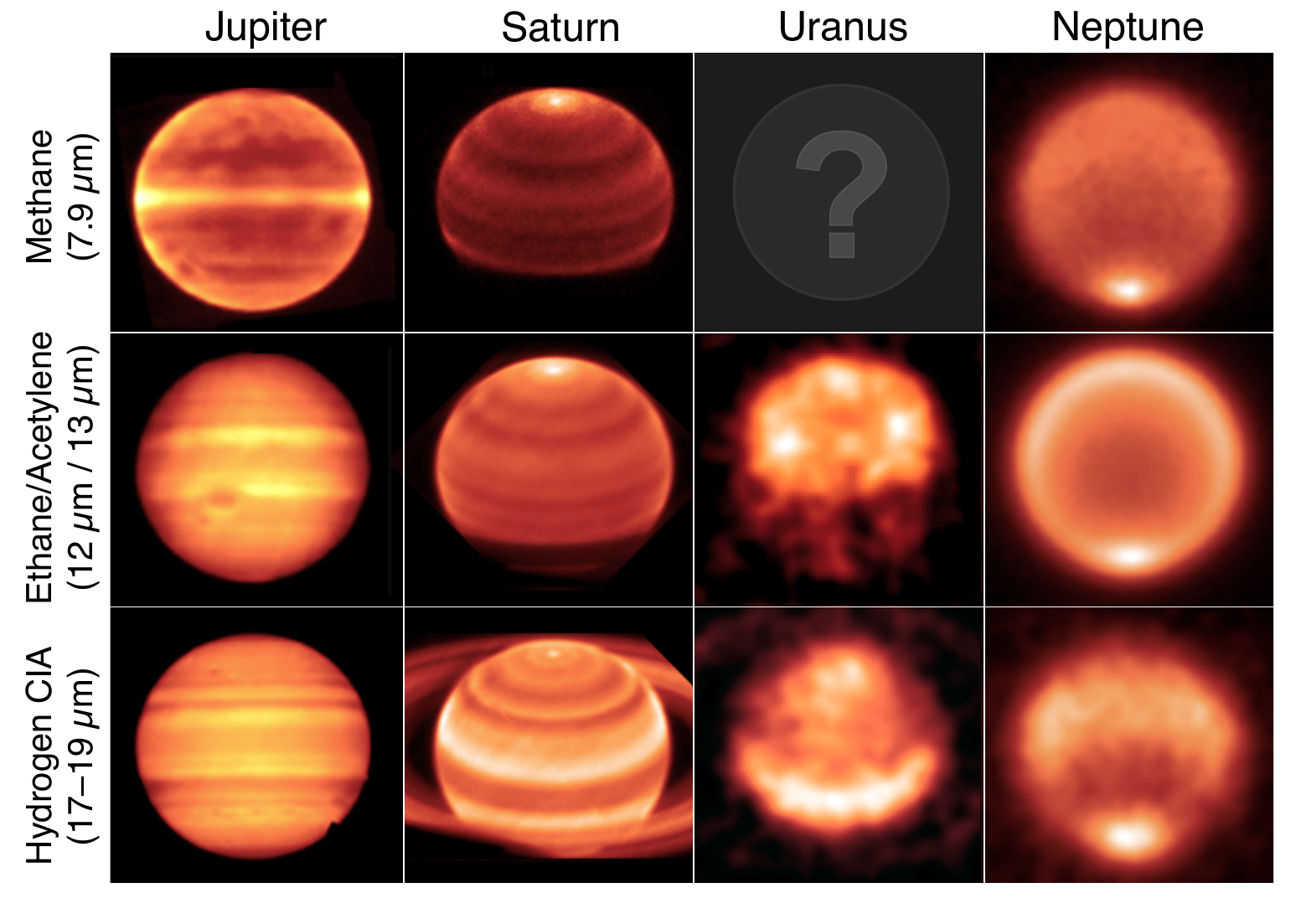}
	\caption{Mid-infrared images of the giant planets from ground-based observatories at three different wavelengths regions, each primarily sensitive to different molecules and pressures: stratospheric methane (centered at $\sim$7.9\,$\upmu$m); stratospheric ethane ($\sim$12.2\,$\upmu$m, relevant to Jupiter, Saturn, and Neptune) and acetylene ($\sim$13\,$\upmu$m, relevant to Uranus); and tropospheric hydrogen ($\sim$17--19\,$\upmu$m). Images have been rotated so that north is up in all cases.  Note that Uranus appears remarkably different in structure in its stratospheric emission compared to other planets.  Furthermore, note that Uranus images are of starkly poorer quality owing to Uranus' weaker emission.  Images of Uranus at \mbox{7.9-$\upmu$m} do not exist in the literature, given poorer telluric transmission and Uranus' particularly weak emission at these wavelengths.  Images are from the following sources: Jupiter from IRTF-MIRSI in 2010  \citep{antunano2021cycles}; Saturn from VLT-VISIR in 2016  \citep{blake2023saturn}; Uranus from VLT-VISIR in 2018  \citep{roman2020uranus}; Neptune from VLT-VISIR, averaged from images dating between 2008 and 2018   \citep{roman2022subseasonal}. \label{fig:imgs}} %MDPI: Figure 10 was moved after its first citation, please confirm; MTR: Readjusted
\end{figure}

The Solar System planets vary significantly in observed structure at mid-infrared wavelengths, as can be seen in the representative examples of mid-infrared images shown in Figure \ref{fig:imgs}. Filtered images are shown in three typical mid-infrared passbands for each planet. Each of these filters senses radiation from a different wavelength and is thus associated with different molecular transitions and pressure levels in the atmosphere.  The Q-band is represented by the images with filtered bandpasses around 18--19 $\upmu$m. These sense thermal emission from the upper troposphere to the lower stratosphere that results from the collision-induced hydrogen-helium continuum. The 12--13-$\upmu$m filters are centered on wavelengths dominated by ethane and/or acetylene emission lines originating from the stratospheres. The 7.9-$\upmu$m filters are sensing emissions from stratospheric methane. 

In general, the measured mid-infrared radiances are dependent on the abundance of the emitting gas as well as the temperature of the gas. In all cases, hydrogen and helium are assumed to be uniformly well mixed throughout the atmosphere below the homopause, and so the observed spatial structure can be explained by spatially varying temperatures.  On Jupiter and Saturn, methane is likewise considered well mixed, and thus the 7.9-$\upmu$m observations are again indicative of temperatures structure  \citep{fletcher2017moist,antunano2020characterizing}, but at lower pressures. However, on Uranus and Neptune, it is cold enough for methane to condense in the troposphere, and therefore methane cannot necessarily be assumed to be uniformly well mixed  \citep{greathouse2011spatially,roman2022subseasonal}. Similarly, stratospheric ethane and acetylene are disequilibrium species, with sources and sinks dependent on photochemistry and temperatures, and so these hydrocarbons are not expected to be uniformly well mixed in pressure or latitude on any of the planets. For these potentially variable gases, the cause of the structure is inherently ambiguous, and interpretation of the radiances requires independent knowledge of the temperatures or assumptions regarding the gaseous distributions. Hence, temperatures derived from thermal observations, particularly from imaging and low-resolution spectra, are inherently subject to large degeneracies with chemical composition (and sometimes cloud opacity), resulting in potentially large uncertainties.

\subsubsection{Spatial Structure of Jupiter and Saturn}\label{ref:jupsatim}
\begin{figure}[H]
	\includegraphics[width=.90\linewidth, trim=.0in .0in .0in .0in, clip]{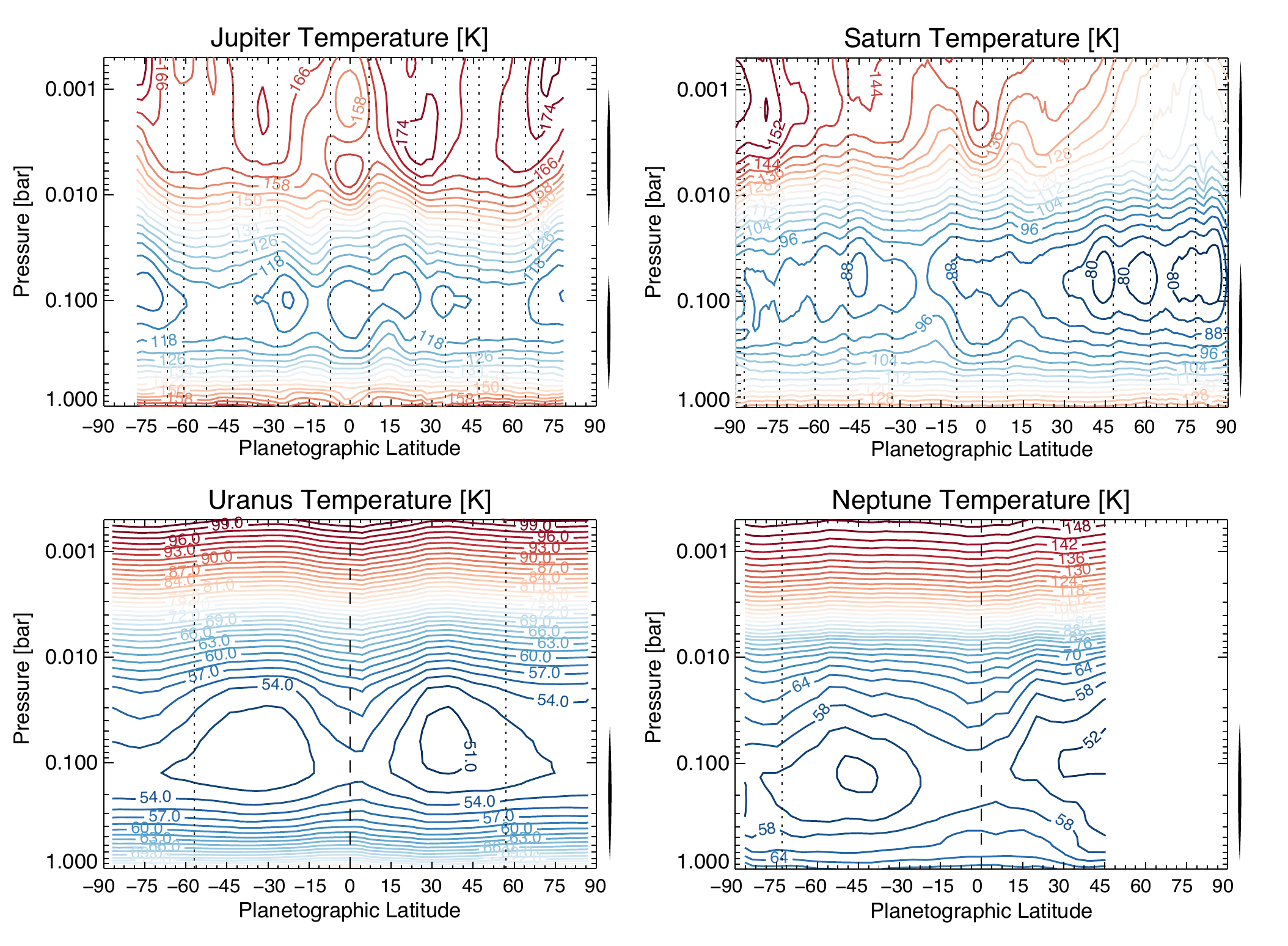}
	\caption{Contours depicting retrieved temperatures versus latitude and pressure for each of giant planets, reproduced from Fletcher et al.  \citep{fletcher2020beltzones}. Colors suggest the transition from warmer (redder) to colder (bluer) temperatures. Temperature data for Jupiter are from the Cassini-CIRS Jupiter flyby in 2000  \citep{fletcher2009phosphine}. Data for Saturn are from Cassini-CIRS while in orbit around the planet, dating between 2006 and 2010  \citep{fletcher2017saturndisruption}. Temperature data from Uranus  \citep{flasar1987voyager,orton2015thermal} and Neptune  \citep{fletcher2014neptune} are from the Voyager 2 flybys in 1986 and 1989, respectively.  The vertical lines to the right of each plot indicate the pressures at which temperatures are constrained by the observations; outside these pressure ranges, temperatures simply relax to an assumed starting profile  \citep{fletcher2020beltzones}. Vertical dotted and dashed lines indicate the position of prograde and retrograde zonal jets, respectively, (from  \citep{porco2003cassini,garcia2011saturn,sanchez2019gas}). 
		Zonal winds and temperatures are in geostrophic balance.\label{fig:temp2d}}   %MDPI: Please change the hyphen (-) into minus sign ($-$, "U+2212"), e.g., "-1" should be "$-$1". MTR: corrected
\end{figure}

Jupiter and Saturn show distinct zonal banding across the mid-infrared, indicative of a complex temperature structure associated with belt-zone \mbox{dynamics  \citep{orton1975thermal,hanel1979infraredjupitervoy1,fletcher2020beltzones,dowling2021emoticons}}. Temperature structures retrieved from spatially resolved spectra are shown in Figure \ref{fig:temp2d}. Regions that appear brighter in thermal infrared emission (see Figures \ref{fig:5micron} and \ref{fig:imgs}) are warmer with thinner clouds, whereas darker areas are colder with thicker clouds. The mechanism behind these regional temperature differences has been interpreted as evidence of adiabatic warming and cooling associated with sinking and rising currents of gas, respectively,  \citep{conrath1983thermalSaturn, gierasch1986zonal, conrath1990temperature,conrath1998thermal, fletcher2020beltzones}. However, it has been argued that the temperature anomalies can be sustained dynamically given cyclonic/anticyclonic zonal shear and the strong vertical stability of the tropopause  \citep{dowling2021emoticons}. In this interpretation, pressure differences between cyclonic and anticyclonic shear regions lead to temperature differences, given constraints on the column thickness imposed by the static stability of the tropopause. However, upwelling and downwelling may still be necessary to explain evidence of chemical disequilibrium, including that of ortho-para hydrogen, which suggests equatorial upwelling on Jupiter and Saturn  \citep{conrath1983evidence,conrath1984global, fletcher2016mid,fletcher2017jupiter,de2021sofia}.

The meridional temperature gradients imply vertical wind shear by the geostrophic thermal wind balance, and the regions of maximum gradients appear well correlated with the latitudes of localized peaks in the zonal winds (i.e., zonal jets) detected by cloud tracking~\citep{porco2003cassini,Tollefson2017jupiterzonal,garcia2011saturn,sromovsky2001neptune,sromovsky2005dynamics}. The vertical motions and shears implied by the temperature field must also be balanced by meridional winds, and Cassini-CIRS observations evidence of this meridional transport in chemical tracers (e.g.,  C$_2$H$_2$, C$_2$H$_6$, C$_3$H$_8$) on Saturn  \citep{greathouse2005meridional, hesman2009saturn,guerlet2010meridional} and Jupiter  \citep{liang2005meridional,zhang2013stratospheric}. Distributions of ammonia  \citep{achterberg2006cassini} on Jupiter and phosphine  \citep{fletcher2009phosphine} on both Jupiter and Saturn also show signs of dynamical motions, with maximum abundances in the cool equatorial zone and reduced abundances in the adjacent warm belts. This is consistent with the picture suggested by the temperature field, with strong uplift in the equatorial zone and descent in the neighboring belts at the top of the troposphere.  As these results demonstrate, the mid-infrared measurements provide an independent diagnostic of the winds and dynamics, beyond which visual imaging of aerosol scattering alone can provide. 

%Likewise, high equatorial para-hydrogen fractions, indicative of disequilibrium with the local temperatures, further suggests upwelling at the equator of both planets  \citep{conrath1983thermal,conrath1998thermal,fletcher2016mid,fletcher2017parajupiter}.

This full picture of the gas giant circulations becomes more complicated when one also considers the distribution of storms, deep ammonia, and microwave radiances---all of which potentially point towards deeper, vertically coincident, but directionally opposite circulation cells (``stacked'' circulation cells) on Jupiter  \citep{orsolini1993model,ingersoll2000moist,ingersoll2021jupiter,duer2021evidence,fletcher2021jupiter}. A discussion of this circulation is beyond the scope of this review, but see Fletcher et al.,  \citep{fletcher2020beltzones} for a comprehensive review.   

Saturn also displays enhanced emission at its poles, which measures 4–7 K warmer than the surrounding latitudes  \citep{achterberg2018thermal,fletcher2018hexagon}. As can be seen in the consistency across three filtered images in Figure \ref{fig:imgs}, the feature extends from the upper troposphere into the stratosphere. The enhanced emission implies downwelling and adiabatic warming, consistent with the local reduction in phosphine  \citep{fletcher2009phosphine}. Observations over time have shown that this is a seasonally varying feature, as discussed in Section \ref{sec:satvar}.

\subsubsection{Uranus and Neptune}\label{ref:urnnepim}
In the case of Uranus and Neptune, the thermal structures appear, at first glance, less complex. On both planets, the equators and poles appear relatively more radiant than do the mid-latitudes in the Q-band images (18--19-$\upmu$m)  \citep{hammel2005new,hammel2006mid,orton2007evidencehotspot,orton2012recovery,orton2015thermal, fletcher2014neptune,roman2020uranus,roman2022subseasonal}. This is consistent with tropopause pressures (40--200 mbar) being colder (roughly 3--6 K) compared to the warmer equator and poles  \citep{flasar1987voyager,conrath1989neptune,conrath1990temperature,orton2015thermal,fletcher2014neptune,roman2020uranus,roman2022subseasonal}. The stratospheres of the Ice Giants, however, appear significantly different in structure compared to each other and their tropospheres. 

Neptune possesses signs of faint banding at 7.9 $\upmu$m and strong limb-brightening at \mbox{12 $\upmu$m}, but only slightly enhanced equatorial brightening  \citep{sinclair2020spatial, roman2022subseasonal}. The limb brightening can be explained by temperature and ethane profiles that increase with height at the range of pressures sensed  \citep{fletcher2014neptune,moses2018seasonal,roman2022subseasonal}, in contrast to the decreasing profile.  However, the banding, if truly present, appears somewhat more complicated than the temperature structure below. With some squinting, one may even argue that 7.9 $\upmu$m images of Neptune appear vaguely more similar to those of Saturn, with its strong polar vortex and banding, only degraded by poorer spatial resolution. With slightly weaker radiances at mid-latitudes compared to the equator and pole, it is possible that we are simply seeing an extension of the upper tropospheric circulation imprinted upon a more complex stratospheric temperature and/or chemical structure, but this cannot be conclusively determined with existing data  \citep{dePater2014neptune,roman2022subseasonal}. 

\begin{figure}[H]
	\includegraphics[width=1.\linewidth, trim=.0in 0in .1in .3in, clip]{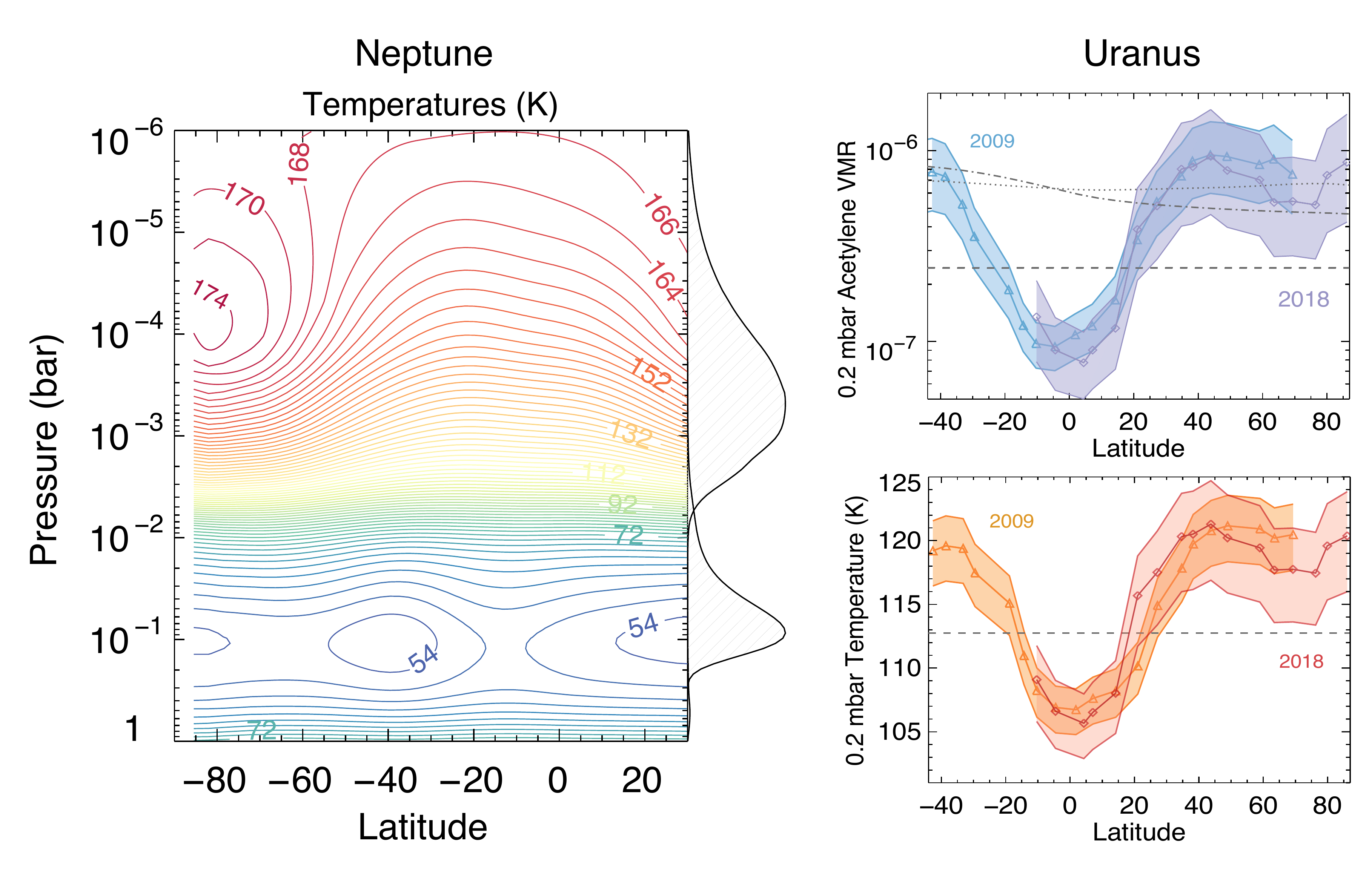}
	\caption{Retrieved stratospheric properties from ground-based images of the Ice Giants.  Left: Neptune's temperature structure, retrieved from 2020 VLT-VISIR imaging data  \citep{roman2022subseasonal}. Temperatures are indicated by the colored contours at 2 K intervals. The heights constrained by the data are suggested by the vertical curves on the right, with maxima contributions peaking near 100 and 0.5. A warm polar vortex is evident at south polar (planetocentric) latitudes. Right: Meridional gradients in C$_2$H$_2$ (top) and temperature (bottom), consistent with Uranus' observed stratospheric radiances (see Figure \ref{fig:uranusstratimg}).  Current data cannot differentiate between the two potential extreme solutions, given the ambiguous nature of the stratospheric emission  \citep{roman2020uranus}. \label{fig:icegiant_strats}}
\end{figure}

Observations of Neptune's hydrogen quadrupole emission (17.03-$\upmu$m H$_2$ S(1)) suggest that Neptune's stratospheric emission structure is primarily owing to latitudinal gradients in its temperature field  \citep{roman2022subseasonal}. Assuming that the atmospheric composition is uniform with latitude, retrievals of atmospheric temperatures reveal a strong meridional gradient, with a 30 K difference between the cool mid-latitudes and the warm polar vortex at 0.5 mbar in 2020 (see Figure \ref{fig:icegiant_strats}). As discussed in Section \ref{sec:nepvar}, this temperature structure appears variable in time. 

\begin{figure}[H]
	\includegraphics[width=.94\linewidth, trim=.0in .0in .0in .0in, clip]{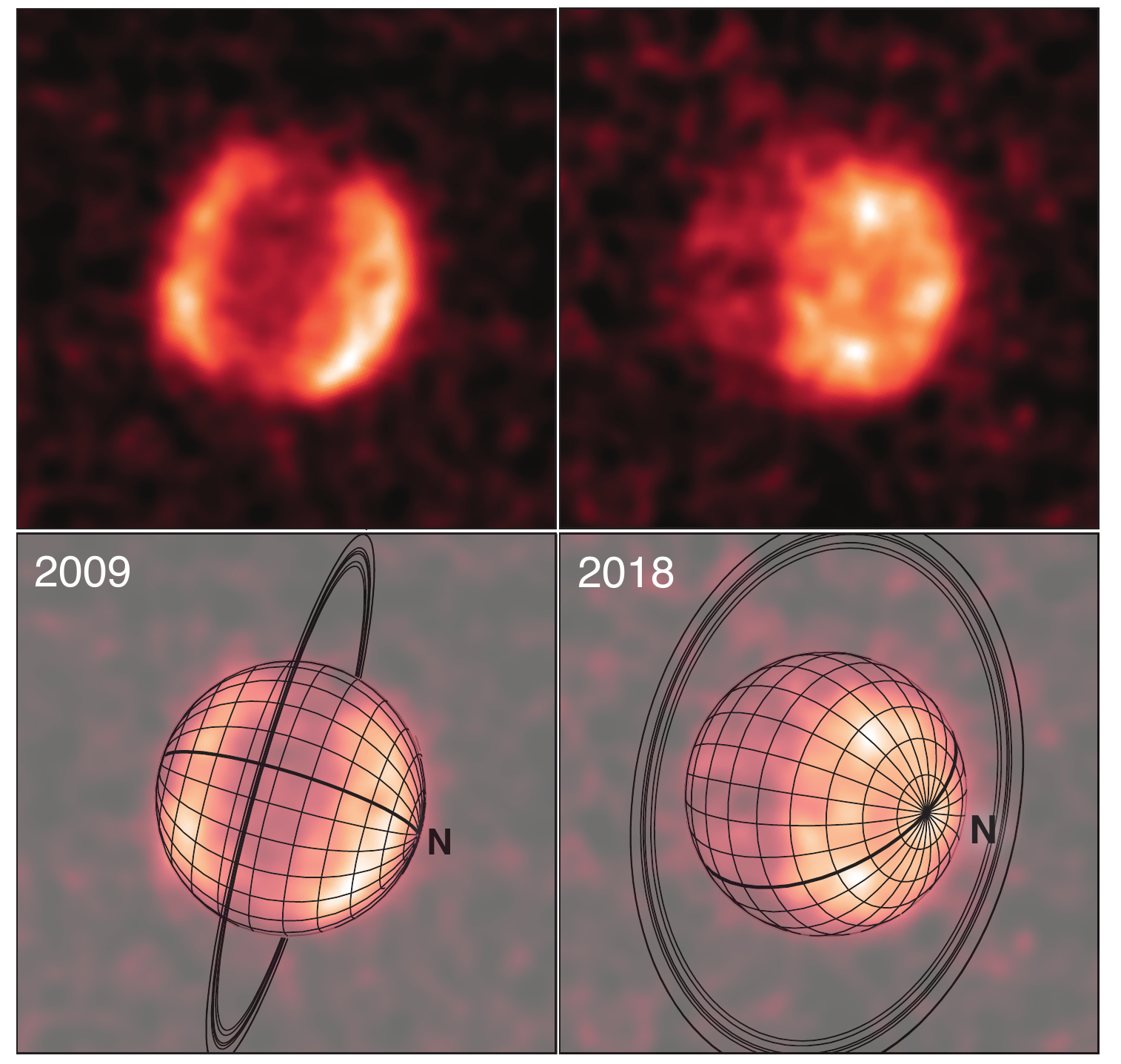}
	\caption{Uranus' stratosphere at 13 $\upmu$m, as seen from VLT-VISIR in 2009 (\textbf{left}) and 2018 (\textbf{right}). Differences in the geometry of the observations are illustrated in the bottom panels. The cause and precise spatial distribution of the enhanced radiance at high latitudes is unclear  \citep{roman2020uranus}. \label{fig:uranusstratimg}}
\end{figure}

\textls[-15]{Finally, and most peculiar of all, Uranus' stratosphere appears completely different from all other giant planets. Uranus' lower stratosphere is very cold and relatively \mbox{dry  \citep{orton2014a,orton2014mid,moses2018seasonal,moses2020icegiantchem},} and as such, no methane-sensing images (7.9 $\upmu$m) currently exist (see Figure \ref{fig:imgs}). However, a few images at 13 $\upmu$m, sensitive to stratospheric C$_2$H$_2$, do exist, and they show excess radiance at high latitudes in the northern and southern \mbox{hemispheres  \citep{orton2018neii,orton2019spatial,roman2020uranus}} (see \mbox{Figure \ref{fig:uranusstratimg}})}. From existing data, it cannot be determined whether these greater high-latitude radiances are due to warmer temperatures or an enhancement in C$_2$H$_2$ (see Figure \ref{fig:icegiant_strats}).  Additionally, the peak latitude of this radiance cannot be strongly constrained given the low signal-to-noise ratio (SNR) of the data. It is tentatively placed at 40$^{\circ}$ latitude, but it may remain constant poleward of this value, depending on the amount of limb-brightening \mbox{present  \citep{roman2020uranus}}. The determination of the distribution is significant. A peak at 40$^{\circ}$ would coincide with the latitudes of temperature minima and assumed maxima upwelling in the upper troposphere, implying a dynamical connection from below. This could be in the form of a vertically coincident but opposite circulation cell, or, in contrast, an extension of the existing upper-tropospheric circulation simply supplying excess hydrocarbons to the local stratosphere.  However, a uniform distribution north of 40$^{\circ}$ would require a completely different explanation. The latter would imply either a separate and somewhat independent circulation, or simply that a completely different mechanism (e.g.,  annual radiative heating, photochemistry, or breaking waves) is shaping the stratospheric radiance  \citep{roman2020uranus}. In any case, the lack of data is limiting our ability to understand the stratospheric dynamics and/or chemistry of Uranus. Fortunately, JWST should soon provide the data necessary to make considerable advances in our understanding of Uranus' stratosphere.

\subsection{Temporal Variability}\label{sec:temporalvar}

The atmospheres of the giant planets exhibit significant variation at visible and near-infrared wavelengths, where we observe sunlight scattered and/or absorbed by gases, clouds, and hazes  \citep{peek1981planet,rogers1995giant,simon2010long,karkoschka2005saturn,lockwood2002photometric,lockwood2006photometric,sromovsky2005dynamics,sromovsky2009uranus,sromovsky2019methane,roman2018aerosols,lockwood2019final,karkoschka2011neptune,huesoneptunelonglived2017,wong2018new,molter2019neptune,hueso2019atmospheric} (see Simon et al.  \citep{simon2022giantvisreview} for a review). Corresponding variations in atmospheric temperatures and chemistry may naturally be expected.  With decades of mid-infrared observations now available, investigations of temporal variability at thermal wavelengths have revealed intriguing findings in recent years. 

In general, many potential sources of temporal variability exist in planetary atmospheres, acting over a wide range of timescales  \citep{mitchell1976overviewvariability}. We can divide these sources into two basic groups, categorized as either internal or external mechanisms. Internal mechanisms include meteorological phenomena and generally stochastic processes within the atmosphere that are poorly understood in the giant planets, whereas external mechanisms act upon the atmosphere and may be considered more deterministic\footnote{Impactors may be considered a notably stochastic exception}. The latter category includes solar energy incident upon the atmosphere, the effects of which can be assessed with seasonal \mbox{models   \citep{wallace1983seasonal,conrath1990temperature,greathouse2008general,moses2005photochemistry,moses2018seasonal}}. 

For planets with significant axial tilts\footnote{The giant planet axial tilts are 3.12$^\circ$, 26.73$^\circ$, 97.77$^\circ$, and 28.33$^\circ$, for Jupiter, Saturn, Uranus, and Neptune, respectively, where the axial tilt is defined as the angle between the direction of the positive pole and the normal to the orbital plane. Note that this differs from the definition adopted by the International Astronomical Union (IAU), which defines Uranus' north pole as the one that lies on the north side of the Solar System's invariable plane, thus placing Uranus's tilt at \mbox{82.23$^\circ$  \citep{seidelmann1992explanatory,meeus1997equinoxes}}.} the daily mean insolation (per unit area) varies seasonally across the disk, with the greatest variation at higher latitudes. The period of this cyclic variation is determined by the tropical orbital period of the planet\footnote{Tropical orbital periods are 11.86, 29.42, 83.75, and 163.72 years for Jupiter, Saturn, Uranus, and Neptune, respectively,  \citep{seidelmann1992explanatory,Astrophysical_Quantities2002}. It is defined as the period of time that the Sun takes to return to the same position in the sky as viewed from the planet}. With a 98$^{\circ}$ axial tilt and 84-year orbit, Uranus arguably serves as the most extreme example of variable seasonal forcing, with much of the planet experiencing decades of uninterrupted summer daylight and winter darkness   \citep{pearl1990albedo,moses2018seasonal}.  Although solar fluxes are weak in the outer Solar System, modeling suggests seasonal variation in temperatures and chemistry are likely  \citep{moses2005photochemistry,moses2018seasonal}, and, in the case of Saturn, well documented by observations~\citep{fletcher2015seasonal} (see Section \ref{sec:satvar}). 

In addition to these larger changes in seasonal forcing, the Sun is intrinsically variable over the course of a roughly 11-year solar cycle. While the total solar irradiance differs by little more than 0.1\% over a typical solar cycle  \citep{Kopp2019solar}, variation in far-ultraviolet (e.g., \mbox{121.57 nm} Lyman-$\alpha$ irradiance) can exceed 40\%. Such high-energy photons are the main drivers behind methane photochemistry, and so modulation in the UV flux can potentially produce observable variation in photochemistry if the reaction timescales are sufficiently short  \citep{moses2005photochemistry}.

\begin{figure}[H]
	\includegraphics [width=1.\linewidth, trim=.0in .0in .0in .0in, clip]{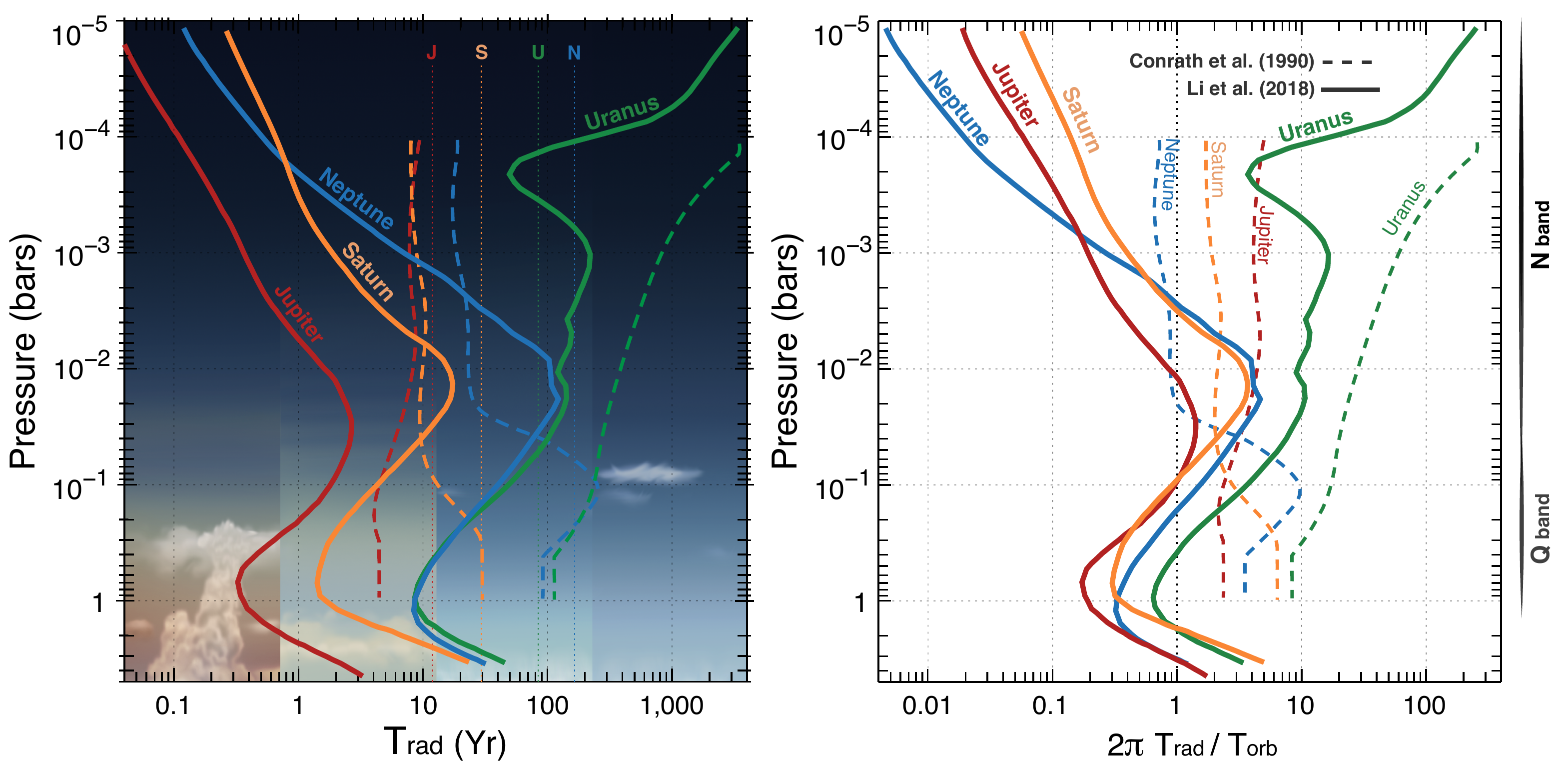}
	\caption{Theoretical radiative time constants for the giant planets.  Plots show these characteristic timescales for each planet over a range of pressures, as derived in two separate studies---dashed lines are Conrath et al.  \citep{conrath1990temperature} while solid lines are from Li et al.  \citep{li2018high}.  The left plot presents the radiative time constant in years, with the orbital periods of each planet indicated by the vertical dotted lines (labeled ``J'' for Jupiter, ``S'' for Saturn, etc.). The right plot expresses the values in terms of a ratio parameter (of a form akin to the resonance behavior of an underdamped harmonic oscillator  \citep{conrath1990temperature}), for which values of order unity or smaller indicate the potential for stronger seasonal responses. The approximate pressures sensed by the N and Q bands are suggested at the far right.  Note that Uranus has the longest radiative time constants throughout the stratosphere. \label{fig:radtcons}}
\end{figure}

The expected extent of the seasonal variation will depend on the change in solar forcing and the capacity of the atmosphere to respond to that change.  Characteristic timescales for the atmospheric responses can be calculated from radiative and chemical models, and by comparing these timescales to the orbital periods, the potential for seasonal changes can be assessed.  Figure \ref{fig:radtcons} illustrates the results of two separate studies, in which radiative time constants were calculated by perturbing the temperature profile and calculating the resultant change in cooling rates  \citep{conrath1990temperature,li2018high}. While significant differences exist between the results (likely owing to the use of updated gaseous absorption coefficients  \citep{orton2007revisedh2h2} and more rigorous radiative-transfer modeling by Li  {et~al.}  \citep{li2018high}), both analyses suggest that Uranus is an outlier, with radiative time constants far longer than the orbital/seasonal timescales, as discussed in Section \ref{sec:urnvar}. However, variation in the stratospheric temperatures of the other giant planets seems likely according to the more recent analysis, and, indeed, this potential appears consistent with observed variability.

Analyses of the characteristic timescales of chemical reactions  \citep{moses2004stratospherejupiter,moses2005photochemistry,moses2018seasonal} and dynamical transport  \citep{conrath1984global,friedson1987seasonal,conrath1990temperature,conrath1998thermal} have similarly been explored to assess the potential for variability.  Uranus again appears relatively sluggish compared to the other planets, with expectations for less seasonal photochemical variation  \citep{moses2018seasonal} and exceedingly long dynamical time constants (estimated at 700 years compared to $\leq$200 for the others)  \citep{conrath1998thermal}.  Chemical variation has been detected in the atmospheres of Jupiter and Saturn  \citep{nixon2007meridional,melin2018assessing,fletcher2011thermalsaturn,cavalie2015photochemical,hue2018photochemistry,fletcher2018saturn}, but Uranus and Neptune remain poorly constrained given their long seasonal timescales.  Likewise, dynamical timescales for all the planets remain highly theoretical and uncertain owing to the obvious challenges of observationally constraining such parameters.

\subsubsection{Jupiter Variability}\label{sec:jupvar}

Multi-wavelength imaging of Jupiter over the past nearly 40 years has revealed surprisingly complex variability in Jupiter's atmosphere,  (e.g.,  \citep{antunano2021cycles}). Data have revealed gradual changes in low latitude temperatures, with little seasonal or short-term variation~\citep{orton1994spatial}.  Emission at 5-$\upmu$m has been used to reveal significant variability in the cloud opacity  \citep{fletcher2018jupiter,antunano2019jupiter, antunano2020characterizing}, while stratospheric temperatures, appear more variable and complicated on shorter timescales  \citep{orton1991thermal,friedson1999_QBO}. 

\begin{figure}[H]
	\includegraphics[width=.9\linewidth, trim=.0in .0in 0.in .0in, clip]{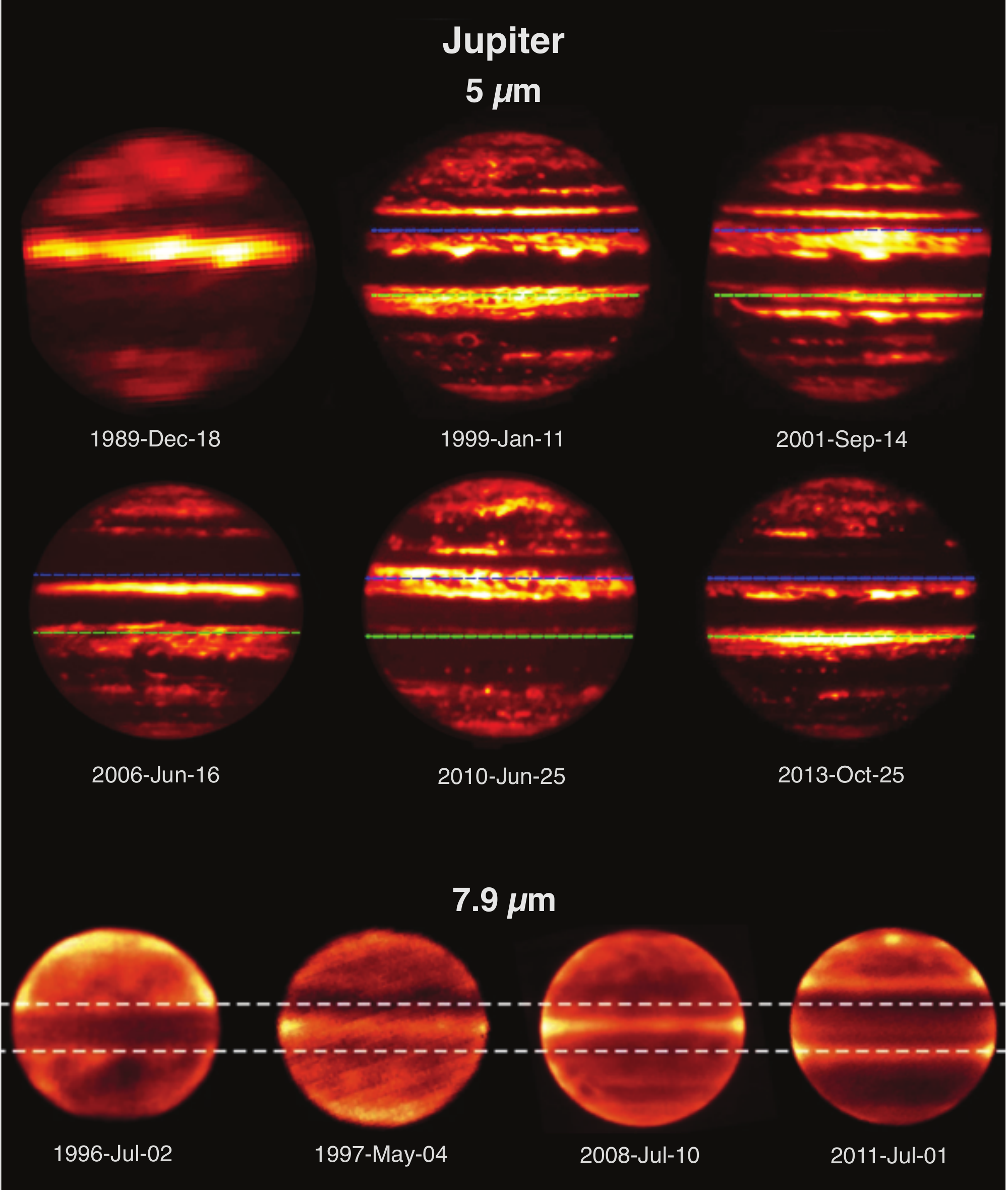}
	\caption{Sequences of mid-infrared images of Jupiter at 5\,$\upmu$m (top) and 7.9\,$\upmu$m (bottom) showing changes over time, adapted from Antu\~{n}ano et al.  \citep{antunano2019jupiter} and Antu\~{n}ano et al.  \citep{antunano2021fluctuations}.  Variation at 5\,$\upmu$m suggests changes in tropospheric temperature and cloud opacity, with large temporal variability mainly at the equatorial and tropical latitudes and less temporal variability at mid-latitudes  \citep{antunano2019jupiter}. Dashed blue and green lines mark 16$^{\circ}$ N and 10$^{\circ}$ S planetocentric latitudes, respectively. Emission at 7.9 $\upmu$m sense stratospheric temperatures (via methane emission), revealing roughly periodic variation associated with the quasi-quadrennial oscillation  \citep{antunano2021fluctuations}. 5-$\upmu$m images are from various instruments on the IRTF, including BOLO-1 (1984)  \citep{orton1994spatial}, NSFCam (1999, 2001)  \citep{ortiz1998evolution}, NSFCam2 (2006)  \citep{fletcher2010thermal}, and SpeX (2010, 2013). 7.9-$\upmu$m images are from IRTF-MIRLIN (1996, 1997) and IRTF-MIRSI (2008, 2011).\label{fig:jupvar}}
\end{figure}

Observations of Jupiter’s stratospheric temperatures via methane at 7.9 $\upmu$m have been used to infer variability between 1980 and 2011  \citep{antunano2021fluctuations} (see Figure \ref{fig:jupvar}). This investigation revealed significantly different periods of oscillation (the quasi-quadrennial oscillation), with a 5.7-year period between 1980 and 1990 and a 3.9-year period between 1996 and 2006. Planetary-scale disturbances in 1992 and 2007 disrupted the predicted quasi-quadrennial oscillation pattern, suggesting that these oscillations are related to vertically propagating waves generated by meteorological sources below  \citep{antunano2021fluctuations}.

Similar studies have revealed surprising apparent correlations (and anti-correlations) between different altitudes and locations.  Equatorial temperature variations in the upper troposphere appear anti-correlated with higher altitudes, in a manner that suggests stratospheric dynamics may also influence the upper tropospheric temperatures below.  Intriguingly, anti-correlations in temperatures have been detected for conjugate latitudes in opposite hemispheres  \citep{antunano2021fluctuations,antunano2021cycles, orton2022unexpected}.

Though the sources of such oscillations are not definitively known, some are thought to be associated with stratospheric winds and temperature oscillations, analogous to Earth’s quasi-biennial oscillation  \citep{orton1991thermal,leovy1991quasiquadrennial,simon2006jupiter}.  Theories suggest wave or eddy-driven meridional winds likely play an important role in modulating the temperatures and winds in the upper troposphere and stratosphere on seasonal and shorter timescales  \citep{greathouse2016tracking}, and analyses of the thermal variability could potentially be used to estimate variation in the mechanical \mbox{forcing  \citep{simon2006jupiter}}.

\subsubsection{Saturn Variability}\label{sec:satvar}

\begin{figure}[H]
	\includegraphics[width=.8\textwidth, trim=.0in .0in .0in .0in, clip]{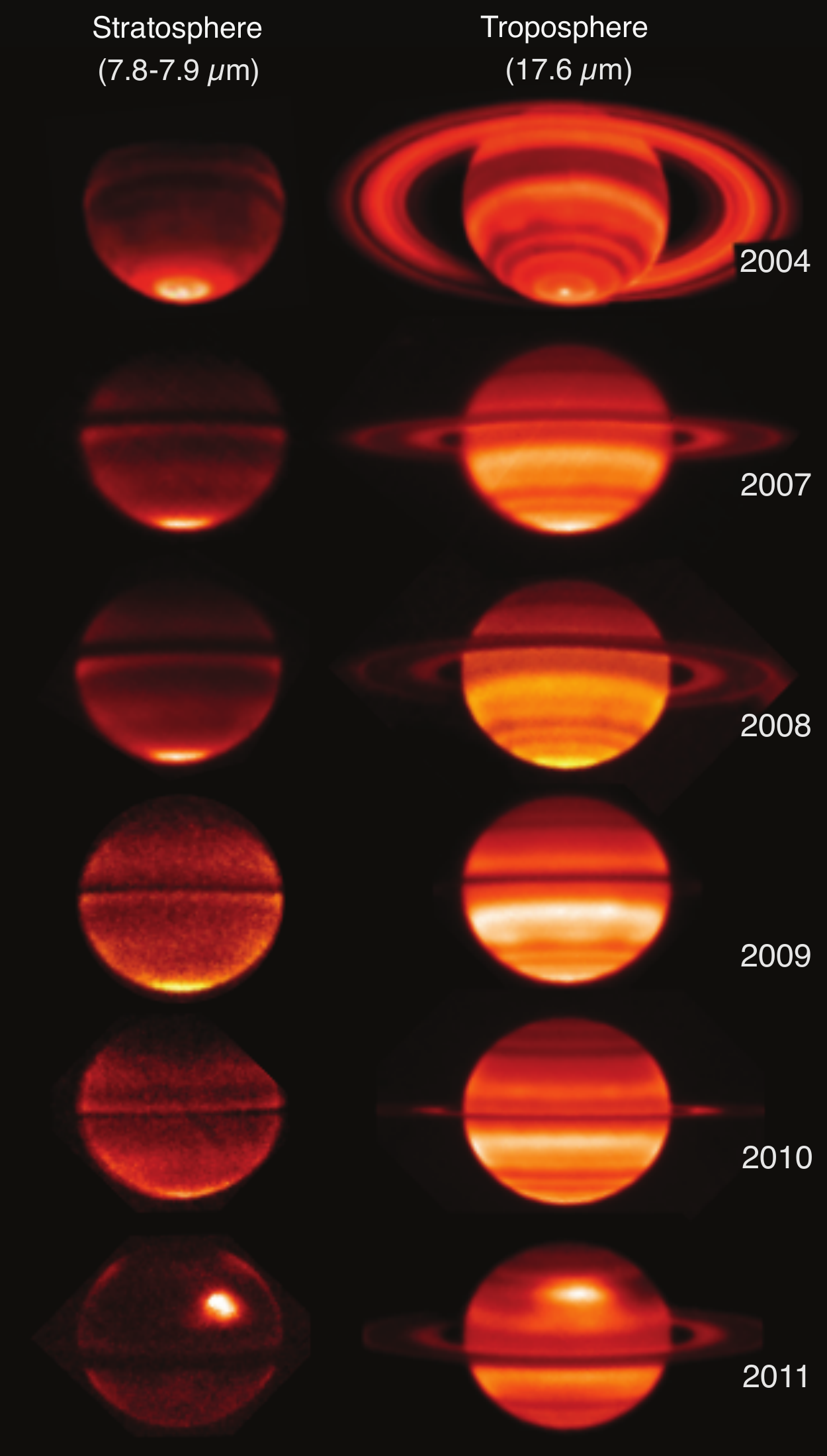}
	\caption{Saturn images showing changes between 2004 and 2012, adapted from  \citep{blake2023saturn}. Images sense stratospheric temperatures via methane emission at 7.8 $\upmu$m (\textbf{left}) and tropospheric temperatures via collision-induced hydrogen at 17.6 $\upmu$m (\textbf{right}). Images are from Keck-LWS (2004)~\citep{orton2005saturn}, Subaru-COMICS (2007), and VLT-VISIR (2008-2012). Note the prominent warm spot associated with a remarkable storm in the northern hemisphere in 2011 \citep{fischer2011giant, fletcher2012originsaturnvortex}. \label{fig:figsatmontage}}
\end{figure}

\begin{figure}[H]
	\includegraphics[width=.96\linewidth, trim=.0in .0in .0in .0in, clip]{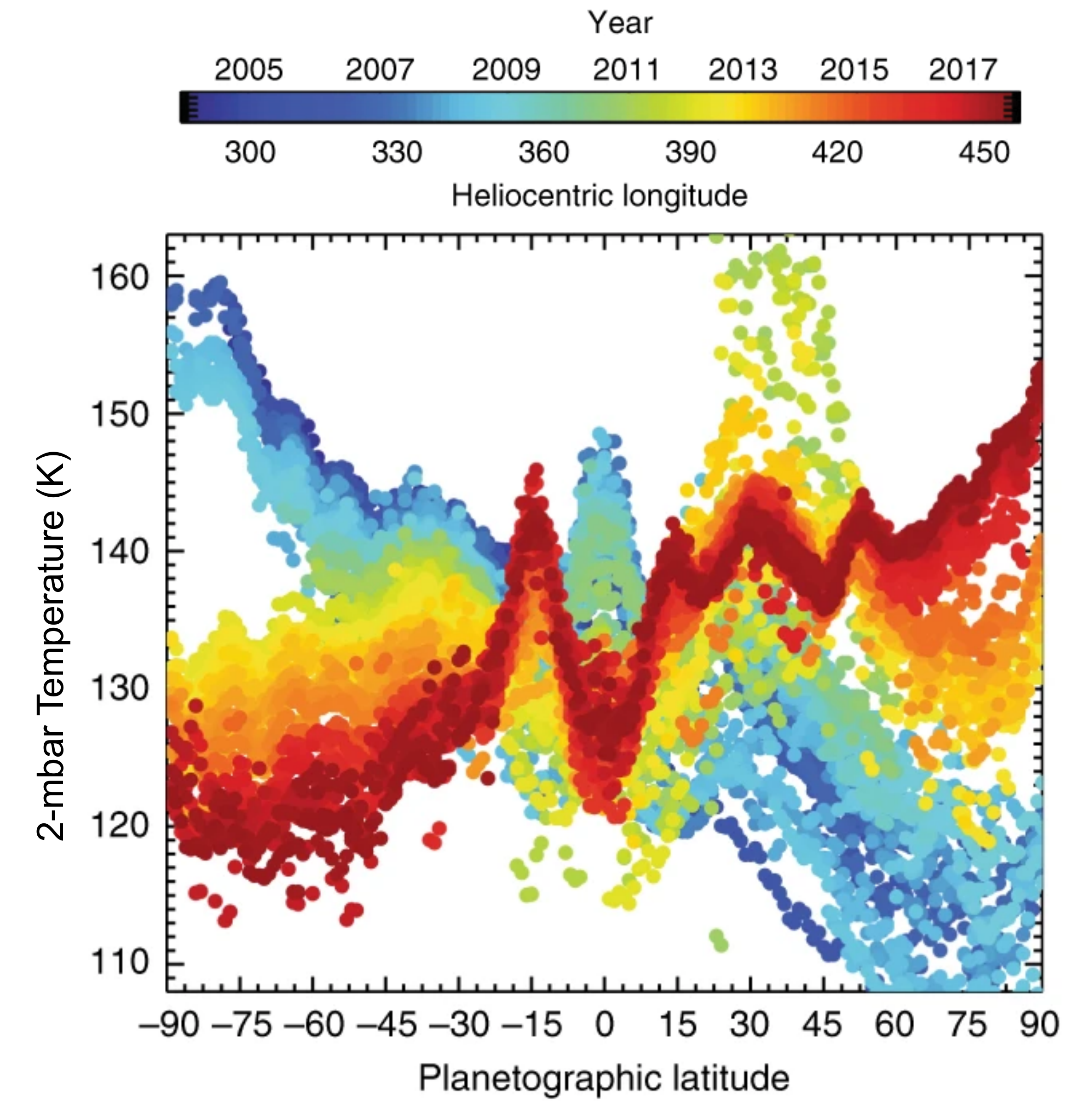}
	\caption{Retrieved temperatures of Saturn's stratosphere at 2 mbar versus latitude over the entire Cassini mission, adapted from Fletcher et al.,  \citep{fletcher2018hexagon}. The years and heliocentric longitudes--indicating the seasonal phase, with 270 and 360 marking the northern winter solstice and spring equinox, respectively--are indicated by the color bar.  \label{fig:figsatvar}}
\end{figure}

Unlike Jupiter, Saturn has a significant axial tilt (26.73$^{\circ}$), as illustrated by the changing views from Earth seen in Figure \ref{fig:figsatmontage}. The resulting seasonal variation in sunlight over a Saturnian year (29.4 Earth years) dominates Saturn's temporal variability in the mid-infrared.  The Cassini-Huygens mission orbited Saturn for 13 years---enough to gain unprecedented detail of how the planet changed over the course of nearly two seasons. Cassini-CIRS observed Saturn’s northern mid-latitude stratosphere warming by 6--10 K as this region emerged from ring-shadow in spring, while the southern mid-latitudes cooled by \mbox{4--6 K  \citep{fletcher2010seasonal}} (see Figure \ref{fig:figsatvar}). The tropospheric temperatures also changed, but to a lesser degree, consistent with theoretical expectations of larger thermal inertia and longer radiative time constants. The fall and winter hemispheres also saw significant depletion in acetylene, consistent with seasonal photochemical modeling  \citep{moses2005photochemistry, hue20162d}. %MDPI: Figure should be cited in order. please cited Figure 16 before Figure 17, MTR: Resolved

As part of Saturn's seasonal cycle, its polar stratosphere sees the development of a warm circumpolar vortex that peaks in the summer and dissipates in the winter. Cassini-CIRS observed the dissipation of Saturn's southern polar vortex in southern mid-autumn (2012)  \citep{fletcher2015seasonal,fletcher2018hexagon}, followed by the eventual formation of the northern polar vortex in late northern spring (2015)  \citep{fletcher2018hexagon}. The northern feature was associated with warmer temperatures poleward of $\sim$75$^{\circ}$ planetographic latitude. Interestingly, this feature exhibited a hexagonal boundary, echoing the hexagonal Rossby wave made visible in the clouds far below. This suggests a dynamical link between the features separated by 300 km in height  \citep{fletcher2018hexagon}. A comprehensive review of Saturn's seasonal changes during the Cassini era can be found in Fletcher et al.  \citep{fletcher2018saturn}.

A recent multi-decadal study of ground-based mid-infrared imaging similarly found seasonal temperature changes of $\sim$30 K in the stratosphere and $\sim$10 K in the upper troposphere, consistent with Cassini observations and predictions from radiative climate models \citep{blake2023saturn}. The most recent observations from VLT-VISIR show warming is continuuing at the northern summer polar stratosphere However, comparison of $\sim$7.9 $\upmu$m-imaging revealed evidence of inter-annual variations at equatorial latitudes.  Variations on these timescales are inconsistent with the strictly semi-annual 15-year equatorial stratospheric oscillation \citep{guerlet2011evolution,bardet2022joint}, suggesting the oscillation’s period is either intrinsically variable and/or subject to disruption by storms or other meteorological phenomenon.  

Aside from seasonal phenomena, mid-infrared observations have also notably detected warm stratospheric features associated with an immense northern-hemisphere storm that appeared in December 2010. The storm was observed to produce enormous changes in stratospheric temperatures and chemistry, warming the localized region by 80 K compared with its surroundings at 2 mbar  \citep{fletcher2011thermalsaturn}. The stratospheric warm “beacons” eventually evolved into a stratospheric anticyclonic vortex in 2011  \citep{fletcher2011thermalsaturn,fletcher2012originsaturnvortex} (see Figure \ref{fig:figsatmontage}). Cassini-CIRS observations were compared with chemical models to explain the mid-infrared changes, and it was found that elevated temperatures alone could not explain the enhanced thermal emission from ethane and acetylene.  Downwelling winds, transporting hydrocarbons to higher pressures, were also needed 
to reproduce the CIRS observations. 

%MDPI:  please moved Figure 16 after its first citation or please add citation in previous content

\subsubsection{Uranus Variability}\label{sec:urnvar}
Reviewing all temporal variability detected in the mid-infrared on Uranus is unfortunately a very brief exercise. There is simply very little to compare given the limited amount of mid-infrared data that exists. Furthermore, what does exist appears largely invariant over the short history of these observations relative to the lengthy 21-year seasons on Uranus. 

One might reasonably expect seasonality on Uranus to be interesting given its extreme axial tilt of 98$^{\circ}$, which forces nearly all latitudes into extended periods of total daylight and darkness  \citep{moses2018seasonal}. However, its atmosphere is sluggish vertical mixing, low stratospheric methane abundances, and cold temperatures result in a great thermal inertia that leads to theoretically small seasonal changes and large lags  \citep{flasar1987voyager,conrath1990temperature, li2018high}. The atmospheric temperatures are thus expected to remain close to the annual mean radiative equilibrium values, even though the seasonal amplitude of the radiative forcing is large  \citep{conrath1990temperature}.  There is some discrepancy in the literature over the length of the theoretical radiative time constants as a function of height for the outer planet atmospheres (see Section \ref{sec:temporalvar} and Figure \ref{fig:radtcons}). Conrath et al.  \citep{conrath1990temperature} calculated values of over 130 years in the upper-troposphere and stratosphere, but Li et al.  \citep{li2018high} found them to be significantly shorter---ranging from roughly 10 to 70~years at pressures of 400 to 70 mbar. The latter would suggest the potential for variability, and observations could potentially confirm or refute these theoretical expectations. 

When Voyager-IRIS produced the first temperature maps of Uranus near the time of the southern summer, there were little differences between the summer and winter hemispheric temperatures at the tropopause. The summer pole was no warmer than the winter pole in the tropopause and only marginally warmer in the lower stratosphere~\citep{orton2015thermal}. This indicated that seasonal variation in the upper troposphere was indeed very small.  Subsequent comparisons between Voyager-era temperatures and ground-based imaging acquired over 20 and 32 years later revealed no significant changes in the upper-tropospheric (\mbox{70--400}~mbar) temperatures more than a full season later  \citep{orton2015thermal,roman2020uranus}. Significant temperature changes have yet to be found.  

In the stratosphere, observations are even more limited. Only nine years separate the existing images sensitive to stratospheric emission, and they appear invariant within the considerable uncertainties  \citep{roman2020uranus} (see Figures \ref{fig:icegiant_strats} and \ref{fig:uranusstratimg}). There have been some hints of possible variation in ground-based images  \citep{roman2020uranus} and Spitzer-IRS observations, averaged over different sub-observer longitudes, but these have been interpreted as possible evidence of longitudinal variation, rather than temporal variability  \citep{rowe2021longitudinal}. A lack of temporal variability in the stratosphere would be consistent with the expected long stratospheric radiative time constants  \citep{conrath1990temperature,li2018high} (see Figure \ref{fig:radtcons}).  However, additional mid-infrared observations, repeated frequently over the coming decade, will be needed to determine whether significant temperature or chemical changes actually occur on Uranus.  

\subsubsection{Neptune Variability}\label{sec:nepvar}

Despite its supremely long seasonal timescales (165 year orbit) and great distance from the Sun, Neptune exhibits remarkable variability at mid-infrared wavelengths. Like Uranus, Neptune's temperatures were first mapped by Voyager-IRIS.  Like Uranus, comparisons with subsequent ground-based imaging have shown the upper tropospheric temperatures are largely invariant in time within uncertainties  \citep{fletcher2014neptune, roman2022subseasonal}. The possible exception is at the south pole, which demonstrates possible variability in the troposphere with no obvious pattern  \citep{fletcher2014neptune,roman2022subseasonal}, but it is subject to large uncertainties. However, unlike Uranus, Neptune's stratosphere clearly exhibits considerable variability.

\begin{figure}[H]
	\includegraphics[width=1.0\linewidth, trim=.0in .0in .25in .0in, clip]{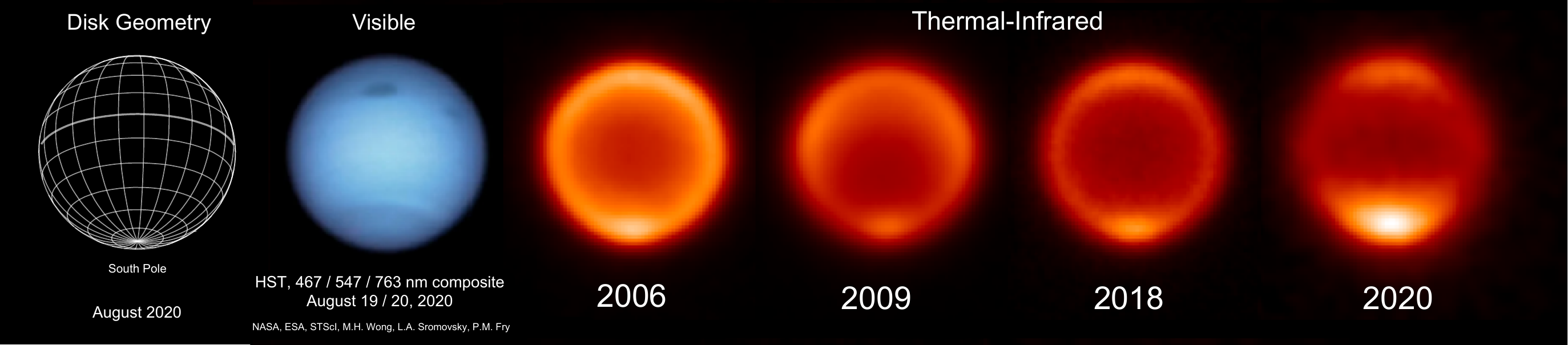}
	\caption{A sequence of mid-infrared images showing the variation of Neptune at roughly 12 $\upmu$m in different years, along with disk geometry and a Hubble Space Telescope (HST) visible image for comparison. The mid-infrared images were taken from VLT-VISIR (2006, 2009, 2018) and Subaru-COMICS (2020)  \citep{roman2022subseasonal}. The sequence shows a global decline in radiances accompanied by dramatic warming at Neptune's south pole between 2018 and 2020. The HST image was taken in 2020, three weeks after the Subaru-COMICS image. (HST Image credit: NASA, ESA, STScI, M.H. Wong (University of California, Berkeley), and L.A. Sromovsky and P.M. Fry (University of Wisconsin-Madison).  \label{fignepvar}}
\end{figure}

A recent analysis of all mid-infrared observations of Neptune existing prior to 2020 has revealed an overall decline in mid-infrared radiances since reliable imaging began in 2003  \citep{roman2022subseasonal} (see Figure \ref{fignepvar}). Combined with spectral data sensitive to atmospheric temperatures via the $\sim$17.03-$\upmu$m H$_2$ S(1) quadrupole emission, these observations indicated that Neptune's disk-integrated temperatures dropped by roughly 8 K in the lower stratosphere  \citep{roman2022subseasonal}. These changes are unexpected, since radiative-seasonal models predicted that temperatures should rise in Neptune's southern hemisphere in early summer  \citep{conrath1990temperature,greathouse2011spatially}. 

\begin{figure}[H]
	\includegraphics[width=1.\linewidth, trim=.1in .0in .1in .1in, clip]{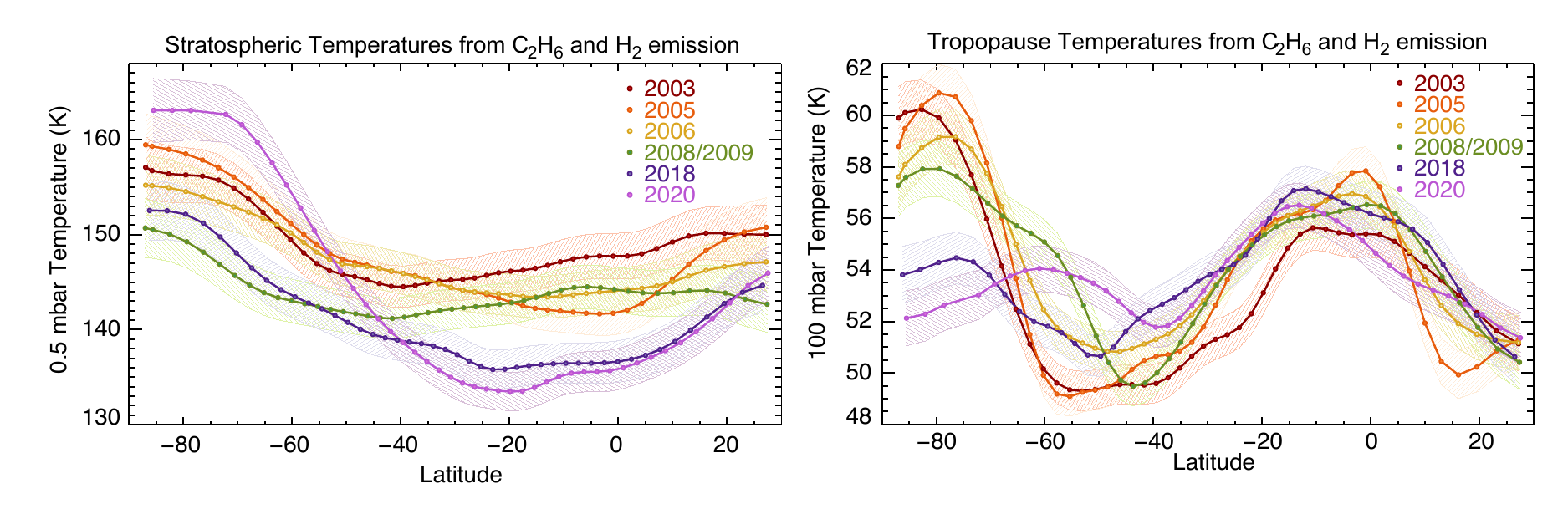}
	\caption{Neptune's temperatures versus planetocentric latitude from ground-based images dating from different years, adapted from Roman et al. \citep{roman2022subseasonal}. Shaded envelopes indicate uncertainties. Temperatures are shown at 0.5 mbar ($\textbf{left}$) and 100 mbar ($\textbf{right}$), corresponding to peaks in the contribution from stratospheric ethane (12.2 $\upmu$m) and tropospheric hydrogen CIA ($\sim$18--25$\upmu$m).  The stratospheric temperatures vary in time, with brightening at the pole in recent years.  Tropospheric temperatures are largely invariant, except for the south pole. Data are from Keck-LWS (2003), Gemini-N-Michelle (2005), VLT-VISIR (2006--2018), and Subaru-COMICS (2020). \label{fig:neptempvar}}
\end{figure}

While global temperatures dropped, images sensitive to 12-$\upmu$m emission from stratospheric ethane showed a dramatic surge in radiance from Neptune's south pole between 2018 and 2020--again attributed to a rise in temperatures ($\sim$13 K) inferred from nearly contemporaneous H$_2$ S(1) spectra  \citep{roman2022subseasonal}. This warming circumpolar vortex was combined with a drop in temperatures at nearly all other latitudes (see Figure \ref{fig:neptempvar}).  Radiative and chemical models have predicted a gradual brightening of the south pole following the southern summer solstice in 2005  \citep{conrath1990temperature, greathouse2011spatially}, but such rapid change is unexpected.

The cause of these stratospheric temperature changes is currently unknown. Roman et al.  \citep{roman2022subseasonal} speculated that it may be related to seasonal changes in chemistry  \citep{moses2018seasonal}, which alters the cooling rates, but explanations involving solar cycle variations, stratospheric oscillations, and meteorological activity cannot be discounted. With such dramatic and unexpected changes in recent years, regular observations over the next decade will be crucial for understanding the nature and trends shaping the stratospheric variability of Neptune.

\section{Conclusions}\label{sec:conclusions}
From more than a century of remote sensing at mid-IR wavelengths, a remarkably detailed picture of the temperature structure, chemistry, and dynamics of the giant planets has emerged. Many questions and challenges remain, particularly regarding how and why the planets change over time.  

Much of the knowledge written in this review will soon be rewritten. The upcoming Solar System observations of the giant planets by JWST-MIRI have the potential to greatly surpass existing observations and revise our knowledge of the atmospheres of the giant planets, particularly regarding the Ice Giants  \citep{norwood2016giant,moses2018seasonal}. Nonetheless, this brief look into the history and results of mid-infrared remote sensing can hopefully continue to provide insight and inspiration, if simply by considering how far the field has come.

	\vspace{6pt} 
	
	\funding{During the preparation of this manuscript, the author was supported by a European Research Council Consolidator Grant, under the European Union’s Horizons 2020 research and innovation program, grant number 723890. }%this part is necessary, pelase add. {Please add: ``This research received no external funding'' or ``This research was funded by NAME OF FUNDER grant number XXX.'' and  and ``The APC was funded by XXX''. Check carefully that the details given are accurate and use the standard spelling of funding agency names at \url{https://search.crossref.org/funding}, any errors may affect your future funding.}

	\acknowledgments{ I wish to thank Leigh Fletcher and Imke de Pater for offering the opportunity, support, and patience necessary for completing this review.  I also wish to acknowledge Arrate Antu\~{n}ano for her readiness to assist with her expertise on the Gas Giant atmospheres.  
 
	In memory of Peter Jay Gierasch (1940--2023), an insightful scientist and generous advisor, whose many enduring contributions have uniquely shaped the field of planetary atmospheres, as demonstrated throughout this review.}
	
	\dataavailability{No new data were created or analyzed in this study. Data sharing is not applicable to this article. The data presented in this study are available from the original sources, as referenced.}%{Please refer to suggested Data Availability Statements in section “MDPI Research Data Policies” at \href{https://www.mdpi.com/ethics}{https://www.mdpi.com/ethics}.}  In this section, please provide details regarding where data supporting reported results can be found, including links to publicly archived datasets analyzed or generated during the study. Please refer to suggested Data Availability Statements in section “MDPI Research Data Policies” at https://www.mdpi.com/ethics. You might choose to exclude this statement if the study did not report any data.
	
	\conflictsofinterest{The authors declare no conflict of interest.} 

		\reftitle{References}
		
		% Please provide either the correct journal abbreviation (e.g.,  according to the “List of Title Word Abbreviations” http://www.issn.org/services/online-services/access-to-the-ltwa/) or the full name of the journal.
		
		
		\begin{adjustwidth}{-\extralength}{0cm}
			%\centering %% If there is a figure in wide page, please release command \centering
			
			\PublishersNote{}
		\end{adjustwidth}


\begin{thebibliography}{999}
				
				\bibitem[Norwood \em{et~al.}(2016)Norwood, Moses, Fletcher, Orton, Irwin,
				Atreya, Rages, Cavali{\'e}, S{\'a}nchez-Lavega, Hueso,
				et~al.]{norwood2016giant}
				Norwood, J.; Moses, J.; Fletcher, L.N.; Orton, G.; Irwin, P.G.; Atreya, S.;
				Rages, K.; Cavali{\'e}, T.; S{\'a}nchez-Lavega, A.; Hueso, R.;  et~al.
				\newblock Giant planet observations with the james webb space telescope.
				\newblock {\em Publ. Astron. Soc.  Pac.} {\bf
					2016}, {\em 128},~018005.
				
				\bibitem[Guan \em{et~al.}(2021)Guan, Yu, Periyanagounder, Benzigar, Huang, Lin,
				Kim, Singh, Hu, Liu, et~al.]{guan2021recent}
				Guan, X.; Yu, X.; Periyanagounder, D.; Benzigar, M.R.; Huang, J.K.; Lin, C.H.;
				Kim, J.; Singh, S.; Hu, L.; Liu, G.;  et~al.
				\newblock Recent progress in short-to long-wave infrared photodetection using
				2D materials and heterostructures.
				\newblock {\em Adv. Opt. Mater.} {\bf 2021}, {\em 9},~2001708.
				
				\bibitem[Noll \em{et~al.}(2012)Noll, Kausch, Barden, Jones, Szyszka,
				Kimeswenger, and Vinther]{noll2012atmospheric}
				Noll, S.; Kausch, W.; Barden, M.; Jones, A.; Szyszka, C.; Kimeswenger, S.;
				Vinther, J.
				\newblock An atmospheric radiation model for Cerro Paranal-I. The optical
				spectral range.
				\newblock {\em Astron. Astrophys.} {\bf 2012}, {\em 543},~A92.
				
				\bibitem[Jones \em{et~al.}(2013)Jones, Noll, Kausch, Szyszka, and
				Kimeswenger]{jones2013advanced}
				Jones, A.; Noll, S.; Kausch, W.; Szyszka, C.; Kimeswenger, S.
				\newblock An advanced scattered moonlight model for Cerro Paranal.
				\newblock {\em Astron. Astrophys.} {\bf 2013}, {\em 560},~A91.
				
				\bibitem[Fazio(1994)]{fazio1994infrared}
				Fazio, G.
				\newblock Infrared array detectors in astrophysics.
				\newblock {\em Infrared Phys. Technol.} {\bf 1994}, {\em 35},~107--117.
				
				\bibitem[Haller(1994)]{haller1994advanced}
				Haller, E.
				\newblock Advanced far-infrared detectors.
				\newblock {\em Infrared Phys. Technol.} {\bf 1994}, {\em 35},~127--146.
				
				\bibitem[Glass and Glass(1999)]{glass1999handbook}
				Glass, I.S.; Glass, I.
				\newblock {\em Handbook of Infrared Astronomy};  Cambridge University
				Press: Cambridge, UK,  1999; Number~1.%Mdpi: newly added, please confirm
				%MDPI: please confirm all added informations in references part: MTR: Looks fine.
				
				\bibitem[Tokunaga(2002)]{tokunaga2002infrared}
				Tokunaga, A.
				\newblock Infrared astronomy. In {\em Allen’s Astrophysical Quantities};
				Springer:  Berlin/Heidelberg, Germany, %We added the location of publisher. Please confirm
				2002; pp. 143--167.
				
				\bibitem[Rieke(2003)]{rieke2003detection}
				Rieke, G.
				\newblock {\em Detection of Light: From the Ultraviolet to the Submillimeter};
				Cambridge University Press: Cambridge, UK,  2003.
				
				\bibitem[McLean(2012)]{mclean2012infrared}
				McLean, I.S.
				\newblock {\em Infrared Astronomy with Arrays: The Next Generation};  
				Springer Science \& Business Media:  Berlin/Heidelberg, Germany,  2012; Volume 190.
				
				\bibitem[Ives \em{et~al.}(2014)Ives, Finger, Jakob, and
				Beckmann]{ives2014aquarius}
				Ives, D.; Finger, G.; Jakob, G.; Beckmann, U.
				\newblock AQUARIUS: the next generation mid-IR detector for ground-based
				astronomy, an update. In
				\newblock  \emph{High Energy, Optical, and Infrared Detectors for Astronomy VI}; SPIE:  Bellingham, WA USA %MDPI: please add location (city, country) of publisher
				2014; Volume 9154, pp. 489--499.

				
				\bibitem[Larson(1980)]{larson1980infrared}
				Larson, H.P.
				\newblock Infrared spectroscopic observations of the outer planets, their
				satellites, and the asteroids.
				\newblock {\em Annu. Rev. Astron. Astrophys.} {\bf 1980}, {\em
					18},~43--75.
 
				
				\bibitem[Mampaso \em{et~al.}(2004)Mampaso, Prieto, and
				S{\'a}nchez]{mampaso2004infrared}
				Mampaso, A.; Prieto, M.; S{\'a}nchez, F.
				\newblock {\em Infrared Astronomy}; Cambridge University Press: Cambridge, UK,  2004.
				
				\bibitem[Rieke \em{et~al.}(2005)Rieke, Kelly, and Horner]{rieke2005overview}
				Rieke, M.J.; Kelly, D.M.; Horner, S.D.
				\newblock Overview of James Webb Space Telescope and NIRCam's Role. In
				\newblock  \emph{Cryogenic Optical Systems and Instruments XI}; SPIE: Bellingham, WA USA %MDPI: please add location (city, country) of publisher
				2005; Volume
				5904, p. 590401.
				
				\bibitem[Wells \em{et~al.}(2006)Wells, Lee, Oudenhuysen, Hastings, Pel, and
				Glasse]{wells2006miri}
				Wells, M.; Lee, D.; Oudenhuysen, A.; Hastings, P.; Pel, J.W.; Glasse, A.
				\newblock The MIRI medium resolution spectrometer for the James Webb Space
				Telescope. In
				\newblock  \emph{Space Telescopes and Instrumentation I: Optical, Infrared, and
					Millimeter}; SPIE: Bellingham, WA USA %MDPI: please add location (city, country) of publisher
				2006; Volume 6265, \mbox{pp. 358--369}.
				
				\bibitem[Rieke \em{et~al.}(2015)Rieke, Wright, B{\"o}ker, Bouwman, Colina,
				Glasse, Gordon, Greene, G{\"u}del, Henning, et~al.]{rieke2015mid}
				Rieke, G.H.; Wright, G.; B{\"o}ker, T.; Bouwman, J.; Colina, L.; Glasse, A.;
				Gordon, K.; Greene, T.; G{\"u}del, M.; Henning, T.;  et~al.
				\newblock The mid-infrared instrument for the james webb space telescope, i:
				Introduction.
				\newblock {\em Publ. Astron. Soc. Pac.} {\bf
					2015}, {\em 127},~584.

				\bibitem[Encrenaz (2014)Encrenaz]{encrenaz2014infrared}
				Encrenaz, T.
				\newblock Infrared spectroscopy of exoplanets: observational constraints.
                 \newblock {\em Philos. Trans. R. Soc. Lond.}, {\bf 2014}, \emph{372},~20130083.

				\bibitem[Pluriel(2023)Pluriel]{pluriel2023hot}
				Pluriel, W.
				\newblock Hot Exoplanetary Atmospheres in 3D.
				\newblock {\em Remote Sens.} {\bf 2023}, {\em 15},~635.

				\bibitem[De~Pater(1991)]{depatermicrowaves}
				De~Pater, I.
				\newblock The significance of microwave observations for the planets.
				\newblock {\em Phys. Rep.} {\bf 1991}, {\em 200},~1--50.
				
				\bibitem[Encrenaz and Moreno(2002)]{encrenaz2002microwave}
				Encrenaz, T.; Moreno, R.
				\newblock The microwave spectra of planets. In
				\newblock  \emph{AIP Conference Proceedings}; American Institute of Physics: College Park, MD, USA,  2002; 
				Volume 616, pp. 330--337.
				
				\bibitem[de~Pater \em{et~al.}(1991)de~Pater, Romani, and
				Atreya]{depaterh2spossible}
				de~Pater, I.; Romani, P.N.; Atreya, S.K.
				\newblock Possible microwave absorption by H2S gas in Uranus' and Neptune's
				atmospheres.
				\newblock {\em Icarus} {\bf 1991}, {\em 91},~220--233.
				
				\bibitem[Naylor \em{et~al.}(1991)Naylor, Clark, Schultz, and
				Davis]{naylor1991atmospheric}
				Naylor, D.A.; Clark, T.A.; Schultz, A.A.; Davis, G.R.
				\newblock Atmospheric transmission at submillimetre wavelengths from Mauna Kea.
				\newblock {\em Mon. Not. R. Astron. Soc.} {\bf 1991},
				{\em 251},~199--202.

   
				\bibitem[Gebbie \em{et~al.}(1951)Gebbie, Harding, Hilsum, Pryce, and
				Roberts]{gebbie1951atmospheric}
				Gebbie, H.; Harding, W.; Hilsum, C.; Pryce, A.; Roberts, V.
				\newblock Atmospheric transmission in the 1 to 14 $\mu$ region.
				\newblock {\em Proc. R. Soc. Lond. Ser. A 
					Math. Phys. Sci.} {\bf 1951}, {\em 206},~87--107.
				
				\bibitem[Taylor and Yates(1957)]{taylor1957atmospherictrans}
				Taylor, J.H.; Yates, H.W.
				\newblock Atmospheric transmission in the infrared.
				\newblock {\em JOSA} {\bf 1957}, {\em 47},~223--226.


				\bibitem[Smette \em{et~al.}(2015)]{smette2015molecfit}
				Smette, A.; Sana, H.; Noll, S.; Horst, H.; Kausch, W.; Kimeswenger, S.; Barden, M., Szyszka, C.; Jones, A.M.; Gallenne, A.; Vinther, J.
				\newblock Molecfit: A general tool for telluric absorption correction-I. Method and application to ESO instruments.
			    \newblock {\em Astron. Astrophys.} {\bf 2015}, {\em 576},~A77.


    
				\bibitem[Thomas \em{et~al.}(1976)Thomas, Robinson, and
				Hyland]{thomas1976intermediate}
				Thomas, J.; Robinson, G.; Hyland, A.
				\newblock Intermediate Bandwidth Spectrometry in the 10-Micron Region and its
				Interpretation.
				\newblock {\em Mon. Not. R. Astron. Soc.} {\bf 1976},
				{\em 174},~711--723.

				\bibitem[Wiedemann (1996) Wiedemann]{wiedemann1996science} Wiedemann, G.
				\newblock Science with the VLT: high-resolution infrared spectroscopy.
				\newblock {\em The Messegner} {\bf 1996},
				{\em 86},~24--30.

                \bibitem[Kaspar \em{et~al.} (2017) Kasper,  Arsenault,  Käufl,  Jakob,  Fuenteseca,  Riquelme,  Siebenmorgen,  Sterzik,  Zins,  Ageorges,  Gutruf]{kaspar2017near} Kasper, M.; Arsenault, R.; Käufl, H.U.; Jakob, G.; Fuenteseca, E.; Riquelme, M.; Siebenmorgen, R.; Sterzik, M.; Zins, G.; Ageorges, N.; Gutruf, S.
                \newblock NEAR: low-mass planets in  Cen with VISIR. 
                \newblock {\em The Messenger} {\bf 2017}, 
                {\em 169}.



				\bibitem[Kendrew \em{et~al.}(2010)Kendrew, Jolissaint, Brandl, Lenzen, Pantin,
				Glasse, Blommaert, Venema, Siebenmorgen, and Molster]{kendrew2010mid}
				Kendrew, S.; Jolissaint, L.; Brandl, B.; Lenzen, R.; Pantin, E.; Glasse, A.;
				Blommaert, J.; Venema, L.; Siebenmorgen, R.; Molster, F.
				\newblock Mid-infrared astronomy with the E-ELT: performance of METIS. In
				\newblock  \emph{Ground-based and Airborne Instrumentation for Astronomy III}; SPIE:  %MDPI: please add location (city, country) of publisher 
                Bellingham, WA USA
				2010; Volume 7735, pp. 2017--2029.


				\bibitem[Nolle \em{et~al.}(2013) Noll, Kausch, Barden,Jones, Szyszka, Kimeswenger]{noll2013cerro}
				Noll, S.; Kausch, W.; Barden, M.; Jones, A.M.; Szyszka, C.; Kimeswenger, S.
				\newblock The cerro paranal advanced sky model
				\newblock {\em VLT-MAN-ESO-19550-5339} {\bf 2013}, {\em 1.1.1},~50.

                \bibitem[Holzlohner \em{et~al.}(2021) Holzlohner, Kimeswenger,  Kausch, Noll] {holzlohner2021bolometric} Holzlohner, R.; Kimeswenger, S.; Kausch, W.; Noll, S.
                \newblock Bolometric night sky temperature and subcooling of telescope structures.
                \newblock {\em Astron. Astrophys.}  {\bf 2021}, 
                {\em 645},~A32

                \bibitem[Papoular(1983)]{papoularchopnod1983}
				Papoular, R.
				\newblock The processing of infrared sky noise by chopping, nodding and
				filtering.
				\newblock {\em Astron. Astrophys.} {\bf 1983}, {\em 117},~46--52.

				\bibitem[Roman \em{et~al.}(2020)Roman, Fletcher, Orton, Rowe-Gurney, and
				Irwin]{roman2020uranus}
				Roman, M.T.; Fletcher, L.N.; Orton, G.S.; Rowe-Gurney, N.; Irwin, P.G.
				\newblock Uranus in northern midspring: persistent atmospheric temperatures and
				circulations inferred from thermal imaging.
				\newblock {\em Astron. J.} {\bf 2020}, {\em 159},~45.
    
				\bibitem[Violle(1874)]{violle1874temperaturesun}
				Violle, J.
				\newblock On the temperature of the sun.
				\newblock {\em  Lond. Edinb.   Dublin Philos. Mag.  
					J. Sci.} {\bf 1874}, {\em 48},~395--398.
				
				\bibitem[Langley(1878)]{langley1878temperaturesun}
				Langley, S.P.
				\newblock On the Temperature of the Sun. In
				\newblock  \emph{Proceedings of the American Academy of Arts and Sciences}; JSTOR: Cambridge, MA USA  %MDPI: please add location (city, country) of publisher
				1878; Volume~14, pp. 106--113.
				
				\bibitem[Moses \em{et~al.}(2004)Moses, Fouchet, Yelle, Friedson, Orton,
				B{\'e}zard, Drossart, Gladstone, Kostiuk, and
				Livengood]{moses2004stratospherejupiter}
				Moses, J.I.; Fouchet, T.; Yelle, R.V.; Friedson, A.J.; Orton, G.S.; B{\'e}zard,
				B.; Drossart, P.; Gladstone, G.R.; Kostiuk, T.; Livengood, T.A.
				\newblock The stratosphere of Jupiter. In
				\newblock {\em Jupiter: Planet, Satellites and Magnetosphere}; Cambridge, UK {2004}; pp.   129--157. %MDPI: please add publisher and publisher's location (city, country)
				
				\bibitem[Moses \em{et~al.}(2005)Moses, Fouchet, B{\'e}zard, Gladstone,
				Lellouch, and Feuchtgruber]{moses2005photochemistry}
				Moses, J.; Fouchet, T.; B{\'e}zard, B.; Gladstone, G.; Lellouch, E.;
				Feuchtgruber, H.
				\newblock Photochemistry and diffusion in Jupiter's stratosphere: constraints
				from ISO observations and comparisons with other giant planets.
				\newblock {\em J. Geophys. Res. Planets} {\bf 2005}, {\em 110}. https://doi.org/10.1029/2005JE002411.
				
				\bibitem[Fouchet \em{et~al.}(2009)Fouchet, Moses, and
				Conrath]{fouchet2009saturn}
				Fouchet, T.; Moses, J.I.; Conrath, B.J.
				\newblock Saturn: Composition and chemistry. In
				\newblock {\em Saturn from Cassini-Huygens}; Springer Dordrecht {\bf 2009}; pp. 83--112. %MDPI: please add publisher and publisher's location (city, country)
				
				
				\bibitem[Moses \em{et~al.}(2020)Moses, Cavali{\'e}, Fletcher, and
				Roman]{moses2020icegiantchem}
				Moses, J.; Cavali{\'e}, T.; Fletcher, L.; Roman, M.
				\newblock Atmospheric chemistry on Uranus and Neptune.
				\newblock {\em Philos. Trans. R. Soc. A} {\bf 2020},
				{\em 378},~20190477.
				
				\bibitem[Irwin \em{et~al.}(2008)Irwin, Teanby, De~Kok, Fletcher, Howett, Tsang,
				Wilson, Calcutt, Nixon, and Parrish]{irwin2008nemesis}
				Irwin, P.; Teanby, N.; De~Kok, R.; Fletcher, L.; Howett, C.; Tsang, C.; Wilson,
				C.; Calcutt, S.; Nixon, C.; Parrish, P.
				\newblock The NEMESIS planetary atmosphere radiative transfer and retrieval
				tool.
				\newblock {\em J. Quant. Spectrosc. Radiat. Transf.}
				{\bf 2008}, {\em 109},~1136--1150.
				
				\bibitem[Kuiper(1949)]{kuiper1949new}
				Kuiper, G.P.
				\newblock New absorptions in the uranian atmosphere.
				\newblock {\em Astrophys. J.} {\bf 1949}, {\em 109},~540--541.
				
				\bibitem[Weidenschilling and Lewis(1973)]{weidenschilling1973atmospheric}
				Weidenschilling, S.; Lewis, J.
				\newblock Atmospheric and cloud structures of the Jovian planets.
				\newblock {\em Icarus} {\bf 1973}, {\em 20},~465--476.
				
				\bibitem[Orton and Ingersoll(1980)]{orton1980saturn}
				Orton, G.S.; Ingersoll, A.P.
				\newblock Saturn's atmospheric temperature structure and heat budget.
				\newblock {\em J. Geophys. Res. Space Phys.} {\bf 1980}, {\em
					85},~5871--5881.
				
				\bibitem[Conrath and Pirraglia(1983)]{conrath1983thermalSaturn}
				Conrath, B.; Pirraglia, J.
				\newblock Thermal structure of Saturn from Voyager infrared measurements:Implications for atmospheric dynamics. 
				\newblock {\em Icarus} {\bf 1983}, {\em 53},~286--292.
				
				\bibitem[Fletcher \em{et~al.}(2009)Fletcher, Orton, Teanby, Irwin, and
				Bjoraker]{fletcher2009methane}
				Fletcher, L.; Orton, G.; Teanby, N.; Irwin, P.; Bjoraker, G.
				\newblock Methane and its isotopologues on Saturn from Cassini/CIRS
				observations.
				\newblock {\em Icarus} {\bf 2009}, {\em 199},~351--367.
				
				\bibitem[Fletcher \em{et~al.}(2017)Fletcher, de~Pater, Reach, Wong, Orton,
				Irwin, and Gehrz]{fletcher2017jupiter}
				Fletcher, L.N.; de~Pater, I.; Reach, W.; Wong, M.; Orton, G.; Irwin, P.; Gehrz,
				R.
				\newblock Jupiter’s para-H$_2$ distribution from SOFIA/FORCAST and Voyager/IRIS
				17--37 $\upmu$m spectroscopy. %MDPI: Refs. 37 and 281 are duplicated. Please remove one of them and rearrange all the references to appear in numerical order. Please ensure that there are no duplicated references.
				\newblock {\em Icarus} {\bf 2017}, {\em 286},~223--240.
				
				\bibitem[Burgdorf \em{et~al.}(2006)Burgdorf, Orton, van Cleve, Meadows, and
				Houck]{burgdorf2006detection}
				Burgdorf, M.; Orton, G.; van Cleve, J.; Meadows, V.; Houck, J.
				\newblock Detection of new hydrocarbons in Uranus' atmosphere by infrared
				spectroscopy.
				\newblock {\em Icarus} {\bf 2006}, {\em 184},~634--637.
				
				\bibitem[Feuchtgruber \em{et~al.}(1997)Feuchtgruber, Lellouch, de~Graauw,
				B{\'e}zard, Encrenaz, and Griffin]{feuchtgruber1997external}
				Feuchtgruber, H.; Lellouch, E.; de~Graauw, T.; B{\'e}zard, B.; Encrenaz, T.;
				Griffin, M.
				\newblock External supply of oxygen to the atmospheres of the giant planets.
				\newblock {\em Nature} {\bf 1997}, {\em 389},~159--162.
				
				\bibitem[Orton(1975)]{orton1975thermal}
				Orton, G.S.
				\newblock The thermal structure of Jupiter II. Observations and analysis of
				8--14 micron radiation.
				\newblock {\em Icarus} {\bf 1975}, {\em 26},~142--158.
				
				\bibitem[Orton \em{et~al.}(1991)Orton, Friedson, Baines, Martin, West,
				Caldwell, Hammel, Bergstralh, Malcom, Golisch, et~al.]{orton1991thermal}
				Orton, G.S.; Friedson, A.J.; Baines, K.H.; Martin, T.Z.; West, R.A.; Caldwell,
				J.; Hammel, H.B.; Bergstralh, J.T.; Malcom, M.E.; Golisch, W.F.;  et~al.
				\newblock Thermal maps of Jupiter: Spatial organization and time dependence of
				stratospheric temperatures, 1980 to 1990.
				\newblock {\em Science} {\bf 1991}, {\em 252},~537--542.
				
				\bibitem[Fletcher \em{et~al.}(2010)Fletcher, Orton, Mousis, Yanamandra-Fisher,
				Parrish, Irwin, Fisher, Vanzi, Fujiyoshi, Fuse, et~al.]{fletcher2010thermal}
				Fletcher, L.N.; Orton, G.; Mousis, O.; Yanamandra-Fisher, P.; Parrish, P.;
				Irwin, P.; Fisher, B.; Vanzi, L.; Fujiyoshi, T.; Fuse, T.;  et~al.
				\newblock Thermal structure and composition of Jupiter’s Great Red Spot from
				high-resolution thermal imaging.
				\newblock {\em Icarus} {\bf 2010}, {\em 208},~306--328.
				
				\bibitem[Fletcher \em{et~al.}(2016)Fletcher, Greathouse, Orton, Sinclair,
				Giles, Irwin, and Encrenaz]{fletcher2016mid}
				Fletcher, L.N.; Greathouse, T.; Orton, G.; Sinclair, J.; Giles, R.; Irwin, P.;
				Encrenaz, T.
				\newblock Mid-infrared mapping of Jupiter’s temperatures, aerosol opacity and
				chemical distributions with IRTF/TEXES.
				\newblock {\em Icarus} {\bf 2016}, {\em 278},~128--161.
				
				\bibitem[Fletcher \em{et~al.}(2017)Fletcher, Orton, Rogers, Giles, Payne,
				Irwin, and Vedovato]{fletcher2017moist}
				Fletcher, L.N.; Orton, G.; Rogers, J.; Giles, R.; Payne, A.; Irwin, P.;
				Vedovato, M.
				\newblock Moist convection and the 2010--2011 revival of Jupiter’s South
				Equatorial Belt.
				\newblock {\em Icarus} {\bf 2017}, {\em 286},~94--117.
				
				\bibitem[Antu{\~n}ano \em{et~al.}(2020)Antu{\~n}ano, Fletcher, Orton, Toledo,
				Melin, Roman, Sinclair, Donnelly, Morton, and
				Selves]{antunano2020characterizing}
				Antu{\~n}ano, A.; Fletcher, L.N.; Orton, G.S.; Toledo, D.; Melin, H.; Roman,
				M.T.; Sinclair, J.A.; Donnelly, P.T.; Morton, E.K.; Selves, P.
				\newblock Characterizing temperature and aerosol variability during Jupiter's
				2006--2007 Equatorial Zone disturbance.
				\newblock {\em J. Geophys. Res. Planets} {\bf 2020}, {\em
					125},~e2020JE006413.
				
				\bibitem[Courtin \em{et~al.}(1984)Courtin, Gautier, Marten, B{\'e}zard, and
				Hanel]{courtin1984compositionsaturn}
				Courtin, R.; Gautier, D.; Marten, A.; B{\'e}zard, B.; Hanel, R.
				\newblock The composition of Saturn's atmosphere at northern temperate
				latitudes from Voyager IRIS spectra-NH3, PH3, C2H2, C2H6, CH3D, CH4, and the
				Saturnian D/H isotopic ratio.
				\newblock {\em Astrophys. J.} {\bf 1984}, {\em 287},~899--916.
				
				\bibitem[Noll \em{et~al.}(1986)Noll, Knacke, Tokunaga, Lacy, Beck, and
				Serabyn]{noll1986abundances}
				Noll, K.; Knacke, R.; Tokunaga, A.; Lacy, J.; Beck, S.; Serabyn, E.
				\newblock The abundances of ethane and acetylene in the atmospheres of Jupiter
				and Saturn.
				\newblock {\em Icarus} {\bf 1986}, {\em 65},~257--263.
				
				\bibitem[Sada \em{et~al.}(1996)Sada, McCabe, Bjoraker, Jennings, and
				Reuter]{sada199613c}
				Sada, P.V.; McCabe, G.H.; Bjoraker, G.L.; Jennings, D.E.; Reuter, D.C.
				\newblock 13C-ethane in the atmospheres of Jupiter and Saturn.
				\newblock {\em Astrophys. J.} {\bf 1996}, {\em 472},~903.
				
				\bibitem[Fletcher \em{et~al.}(2007)Fletcher, Irwin, Teanby, Orton, Parrish,
				Calcutt, Bowles, de~Kok, Howett, and Taylor]{fletcher2007meridional}
				Fletcher, L.; Irwin, P.; Teanby, N.; Orton, G.; Parrish, P.; Calcutt, S.;
				Bowles, N.; de~Kok, R.; Howett, C.; Taylor, F.
				\newblock The meridional phosphine distribution in Saturn's upper troposphere
				from Cassini/CIRS observations.
				\newblock {\em Icarus} {\bf 2007}, {\em 188},~72--88.
				
				\bibitem[Fletcher \em{et~al.}(2008)Fletcher, Irwin, Orton, Teanby, Achterberg,
				Bjoraker, Read, Simon-Miller, Howett, de~Kok,
				et~al.]{fletcher2008temperature}
				Fletcher, L.; Irwin, P.; Orton, G.; Teanby, N.; Achterberg, R.; Bjoraker, G.;
				Read, P.; Simon-Miller, A.; Howett, C.; de~Kok, R.;  et~al.
				\newblock Temperature and composition of Saturn's polar hot spots and hexagon.
				\newblock {\em Science} {\bf 2008}, {\em 319},~79--81.
				
				\bibitem[Hesman \em{et~al.}(2009)Hesman, Jennings, Sada, Bjoraker, Achterberg,
				Simon-Miller, Anderson, Boyle, Nixon, Fletcher, et~al.]{hesman2009saturn}
				Hesman, B.E.; Jennings, D.E.; Sada, P.V.; Bjoraker, G.L.; Achterberg, R.K.;
				Simon-Miller, A.A.; Anderson, C.M.; Boyle, R.J.; Nixon, C.A.; Fletcher, L.N.;
				et~al.
				\newblock Saturn's latitudinal C2H2 and C2H6 abundance profiles from
				Cassini/CIRS and ground-based observations.
				\newblock {\em Icarus} {\bf 2009}, {\em 202},~249--259.
				
				\bibitem[Fletcher \em{et~al.}(2009)Fletcher, Orton, Yanamandra-Fisher, Fisher,
				Parrish, and Irwin]{fletcher2009retrievals}
				Fletcher, L.; Orton, G.; Yanamandra-Fisher, P.; Fisher, B.; Parrish, P.; Irwin,
				P.
				\newblock Retrievals of atmospheric variables on the gas giants from
				ground-based mid-infrared imaging.
				\newblock {\em Icarus} {\bf 2009}, {\em 200},~154--175.
				
				\bibitem[Fletcher \em{et~al.}(2012)Fletcher, Hesman, Achterberg, Irwin,
				Bjoraker, Gorius, Hurley, Sinclair, Orton, Legarreta,
				et~al.]{fletcher2012originsaturnvortex}
				Fletcher, L.N.; Hesman, B.; Achterberg, R.; Irwin, P.; Bjoraker, G.; Gorius,
				N.; Hurley, J.; Sinclair, J.; Orton, G.; Legarreta, J.;  et~al.
				\newblock The origin and evolution of Saturn’s 2011--2012 stratospheric
				vortex.
				\newblock {\em Icarus} {\bf 2012}, {\em 221},~560--586.
				
				\bibitem[Fletcher \em{et~al.}(2018)Fletcher, Greathouse, Guerlet, Moses, and
				West]{fletcher2018saturn}
				Fletcher, L.N.; Greathouse, T.K.; Guerlet, S.; Moses, J.I.; West, R.A.
				\newblock Saturn’s Seasonally Changing Atmosphere. In
				\newblock {\em Saturn in the 21st Century}; {2018}; Cambridge, UK {Volume 20}, p.~251. %MDPI: please add publisher and publisher's location (city, country)
				
				\bibitem[Blake \em{et~al.}(2021)Blake, Fletcher, Greathouse, Orton, Melin,
				Roman, Antu{\~n}ano, Donnelly, Rowe-Gurney, and King]{blake2021refining}
				Blake, J.S.; Fletcher, L.N.; Greathouse, T.K.; Orton, G.S.; Melin, H.; Roman,
				M.T.; Antu{\~n}ano, A.; Donnelly, P.T.; Rowe-Gurney, N.; King, O.
				\newblock Refining Saturn’s deuterium-hydrogen ratio via IRTF/TEXES
				spectroscopy.
				\newblock {\em Astron. Astrophys.} {\bf 2021}, {\em 653},~A66.
				
				\bibitem[Blake \em{et~al.}(2023)Blake, Fletcher, Orton, Antu{\~n}ano, Roman,
				Kasaba, Fujiyoshi, Melin, Bardet, Sinclair, et~al.]{blake2023saturn}
				Blake, J.S.; Fletcher, L.N.; Orton, G.S.; Antu{\~n}ano, A.; Roman, M.T.;
				Kasaba, Y.; Fujiyoshi, T.; Melin, H.; Bardet, D.; Sinclair, J.A.;  et~al.
				\newblock Saturn’s seasonal variability from four decades of ground-based
				mid-infrared observations.
				\newblock {\em Icarus} {\bf 2023}, {\em 392},~115347.
				
				\bibitem[Tokunaga \em{et~al.}(1983)Tokunaga, Orton, and
				Caldwell]{tokunaga1983new}
				Tokunaga, A.; Orton, G.; Caldwell, J.
				\newblock New observational constraints on the temperature inversions of Uranus
				and Neptune.
				\newblock {\em Icarus} {\bf 1983}, {\em 53},~141--146.
				
				\bibitem[Orton \em{et~al.}(1983)Orton, Tokunaga, and
				Caldwell]{orton1983observational}
				Orton, G.S.; Tokunaga, A.T.; Caldwell, J.
				\newblock Observational constraints on the atmospheres of Uranus and Neptune
				from new measurements near 10 $\upmu$m.
				\newblock {\em Icarus} {\bf 1983}, {\em 56},~147--164.
				
				\bibitem[Orton \em{et~al.}(1992)Orton, Lacy, Achtermann, Parmar, and
				Blass]{orton1992thermalneptune}
				Orton, G.S.; Lacy, J.H.; Achtermann, J.M.; Parmar, P.; Blass, W.E.
				\newblock Thermal spectroscopy of Neptune: The stratospheric temperature,
				hydrocarbon abundances, and isotopic ratios.
				\newblock {\em Icarus} {\bf 1992}, {\em 100},~541--555.
				
				\bibitem[Fletcher \em{et~al.}(2010)Fletcher, Drossart, Burgdorf, Orton, and
				Encrenaz]{fletcher2010neptune}
				Fletcher, L.N.; Drossart, P.; Burgdorf, M.; Orton, G.; Encrenaz, T.
				\newblock Neptune's atmospheric composition from AKARI infrared spectroscopy.
				\newblock {\em Astron. Astrophys.} {\bf 2010}, {\em 514},~A17.
				
				\bibitem[Greathouse \em{et~al.}(2011)Greathouse, Richter, Lacy, Moses, Orton,
				Encrenaz, Hammel, and Jaffe]{greathouse2011spatially}
				Greathouse, T.K.; Richter, M.; Lacy, J.; Moses, J.; Orton, G.; Encrenaz, T.;
				Hammel, H.; Jaffe, D.
				\newblock A spatially resolved high spectral resolution study of Neptune’s
				stratosphere.
				\newblock {\em Icarus} {\bf 2011}, {\em 214},~606--621.
				
				\bibitem[Orton \em{et~al.}(2014)Orton, Moses, Fletcher, Mainzer, Hines, Hammel,
				Martin-Torres, Burgdorf, Merlet, and Line]{orton2014mid}
				Orton, G.S.; Moses, J.I.; Fletcher, L.N.; Mainzer, A.K.; Hines, D.; Hammel,
				H.B.; Martin-Torres, J.; Burgdorf, M.; Merlet, C.; Line, M.R.
				\newblock Mid-infrared spectroscopy of Uranus from the Spitzer infrared
				spectrometer: 2. Determination of the mean composition of the upper
				troposphere and stratosphere.
				\newblock {\em Icarus} {\bf 2014}, {\em 243},~471--493.
				
				\bibitem[Fletcher \em{et~al.}(2014)Fletcher, de~Pater, Orton, Hammel, Sitko,
				and Irwin]{fletcher2014neptune}
				Fletcher, L.N.; de~Pater, I.; Orton, G.S.; Hammel, H.B.; Sitko, M.L.; Irwin,
				P.G.
				\newblock Neptune at summer solstice: zonal mean temperatures from ground-based
				observations, 2003--2007.
				\newblock {\em Icarus} {\bf 2014}, {\em 231},~146--167.
				
				\bibitem[de~Pater \em{et~al.}(2014)de~Pater, Fletcher, Luszcz-Cook, DeBoer,
				Butler, Hammel, Sitko, Orton, and Marcus]{dePater2014neptune}
				de~Pater, I.; Fletcher, L.N.; Luszcz-Cook, S.; DeBoer, D.; Butler, B.; Hammel,
				H.B.; Sitko, M.L.; Orton, G.; Marcus, P.S.
				\newblock Neptune’s global circulation deduced from multi-wavelength
				observations.
				\newblock {\em Icarus} {\bf 2014}, {\em 237},~211--238.
				
				
				\bibitem[Roman \em{et~al.}(2022)Roman, Fletcher, Orton, Greathouse, Moses,
				Rowe-Gurney, Irwin, Antu{\~n}ano, Sinclair, Kasaba,
				et~al.]{roman2022subseasonal}
				Roman, M.T.; Fletcher, L.N.; Orton, G.S.; Greathouse, T.K.; Moses, J.I.;
				Rowe-Gurney, N.; Irwin, P.G.; Antu{\~n}ano, A.; Sinclair, J.; Kasaba, Y.;
				et~al.
				\newblock Subseasonal Variation in Neptune’s Mid-infrared Emission.
				\newblock {\em  Planet. Sci. J.} {\bf 2022}, {\em 3},~78.
				
				\bibitem[de~Pater \em{et~al.}(2010)de~Pater, Fletcher, P{\'e}rez-Hoyos, Hammel,
				Orton, Wong, Luszcz-Cook, S{\'a}nchez-Lavega, and
				Boslough]{depater2010multijupiter}
				de~Pater, I.; Fletcher, L.N.; P{\'e}rez-Hoyos, S.; Hammel, H.B.; Orton, G.S.;
				Wong, M.H.; Luszcz-Cook, S.; S{\'a}nchez-Lavega, A.; Boslough, M.
				\newblock A multi-wavelength study of the 2009 impact on Jupiter: Comparison of
				high resolution images from Gemini, Keck and HST.
				\newblock {\em Icarus} {\bf 2010}, {\em 210},~722--741.
				
				\bibitem[Orton \em{et~al.}(2014)Orton, Fletcher, Moses, Mainzer, Hines, Hammel,
				Martin-Torres, Burgdorf, Merlet, and Line]{orton2014a}
				Orton, G.S.; Fletcher, L.N.; Moses, J.I.; Mainzer, A.K.; Hines, D.; Hammel,
				H.B.; Martin-Torres, F.J.; Burgdorf, M.; Merlet, C.; Line, M.R.
				\newblock Mid-infrared spectroscopy of Uranus from the Spitzer Infrared
				Spectrometer: 1. Determination of the mean temperature structure of the upper
				troposphere and stratosphere.
				\newblock {\em Icarus} {\bf 2014}, {\em 243},~494--513.
				
				\bibitem[Orton \em{et~al.}(2015)Orton, Fletcher, Encrenaz, Leyrat, Roe,
				Fujiyoshi, and Pantin]{orton2015thermal}
				Orton, G.S.; Fletcher, L.N.; Encrenaz, T.; Leyrat, C.; Roe, H.G.; Fujiyoshi,
				T.; Pantin, E.
				\newblock Thermal imaging of Uranus: Upper-tropospheric temperatures one season
				after Voyager.
				\newblock {\em Icarus} {\bf 2015}, {\em 260},~94--102.
				
				\bibitem[Fletcher \em{et~al.}(2018)Fletcher, Orton, Sinclair, Guerlet, Read,
				Antu{\~n}ano, Achterberg, Flasar, Irwin, Bjoraker,
				et~al.]{fletcher2018hexagon}
				Fletcher, L.; Orton, G.; Sinclair, J.; Guerlet, S.; Read, P.; Antu{\~n}ano, A.;
				Achterberg, R.; Flasar, F.; Irwin, P.; Bjoraker, G.;  et~al.
				\newblock A hexagon in Saturn’s northern stratosphere surrounding the
					emerging summertime polar vortex. %MDPI: Refs. 70 and 248 are duplicated. Please remove one of them and rearrange all the references to appear in numerical order. Please ensure that there are no duplicated references.
				\newblock {\em Nat. Commun.} {\bf 2018}, {\em 9},~3564.
				
				\bibitem[Fletcher \em{et~al.}(2020)Fletcher, Orton, Greathouse, Rogers, Zhang,
				Oyafuso, Eichst{\"a}dt, Melin, Li, Levin, et~al.]{fletcher2020jupiter}
				Fletcher, L.N.; Orton, G.S.; Greathouse, T.K.; Rogers, J.H.; Zhang, Z.;
				Oyafuso, F.A.; Eichst{\"a}dt, G.; Melin, H.; Li, C.; Levin, S.M.;  et~al.
				\newblock Jupiter's equatorial plumes and hot spots: Spectral mapping from
				Gemini/TEXES and Juno/MWR.
				\newblock {\em J. Geophys. Res. Planets} {\bf 2020}, {\em
					125},~e2020JE006399.
				
				\bibitem[Gillett \em{et~al.}(1969)Gillett, Low, and Stein]{gillett1969jupspec}
				Gillett, F.; Low, F.; Stein, W.
				\newblock The 2.8--14-micron spectrum of Jupiter.
				\newblock {\em Astrophys. J.} {\bf 1969}, {\em 157},~925.
				
				\bibitem[Westphal(1969)]{westphal1969observations}
				Westphal, J.
				\newblock Observations of localised 5-micron radiation from Jupiter.
				\newblock {\em Astrophys. J.} {\bf 1969}, {\em 157},~L63--L64.
				
				\bibitem[Keay \em{et~al.}(1973)Keay, Low, Rieke, and Minton]{keay1973high}
				Keay, C.; Low, F.; Rieke, G.; Minton, R.
				\newblock High-resolution maps of Jupiter at 5 microns.
				\newblock {\em Astrophys. J.} {\bf 1973}, {\em 183},~1063--1074.
				
				\bibitem[Westphal \em{et~al.}(1974)Westphal, Matthews, and
				Terrile]{westphal1974five}
				Westphal, J.; Matthews, K.; Terrile, R.J.
				\newblock Five-micron pictures of Jupiter.
				\newblock {\em Astrophys. J.} {\bf 1974}, {\em 188},~L111--L112.
				
				\bibitem[Fink \em{et~al.}(1978)Fink, Larson, and Treffers]{fink1978germane}
				Fink, U.; Larson, H.P.; Treffers, R.R.
				\newblock Germane in the atmosphere of Jupiter.
				\newblock {\em Icarus} {\bf 1978}, {\em 34},~344--354.
				
				\bibitem[Bjoraker \em{et~al.}(1986)Bjoraker, Larson, and
				Kunde]{bjoraker1986gas}
				Bjoraker, G.L.; Larson, H.P.; Kunde, V.G.
				\newblock The gas composition of Jupiter derived from 5-$\upmu$m airborne
				spectroscopic observations.
				\newblock {\em Icarus} {\bf 1986}, {\em 66},~579--609.
				
				\bibitem[Bjoraker \em{et~al.}(2015)Bjoraker, Wong, De~Pater, and
				{\'A}d{\'a}mkovics]{bjoraker2015jupiter}
				Bjoraker, G.; Wong, M.; De~Pater, I.; {\'A}d{\'a}mkovics, M.
				\newblock Jupiter’s deep cloud structure revealed using Keck observations of
				spectrally resolved line shapes.
				\newblock {\em Astrophys. J.} {\bf 2015}, {\em 810},~122.
				
				\bibitem[Bjoraker \em{et~al.}(2018)Bjoraker, Wong, de~Pater, Hewagama,
				{\'A}d{\'a}mkovics, and Orton]{bjoraker2018gas}
				Bjoraker, G.L.; Wong, M.H.; de~Pater, I.; Hewagama, T.; {\'A}d{\'a}mkovics, M.;
				Orton, G.S.
				\newblock The gas composition and deep cloud structure of Jupiter's Great Red
				Spot.
				\newblock {\em Astron. J.} {\bf 2018}, {\em 156},~101.
				
				\bibitem[Bjoraker(2020)]{bjoraker2020jupiter}
				Bjoraker, G.L.
				\newblock Jupiter’s elusive water.
				\newblock {\em Nat. Astron.} {\bf 2020}, {\em 4},~558--559.
				
				\bibitem[Bjoraker \em{et~al.}(2022)Bjoraker, Wong, de~Pater, Hewagama, and
				{\'A}d{\'a}mkovics]{bjoraker2022spatial}
				Bjoraker, G.L.; Wong, M.H.; de~Pater, I.; Hewagama, T.; {\'A}d{\'a}mkovics, M.
				\newblock The Spatial Variation of Water Clouds, NH3, and H2O on Jupiter Using
				Keck Data at 5 Microns.
				\newblock {\em Remote Sens.} {\bf 2022}, {\em 14},~4567.
				
				\bibitem[Wong \em{et~al.}(2023)Wong, Bjoraker, Goullaud, Stephens, Luszcz-Cook,
				Atreya, de~Pater, and Brown]{wong2023deep}
				Wong, M.H.; Bjoraker, G.L.; Goullaud, C.; Stephens, A.W.; Luszcz-Cook, S.H.;
				Atreya, S.K.; de~Pater, I.; Brown, S.T.
				\newblock Deep Clouds on Jupiter.
				\newblock {\em Remote Sens.} {\bf 2023}, {\em 15},~702.
				
				\bibitem[Momary \em{et~al.}(2006)Momary, Baines, Team,
				et~al.]{momary2006zoology}
				Momary, T.W.; Baines, K.; Cassini/VIMS Science Team.
				\newblock The zoology of Saturn: The bizarre features unveiled by the 5 micron
				eyes of Cassini/VIMS. In
				\newblock  \emph{AAS/Division for Planetary Sciences Meeting Abstracts\# 38 AAS Washington, DC USA};  2006; 
				Volume~38, pp. 11--21. %MDPI: please add publisher and publisher's location (city, country)
				
				\bibitem[Bjoraker \em{et~al.}(2006)Bjoraker, Chanover, Glenar, and
				Hewagama]{bjoraker2006ammonia}
				Bjoraker, G.; Chanover, N.; Glenar, D.; Hewagama, T.
				\newblock Ammonia, phosphine, and cloud structure on Saturn derived from
				5-micron spectra. In
				\newblock  \emph{AAS/Division for Planetary Sciences Meeting Abstracts\# 38 AAS Washington, DC USA};  2006,
				Volume~38, p. 488. %MDPI: please add publisher and publisher's location (city, country)
				%MDPI: newly added page, please confirm
				
				\bibitem[Bjoraker \em{et~al.}(2007)Bjoraker, Chanover, Glenar, and
				Hewagama]{bjoraker2007saturn}
				Bjoraker, G.; Chanover, N.; Glenar, D.; Hewagama, T.
				\newblock Saturn's deep cloud structure derived from 5-micron spectra. In
				\newblock  \emph{AGU Fall Meeting Abstracts};  2007; Volume 2007, p. P31A-0184. %MDPI: please add publisher and publisher's location (city, country)
				
				\bibitem[Fletcher \em{et~al.}(2011)Fletcher, Baines, Momary, Showman, Irwin,
				Orton, Roos-Serote, and Merlet]{fletcher2011saturn}
				Fletcher, L.N.; Baines, K.H.; Momary, T.W.; Showman, A.P.; Irwin, P.G.; Orton,
				G.S.; Roos-Serote, M.; Merlet, C.
				\newblock Saturn’s tropospheric composition and clouds from Cassini/VIMS
				4.6--5.1 $\upmu$m nightside spectroscopy.
				\newblock {\em Icarus} {\bf 2011}, {\em 214},~510--533.
				
				\bibitem[Barstow \em{et~al.}(2016)Barstow, Irwin, Fletcher, Giles, and
				Merlet]{barstow2016probing}
				Barstow, J.K.; Irwin, P.G.; Fletcher, L.N.; Giles, R.S.; Merlet, C.
				\newblock Probing Saturn’s tropospheric cloud with Cassini/VIMS.
				\newblock {\em Icarus} {\bf 2016}, {\em 271},~400--417.
				
				\bibitem[Yanamandra-Fisher \em{et~al.}(2015)Yanamandra-Fisher, Gutierrez,
				Payne, Orton, and Sinclair]{yanamandra2015probing}
				Yanamandra-Fisher, P.A.; Gutierrez, S.M.; Payne, A.; Orton, G.S.; Sinclair, J.
				\newblock Probing the Depths of Jupiter and Saturn at Five-Microns. In
				\newblock  \emph{AAS/Division for Planetary Sciences Meeting Abstracts\# 47};  2015;
				Volume~47, pp. 311--335. %MDPI: please add publisher and publisher's location (city, country)
				
				\bibitem[Encrenaz \em{et~al.}(2004)Encrenaz, Lellouch, Drossart, Feuchtgruber,
				Orton, and Atreya]{encrenaz2004first}
				Encrenaz, T.; Lellouch, E.; Drossart, P.; Feuchtgruber, H.; Orton, G.S.;
				Atreya, S.K.
				\newblock First detection of CO in Uranus.
				\newblock {\em Astron. Astrophys.} {\bf 2004}, {\em 413},~L5--L9.
				
				\bibitem[Hodapp \em{et~al.}(2003)Hodapp, Jensen, Irwin, Yamada, Chung,
				Fletcher, Robertson, Hora, Simons, Mays, et~al.]{hodapp2003gemini}
				Hodapp, K.W.; Jensen, J.B.; Irwin, E.M.; Yamada, H.; Chung, R.; Fletcher, K.;
				Robertson, L.; Hora, J.L.; Simons, D.A.; Mays, W.;  et~al.
				\newblock The Gemini Near-Infrared Imager (NIRI).
				\newblock {\em Publ. Astron. Soc. Pac.} {\bf
					2003}, {\em 115},~1388.
				
				\bibitem[gem()]{gemini5micron}
				International Gemini Observatory/NOIRLab/NSF/AURA, M.H. Wong, (UC Berkeley) et
				al. Available online: \url{https://noirlab.edu/public/images/noirlab2116a/} (accessed on 21 December 2022). %MDPI: please add access date (day month year)
				
				\bibitem[Herschel(1800)]{herschel1800xiv}
				Herschel, W.
				\newblock XIV. Experiments on the refrangibility of the invisible rays of the
				sun.
				\newblock {\em Philos. Trans. R. Soc. Lond.} {\bf
					1800}, \emph{90}, 284--292.
				
				\bibitem[Coblentz(1949)]{coblentz_review}
				Coblentz, W.W.
				\newblock Early History of Infrared Spectroradiometry.
				\newblock {\em  Sci. Mon.} {\bf 1949}, {\em 68},~102--107.
				
				\bibitem[Pasachoff(1985)]{pasachoff1985contemporary}
				Pasachoff, J.M.
				\newblock \emph{Contemporary Astronomy}; 
				\newblock {Saunders College Pub: Philadelphia, Pennsylvania}, {1985}.
				
				\bibitem[Walker(2000)]{walker2000briefhistory}
				Walker, H.J.
				\newblock A brief history of infrared astronomy.
				\newblock {\em Astron. Geophys.} {\bf 2000}, {\em 41},~5--10.
				
				\bibitem[Langley(1880)]{langley1880bolometer}
				Langley, S.P.
				\newblock The bolometer and radiant energy.
				\newblock  In
				\newblock  \emph{Proceedings of the American Academy of Arts and Sciences}; JSTOR: Cambridge, MA USA %MDPI: please add location (city, country) of publisher
				1880; Volume~16, pp. 342--358.
				
				\bibitem[Holland \em{et~al.}(2002)Holland, Duncan, and
				Griffin]{holland2002bolometers}
				Holland, W.; Duncan, W.; Griffin, M.
				\newblock Bolometers for submillimeter and millimeter astronomy. In 
				\newblock  \emph{Single-Dish Radio Astronomy: Techniques and Applications};   2002,
				Volume 278, pp. 463--491.
				
				\bibitem[Balog \em{et~al.}(2014)Balog, M{\"u}ller, Nielbock, Altieri, Klaas,
				Blommaert, Linz, Lutz, Mo{\'o}r, Billot, et~al.]{balog2014herschel}
				Balog, Z.; M{\"u}ller, T.; Nielbock, M.; Altieri, B.; Klaas, U.; Blommaert, J.;
				Linz, H.; Lutz, D.; Mo{\'o}r, A.; Billot, N.;  et~al.
				\newblock The Herschel-PACS photometer calibration.
				\newblock {\em Exp. Astron.} {\bf 2014}, {\em 37},~129--160.
				
				\bibitem[Stroke(1967)]{stroke1967diffraction}
				Stroke, G.W.
				\newblock Diffraction gratings.
				\newblock {\em Handbuch der Physik} {\bf 1967}, {\em 29},~426--754.
				
				\bibitem[Palik(1977)]{palik1977history}
				Palik, E.
				\newblock History of far-infrared research. I. The Rubens era.
				\newblock {\em JOSA} {\bf 1977}, {\em 67},~857--865.
				
				\bibitem[Rosse(1870)]{rosse1870radiation}
				Rosse, E.o.
				\newblock On the Radiation of Heat from the Moon.--No. II.
				\newblock {\em Proc. R. Soc. Lond.} {\bf 1870}, {\em
					19},~9--14.
				
				\bibitem[Langley(1889)]{langley1889temperature}
				Langley, S.P.
				\newblock {\em The Temperature of the Moon: From Studies at the Allegheny
					Observatory by SP Langley, With the Assistance of FW Very};  National Acadamey of Sciences, Washington, DC USA:  1889; Volume~7. %MDPI: please add publisher's location (city, country)
				
				\bibitem[Boys(1890)]{boys1890iii}
				Boys, C.V.
				\newblock III. On the heat of the Moon and stars.
				\newblock {\em Proc. R. Soc. Lond.} {\bf 1890}, {\em
					47},~480--499.
				
				\bibitem[Very(1898{\natexlab{a}})]{very1898probable1}
				Very, F.W.
				\newblock The Probable Range of Temperature on the Moon. I.
				\newblock {\em Astrophys. J.} {\bf 1898}, {\em 8},~199.
				
				\bibitem[Very(1898{\natexlab{b}})]{very1898probable2}
				Very, F.W.
				\newblock The Probable Range of Temperature on the Moon. II.
				\newblock {\em Astrophys. J.} {\bf 1898}, {\em 8},~265.
				
				\bibitem[Coblentz(1922)]{coblentz1922further}
				Coblentz, W.
				\newblock Further Measurements of Stellar Temperatures and Planetary Radiation.
				\newblock {\em Proc. Natl. Acad. Sci. USA} {\bf 1922},
				{\em 8},~330--333.
				
				\bibitem[Coblentz and Lampland(1923)]{coblentz1923measurments}
				Coblentz, W.W.; Lampland, C.O.
				\newblock Measurments of planetary radiation.  
				\newblock {\em Lowell Obs. Bull.}  {{\textbf{1923}}}, 
				{\emph{3}}, 91--134. %MDPI: please add publisher and publisher's location (city, country)
				
				\bibitem[Coblentz and Lampland(1924)]{coblentz1924new}
				Coblentz, W.W.; Lampland, C.
				\newblock New measurements of planetary radiation.
				\newblock {\em Science} {\bf 1924}, {\em 60},~295--295.
				
				\bibitem[Coblentz \em{et~al.}(1925)Coblentz, Lampland,
				et~al.]{coblentz1925some}
				Coblentz, W.W.; Lampland, C. 
				\newblock Some measurements of the spectral components of planetary radiation
				and planetary temperatures.
				\newblock {\em J. Frankl. Inst.} {\bf 1925}, {\em
					199},~785--841.
				
				\bibitem[Coblentz \em{et~al.}(1927)Coblentz, Lampland, and
				Menzel]{coblentz1927temperatures}
				Coblentz, W.W.; Lampland, C.; Menzel, D.
				\newblock Temperatures of Mars, 1926, as derived from the Water-Cell
				Transmissions.
				\newblock {\em Publ. Astron. Soc. Pac.} {\bf
					1927}, {\em 39},~97--100.
				
				\bibitem[Pettit and Nicholson(1923)]{pettit1923measurements}
				Pettit, E.; Nicholson, S.B.
				\newblock Measurements of the Radiation from the Planet Mercury.
				\newblock {\em Publ. Astron. Soc. Pac.} {\bf
					1923}, {\em 35},~194--198.
				
				\bibitem[Pettit and Nicholson(1924)]{pettit1924radiation}
				Pettit, E.; Nicholson, S.B.
				\newblock Radiation measures on the planet Mars.
				\newblock {\em Publ. Astron. Soc. Pac.} {\bf
					1924}, {\em 36},~269--272.
				
				\bibitem[Menzel \em{et~al.}(1926)Menzel, Coblentz, and
				Lampland]{menzel1926planetary}
				Menzel, D.; Coblentz, W.; Lampland, C.
				\newblock Planetary temperatures derived from water-cell transmissions.
				\newblock {\em Astrophys. J.} {\bf 1926}, {\em 63},~177--187.
				
				\bibitem[Menzel(1923)]{menzel1923water}
				Menzel, D.H.
				\newblock Water-cell transmissions and planetary temperatures.
				\newblock {\em Astrophys. J.} {\bf 1923}, {\em 58},~65.
				
				\bibitem[Tholen \em{et~al.}(2002)Tholen, Tejfel, and
				Cox]{Astrophysical_Quantities2002}
				Tholen, D.J.; Tejfel, V.G.; Cox, A.N., Planets and Satellites.
				\newblock In {\em Allen's Astrophysical Quantities}; Cox, A.N., Ed.; Springer: New York, NY,  USA, 2002; pp. 293--313.
				\newblock
				{\changeurlcolor{black}\href{https://doi.org/10.1007/978-1-4612-1186-0_12}{\detokenize{https://doi.org/10.1007/978-1-4612-1186-0_12}}}.
				
				\bibitem[Menzel(1930)]{menzel1930hydrogen}
				Menzel, D.H.
				\newblock Hydrogen abundance and the constitution of the giant planets.
				\newblock {\em Publ. Astron. Soc. Pac.} {\bf
					1930}, {\em 42},~228--232.
				
				\bibitem[Sinton and Strong(1960)]{sinton1960radiometric}
				Sinton, W.M.; Strong, J.
				\newblock Radiometric Observations of Venus.
				\newblock {\em Astrophys. J.} {\bf 1960}, {\em 131},~470.
				
				\bibitem[Pfund(1929)]{pfund1929resonance}
				Pfund, A.
				\newblock Resonance radiometry.
				\newblock {\em Science} {\bf 1929}, {\em 69},~71--72.
				
				\bibitem[Kivenson \em{et~al.}(1948)Kivenson, Steinback, and
				Rider]{kivenson1948infra}
				Kivenson, G.; Steinback, R.T.; Rider, M.
				\newblock An Infra-Red Chopped-Radiation Analyzer.
				\newblock {\em JOSA} {\bf 1948}, {\em 38},~1086--1091.
				
				\bibitem[Westphal \em{et~al.}(1963)Westphal, Murray, and
				Martz]{westphal_ir_200in_1963}
				Westphal, J.A.; Murray, B.C.; Martz, D.E.
				\newblock An 8--14 micron infrared astronomical photometer.
				\newblock {\em Appl. Opt.} {\bf 1963}, {\em 2},~749--753.
				
				
				\bibitem[Murray \em{et~al.}(1964)Murray, Wildey, and
				Westphal]{murray1964observations}
				Murray, B.C.; Wildey, R.L.; Westphal, J.A.
				\newblock Observations of Jupiter and the Galilean Satellites at 10 Microns.
				\newblock {\em Astrophys. J.} {\bf 1964}, {\em 139},~986.
				
				\bibitem[Low(1964)]{low1964infraredsaturn}
				Low, F.
				\newblock Infrared Brightness Temperature of Saturn.
				\newblock {\em Astron. J.} {\bf 1964}, {\em 69},~550--551.
				
				\bibitem[Low(1966{\natexlab{a}})]{low1966observations}
				Low, F.J.
				\newblock Observations of Venus, Jupiter, and Saturn at $\lambda$20 $\upmu$.
				\newblock {\em Astron. J.} {\bf 1966}, {\em 71},~391.
				
				\bibitem[Low(1966{\natexlab{b}})]{low1966infrareduranus}
				Low, F.J.
				\newblock The infrared brightness temperature of Uranus.
				\newblock {\em Astrophys. J.} {\bf 1966}, {\em 146},~326--328.
				
				\bibitem[Harwit \em{et~al.}(1966)Harwit, Munutt, Shivanandan, and
				Zajac]{harwit1966results}
				Harwit, M.; Munutt, D.; Shivanandan, K.; Zajac, B.
				\newblock Results of the first infrared astronomical rocket flight.
				\newblock {\em Astron. J.} {\bf 1966}, {\em 71},~1026.
				
				\bibitem[Houck \em{et~al.}(1975)Houck, Pollack, Schaack, Reed, and
				Summers]{houck1975jupiter}
				Houck, J.R.; Pollack, J.B.; Schaack, D.; Reed, R.A.; Summers, A.
				\newblock Jupiter: Its infrared spectrum from 16 to 40 micrometers.
				\newblock {\em Science} {\bf 1975}, {\em 189},~720--722.
				
				\bibitem[Danielson(1966)]{danielson1966stratoscope}
				Danielson, R.E.
				\newblock The infrared spectrum of Jupiter.
				\newblock {\em Astrophys. J.} {\bf 1966}, {\em 143},~949.
				
				\bibitem[Strong(1965)]{strong1965infraredballoon}
				Strong, J.
				\newblock Infrared astronomy by balloon.
				\newblock {\em Sci. Am.} {\bf 1965}, {\em 212},~28--37.
				
				\bibitem[Aumann \em{et~al.}(1969)Aumann, Gillespie~Jr, and
				Low]{aumann1969internaljet}
				Aumann, H.; Gillespie~Jr, C.; Low, F.
				\newblock The internal powers and effective temperatures of Jupiter and Saturn.
				\newblock {\em Astrophys. J.} {\bf 1969}, {\em 157},~L69.
				
				\bibitem[Armstrong \em{et~al.}(1972)Armstrong, Harper~Jr, and
				Low]{armstrong1972far}
				Armstrong, K.; Harper~Jr, D.; Low, F.
				\newblock Far-infrared brightness temperatures of the planets.
				\newblock {\em Astrophys. J.} {\bf 1972}, {\em 178},~L89.
				
				\bibitem[Pearl and Conrath(1991)]{pearl1991albedo}
				Pearl, J.; Conrath, B.
				\newblock The albedo, effective temperature, and energy balance of Neptune, as
				determined from Voyager data.
				\newblock {\em J. Geophys. Res. Space Phys.} {\bf 1991}, {\em
					96},~18921--18930.
				
				\bibitem[Li \em{et~al.}(2018)Li, Jiang, West, Gierasch, Perez-Hoyos,
				Sanchez-Lavega, Fletcher, Fortney, Knowles, Porco,
				et~al.]{li2018jupiterenergy}
				Li, L.; Jiang, X.; West, R.; Gierasch, P.; Perez-Hoyos, S.; Sanchez-Lavega, A.;
				Fletcher, L.; Fortney, J.; Knowles, B.; Porco, C.;  et~al.
				\newblock Less absorbed solar energy and more internal heat for Jupiter.
				\newblock {\em Nat. Commun.} {\bf 2018}, {\em 9},~3709.
				
				\bibitem[Li \em{et~al.}(2010)Li, Conrath, Gierasch, Achterberg, Nixon,
				Simon-Miller, Flasar, Banfield, Baines, West, et~al.]{li2010saturnpower}
				Li, L.; Conrath, B.J.; Gierasch, P.J.; Achterberg, R.K.; Nixon, C.A.;
				Simon-Miller, A.A.; Flasar, F.M.; Banfield, D.; Baines, K.H.; West, R.A.;
				et~al.
				\newblock Saturn's emitted power.
				\newblock {\em J. Geophys. Res. Planets} {\bf 2010}, {\em 115}. https://doi.org/10.1029/2010JE003631.
				
				\bibitem[Simon \em{et~al.}(1972)Simon, Morrison, and
				Cruikshank]{simon1972lowbolometer}
				Simon, T.; Morrison, D.; Cruikshank, D.P.
				\newblock Twenty-micron fluxes of bright stellar standards.
				\newblock {\em Astrophys. J.} {\bf 1972}, {\em 177},~L17.
				
				\bibitem[Morrison and Cruikshank(1973)]{morrison1973temperaturesUandN}
				Morrison, D.; Cruikshank, D.P.
				\newblock Temperatures of Uranus and Neptune at 24 microns.
				\newblock {\em Astrophys. J.} {\bf 1973}, {\em 179},~329--332.
				
				\bibitem[Rieke and Low(1974)]{rieke1974infrared}
				Rieke, G.; Low, F.
				\newblock Infrared measurements of Uranus and Neptune.
				\newblock {\em Astrophys. J.} {\bf 1974}, {\em 193},~L147.
				
				\bibitem[Macy~Jr and Sinton(1977)]{macy1977detection}
				Macy~Jr, W.; Sinton, W.
				\newblock Detection of methane and ethane emission on Neptune but not on
				Uranus.
				\newblock {\em Astrophys. J.} {\bf 1977}, {\em 218},~L79--L81.
				
				\bibitem[Murphy and Trafton(1974)]{murphy1974evidence}
				Murphy, R.; Trafton, L.
				\newblock Evidence for an internal heat source in Neptune.
				\newblock {\em Astrophys. J.} {\bf 1974}, {\em 193},~253--255.
				
				\bibitem[Wright(1976)]{wright1976recalibration}
				Wright, E.
				\newblock Recalibration of the far-infrared brightness temperatures of the
				planets.
				\newblock {\em Astrophys. J.} {\bf 1976}, {\em 210},~250--253.
				
				\bibitem[Loewenstein \em{et~al.}(1977)Loewenstein, Harper, and
				Moseley]{loewenstein1977effective}
				Loewenstein, R.; Harper, D.; Moseley, H.
				\newblock The effective temperature of Neptune.
				\newblock {\em Astrophys. J.} {\bf 1977}, {\em 218},~L145.
				
				\bibitem[Whitcomb \em{et~al.}(1979)Whitcomb, Hildebrand, Keene, Stiening, and
				Harper]{whitcomb1979submillimeter}
				Whitcomb, S.; Hildebrand, R.; Keene, J.; Stiening, R.; Harper, D.
				\newblock Submillimeter brightness temperatures of Venus, Jupiter, Uranus, and
				Neptune.
				\newblock {\em Icarus} {\bf 1979}, {\em 38},~75--80.
				
				\bibitem[Epstein \em{et~al.}(1970)Epstein, Dworetsky, Montgomery, Fogarty, and
				Schorn]{epstein1970mars}
				Epstein, E.E.; Dworetsky, M.M.; Montgomery, J.W.; Fogarty, W.G.; Schorn, R.A.
				\newblock Mars, Jupiter, Saturn, and Uranus: 3.3-mm brightness temperatures and
				a search for variations with time or phase angle.
				\newblock {\em Icarus} {\bf 1970}, {\em 13},~276--281.
				
				\bibitem[Ulich \em{et~al.}(1973)Ulich, Cogdell, and Davis]{ulich1973planetary}
				Ulich, B.; Cogdell, J.; Davis, J.
				\newblock Planetary brightness temperature measurements at 8.6 mm and 3.1 mm wavelengths.
				\newblock {\em Icarus} {\bf 1973}, {\em 19},~59--82.
				
				\bibitem[Werner \em{et~al.}(1978)Werner, Neugebauer, Houck, and
				Hauser]{werner1978one}
				Werner, M.; Neugebauer, G.; Houck, J.; Hauser, M.
				\newblock New values for the 1-mm brightness temperatures of Mercury, Venus, Jupiter, Saturn, Uranus, and Neptune have been determined using Mars as the absolute photometric standard.; 
				\newblock {\em Icarus}; {\bf1978}.{\em 35},~289-296 %MDPI: please add publisher and publisher's location (city, country)


				
				\bibitem[Kellermann(1970)]{kellermann1970thermal}
				Kellermann, K.
				\newblock Thermal radio emission from the major planets.
				\newblock {\em Radio Sci.} {\bf 1970}, {\em 5},~487--493.
				
				\bibitem[Mayer and McCullough(1971)]{mayer1971microwave}
				Mayer, C.; McCullough, T.
				\newblock Microwave radiation of Uranus and Neptune.
				\newblock {\em Icarus} {\bf 1971}, {\em 14},~187--191.
				
				\bibitem[Connes and Connes(1966)]{connes1966near}
				Connes, J.; Connes, P.
				\newblock Near-infrared planetary spectra by Fourier spectroscopy. I.
				Instruments and results.
				\newblock {\em JOSA} {\bf 1966}, {\em 56},~896--910.
				
				\bibitem[Huggins and Huggins(1890)]{huggins1890vii}
				Huggins, W.; Huggins, M.L.
				\newblock VII. Note on the photographic spectra of Uranus and Saturn.
				\newblock {\em Proc. R. Soc. Lond.} {\bf 1890}, {\em
					46},~231--233.
				
				\bibitem[Draper(1879)]{draper1879photographing}
				Draper, H.
				\newblock On photographing the spectra of the stars and planets.
				\newblock {\em Am. J. Sci.} {\bf 1879}, {\em 3},~419--425.
				
				\bibitem[Slipher(1909)]{slipher1909spectra}
				Slipher, V.M.
				\newblock The spectra of the major planets.
				\newblock {\em Lowell Obs. Bull.} {\bf 1909}, {\em 1},~231--238.
				
				\bibitem[Adel and Slipher(1934)]{adel1934constitution}
				Adel, A.; Slipher, V.
				\newblock The constitution of the atmospheres of the giant planets.
				\newblock {\em Phys. Rev.} {\bf 1934}, {\em 46},~902.
				
				\bibitem[Wildt(1932)]{wildt1932absorptionsspektren}
				Wildt, R.
				\newblock Absorptionsspektren und atmosph{\"a}ren der gros; en planeten.
				\newblock {\em Veroeffentlichungen der Universitaets-Sternwarte zu Goettingen}
				{\bf 1932}, {\em 2},~171--2.
				
				\bibitem[Dunham(1933)]{dunham1933note}
				Dunham, T.
				\newblock Note on the spectra of Jupiter and Saturn.
				\newblock {\em Publ. Astron. Soc. Pac.} {\bf
					1933}, {\em 45},~42--44.
				
				\bibitem[Kiess \em{et~al.}(1960)Kiess, Corliss, and Kiess]{kiess1960hydrogen}
				Kiess, C.; Corliss, C.; Kiess, H.K.
				\newblock High-Dispersion Spectra of Jupiter.
				\newblock {\em Astrophys. J.} {\bf 1960}, {\em 132},~221.
				
				\bibitem[Lewis(1969)]{lewis1969observability}
				Lewis, J.S.
				\newblock Observability of spectroscopically active compounds in the atmosphere
				of Jupiter.
				\newblock {\em Icarus} {\bf 1969}, {\em 10},~393--409.
				
				\bibitem[Wildt(1937)]{wildt1937photochemistry}
				Wildt, R.
				\newblock Photochemistry of planetary atmospheres.
				\newblock {\em Astrophys. J.} {\bf 1937}, {\em 86},~321.
				
				\bibitem[Strobel(1969)]{strobel1969photochemistry}
				Strobel, D.F.
				\newblock The photochemistry of methane in the jovian atmosphere.
				\newblock {\em J. Atmos. Sci.} {\bf 1969}, {\em 26},~906--911.
				
				\bibitem[Ridgway(1974)]{ridgway1974jupiter}
				Ridgway, S.
				\newblock Jupiter: Identification of ethane and acetylene.
				\newblock {\em Astrophys. J.} {\bf 1974}, {\em 187},~L41--L43.
				
				\bibitem[Tokunaga \em{et~al.}(1976)Tokunaga, Knacke, and
				Owen]{tokunaga1976ethane}
				Tokunaga, A.; Knacke, R.; Owen, T.
				\newblock Ethane and acetylene abundances in the Jovian atmosphere.
				\newblock {\em Astrophys. J.} {\bf 1976}, {\em 209},~294--301.
				
				\bibitem[Gillett and Forrest(1974)]{gillett1974saturn}
				Gillett, F.; Forrest, W.
				\newblock The 7.5-to 13.5-micron spectrum of Saturn.
				\newblock {\em Astrophys. J.} {\bf 1974}, {\em 187},~L37.
				
				\bibitem[Wark and Hilleary(1969)]{wark1969atmospheric}
				Wark, D.; Hilleary, D.
				\newblock Atmospheric temperature: Successful test of remote probing.
				\newblock {\em Science} {\bf 1969}, {\em 165},~1256--1258.
				
				\bibitem[Ohring(1973)]{ohring1973temperature}
				Ohring, G.
				\newblock The temperature and ammonia profiles in the jovian atmospheres from
				inversion of the jovian emission spectrum.
				\newblock {\em Astrophys. J.} {\bf 1973}, {\em 184},~1027--1040.
				
				\bibitem[Taylor(1972)]{taylor1972temperature}
				Taylor, F.
				\newblock Temperature sounding experiments for the Jovian planets.
				\newblock {\em J. Atmos. Sci.} {\bf 1972}, {\em 29},~950--958.
				
				\bibitem[Rodgers(1976)]{rodgers1976retrieval}
				Rodgers, C.D.
				\newblock Retrieval of atmospheric temperature and composition from remote
				measurements of thermal radiation.
				\newblock {\em Rev. Geophys.} {\bf 1976}, {\em 14},~609--624.
				
				\bibitem[Conrath \em{et~al.}(1998)Conrath, Gierasch, and
				Ustinov]{conrath1998thermal}
				Conrath, B.J.; Gierasch, P.J.; Ustinov, E.A.
				\newblock Thermal Structure and Para Hydrogen Fraction on the Outer Planets
				fromVoyagerIRIS Measurements.
				\newblock {\em Icarus} {\bf 1998}, {\em 135},~501--517.
				
				\bibitem[Connes \em{et~al.}(1968)Connes, Connes, Kaplan, and
				Benedict]{connes1968carbon}
				Connes, P.; Connes, J.; Kaplan, L.; Benedict, W.
				\newblock Carbon monoxide in the Venus atmosphere.
				\newblock {\em Astrophys. J.} {\bf 1968}, {\em 152},~731--743.
				
				\bibitem[Connes \em{et~al.}(1969)Connes, Connes, and Maillard]{connes1969atlas}
				Connes, J.; Connes, P.; Maillard, J.P.
				\newblock \emph{Atlas des Spectres dans le Proche Infrarouge de Venus, Mars, Jupiter
					et Saturn}; 
				\newblock {Centre National de la Recherche Scientifique: Paris, France,} {1969}.
				
				\bibitem[Beer and Taylor(1973)]{beer1973abundance}
				Beer, R.; Taylor, F.W.
				\newblock The abundance of CH3D and the D/H ratio in Jupiter.
				\newblock {\em Astrophys. J.} {\bf 1973}, {\em 179},~309--328.
				
				\bibitem[Encrenaz \em{et~al.}(1978)Encrenaz, Combes, and
				Zeau]{encrenaz1978spectrumnh3}
				Encrenaz, T.; Combes, M.; Zeau, Y.
				\newblock The Spectrum of Jupiter between 10 and 13 $\mu$.
				\newblock {\em Astron. Astrophys.} {\bf 1978}, {\em 70},~29.
				
				\bibitem[Larson \em{et~al.}(1975)Larson, Fink, Treffers, and
				Gautier~III]{larson1975detectionh2o}
				Larson, H.; Fink, U.; Treffers, R.; Gautier~III, T.
				\newblock Detection of water vapor on Jupiter.
				\newblock {\em Astrophys. J.} {\bf 1975}, {\em 197},~L137--L140.
				
				\bibitem[Ridgway \em{et~al.}(1976)Ridgway, Wallace, and Smith]{ridgway1976ph3}
				Ridgway, S.; Wallace, L.; Smith, G.
				\newblock The 800-1200 inverse centimeter absorption spectrum of Jupiter.
				\newblock {\em Astrophys. J.} {\bf 1976}, {\em 207},~1002--1006.
				
				\bibitem[Larson \em{et~al.}(1977)Larson, Treffers, and
				Fink]{larson1977phosphine}
				Larson, H.; Treffers, R.; Fink, U.
				\newblock Phosphine in Jupiter's atmosphere-The evidence from high-altitude
				observations at 5 micrometers.
				\newblock {\em Astrophys. J.} {\bf 1977}, {\em 211},~972--979.
				
				\bibitem[Beer(1975)]{beer1975detectionco}
				Beer, R.
				\newblock Detection of carbon monoxide in Jupiter.
				\newblock {\em Astrophys. J.} {\bf 1975}, {\em 200},~L167--L169.
				
				\bibitem[Larson \em{et~al.}(1978)Larson, Fink, and
				Treffers]{larson1978evidenceco}
				Larson, H.; Fink, U.; Treffers, R.
				\newblock Evidence for CO in Jupiter's atmosphere from airborne spectroscopic
				observations at 5 microns.
				\newblock {\em Astrophys. J.} {\bf 1978}, {\em 219},~1084--1092.
				
				\bibitem[Prinn and Lewis(1975)]{prinn1975phosphine}
				Prinn, R.G.; Lewis, J.S.
				\newblock Phosphine on Jupiter and implications for the Great Red Spot.
				\newblock {\em Science} {\bf 1975}, {\em 190},~274--276.
				
				\bibitem[Prinn and Barshay(1977)]{prinn1977carbon}
				Prinn, R.G.; Barshay, S.S.
				\newblock Carbon monoxide on Jupiter and implications for atmospheric
				convection.
				\newblock {\em Science} {\bf 1977}, {\em 198},~1031--1034.
				
				\bibitem[Fink and Larson(1978)]{fink1978deuteratedsaturn}
				Fink, U.; Larson, H.P.
				\newblock Deuterated methane observed on Saturn.
				\newblock {\em Science} {\bf 1978}, {\em 201},~343--345.
				
				\bibitem[Bregman \em{et~al.}(1975)Bregman, Lester, and
				Rank]{bregman1975observationph3saturn}
				Bregman, J.; Lester, D.; Rank, D.
				\newblock Observation of the nu-squared band of PH3 in the atmosphere of
				Saturn.
				\newblock {\em Astrophys. J.} {\bf 1975}, {\em 202},~L55.
				
				\bibitem[Larson \em{et~al.}(1980)Larson, Fink, Smith, and
				Davis]{larson1980middle}
				Larson, H.; Fink, U.; Smith, H.; Davis, D.
				\newblock The middle-infrared spectrum of Saturn-Evidence for phosphine and
				upper limits to other trace atmospheric constituents.
				\newblock {\em Astrophys. J.} {\bf 1980}, {\em 240},~327--337.
				
				\bibitem[Tokunaga \em{et~al.}(1975)Tokunaga, Knacke, and
				Owen]{tokunaga1975c2h6detection}
				Tokunaga, A.; Knacke, R.; Owen, T.
				\newblock The detection of ethane on Saturn.
				\newblock {\em Astrophys. J.} {\bf 1975}, {\em 197},~L77--L78.
				
				\bibitem[Encrenaz \em{et~al.}(1975)Encrenaz, Combes, Zeau, Vapillon, and
				Berezne]{encrenaz1975tentativec2h4}
				Encrenaz, T.; Combes, M.; Zeau, Y.; Vapillon, L.; Berezne, J.
				\newblock A tentative identification of C$_2$H$_4$ in the spectrum of Saturn.
				\newblock {\em Astron. Astrophys.} {\bf 1975}, {\em 42},~355.
				
				\bibitem[Encrenaz \em{et~al.}(1974)Encrenaz, Owen, and
				Woodman]{encrenaz1974abundancenh3}
				Encrenaz, T.; Owen, T.; Woodman, J.
				\newblock The abundance of ammonia on Jupiter, Saturn and Titan.
				\newblock {\em Astron. Astrophys.} {\bf 1974}, {\em 37},~49--55.
				
				\bibitem[Noll \em{et~al.}(1988)Noll, Knacke, Geballe, and
				Tokunaga]{noll1988evidencesaturngeh4}
				Noll, K.S.; Knacke, R.; Geballe, T.; Tokunaga, A.
				\newblock Evidence for germane in Saturn.
				\newblock {\em Icarus} {\bf 1988}, {\em 75},~409--422.
				
				\bibitem[Chase \em{et~al.}(1974)Chase, Ruiz, Mnch, Neugebauer, Schroeder, and
				Trafton]{chase1974pioneer}
				Chase, S.; Ruiz, R.; Mnch, G.; Neugebauer, G.; Schroeder, M.; Trafton, L.
				\newblock Pioneer 10 infrared radiometer experiment: Preliminary results.
				\newblock {\em Science} {\bf 1974}, {\em 183},~315--317.
				
				\bibitem[Ingersoll \em{et~al.}(1975)Ingersoll, M{\"u}nch, Neugebauer, Diner,
				Orton, Schupler, Schroeder, Chase, Ruiz, and Trafton]{ingersoll1975pioneer}
				Ingersoll, A.; M{\"u}nch, G.; Neugebauer, G.; Diner, D.; Orton, G.; Schupler,
				B.; Schroeder, M.; Chase, S.; Ruiz, R.; Trafton, L.
				\newblock Pioneer 11 infrared radiometer experiment: the global heat balance of
				Jupiter.
				\newblock {\em Science} {\bf 1975}, {\em 188},~472--473.
				
				\bibitem[Ingersoll \em{et~al.}(1976)Ingersoll, M{\"u}nch, Neugebauer, and
				Orton]{ingersoll1976results}
				Ingersoll, A.; M{\"u}nch, G.; Neugebauer, G.; Orton, G.
				\newblock Results of the infrared radiometer experiment on Pioneers 10 and 11. In 
				\newblock  \emph{IAU Colloq. 30: Jupiter: Studies of the Interior, Atmosp here,
					Magnetosphere and Satellites}; Tucson, University of Arizona Press, 1976; pp. 197--205. %MDPI: please add publisher and publisher's location (city, country)
				
				\bibitem[Ingersoll \em{et~al.}(1980)Ingersoll, Orton, M{\"u}nch, Neugebauer,
				and Chase]{ingersoll1980pioneersaturn}
				Ingersoll, A.; Orton, G.; M{\"u}nch, G.; Neugebauer, G.; Chase, S.
				\newblock Pioneer Saturn infrared radiometer: Preliminary results.
				\newblock {\em Science} {\bf 1980}, {\em 207},~439--443.
				
				\bibitem[Hanel \em{et~al.}(1981)Hanel, Conrath, Herath, Kunde, and
				Pirraglia]{hanel1981albedojupiter}
				Hanel, R.; Conrath, B.; Herath, L.; Kunde, V.; Pirraglia, J.
				\newblock Albedo, internal heat, and energy balance of Jupiter: Preliminary
				results of the Voyager infrared investigation.
				\newblock {\em J. Geophys. Res. Space Phys.} {\bf 1981}, {\em
					86},~8705--8712.
				
				\bibitem[Hanel \em{et~al.}(1979)Hanel, Conrath, Flasar, Kunde, Lowman, Maguire,
				Pearl, Pirraglia, Samuelson, Gautier, et~al.]{hanel1979infraredjupitervoy1}
				Hanel, R.; Conrath, B.; Flasar, M.; Kunde, V.; Lowman, P.; Maguire, W.; Pearl,
				J.; Pirraglia, J.; Samuelson, R.; Gautier, D.;  et~al.
				\newblock Infrared observations of the Jovian system from Voyager 1.
				\newblock {\em Science} {\bf 1979}, {\em 204},~972--976.
				
				\bibitem[Kim \em{et~al.}(1985)Kim, Caldwell, Rivolo, Wagener, and
				Orton]{kim1985infrared}
				Kim, S.J.; Caldwell, J.; Rivolo, A.; Wagener, R.; Orton, G.S.
				\newblock Infrared polar brightening on Jupiter: III. Spectrometry from the
				Voyager 1 IRIS experiment.
				\newblock {\em Icarus} {\bf 1985}, {\em 64},~233--248.
				
				\bibitem[Marten \em{et~al.}(1981)Marten, Rouan, Baluteau, Gautier, Conrath,
				Hanel, Kunde, Samuelson, Chedin, and Scott]{marten1981studyammonia}
				Marten, A.; Rouan, D.; Baluteau, J.P.; Gautier, D.; Conrath, B.J.; Hanel, R.A.;
				Kunde, V.; Samuelson, R.; Chedin, A.; Scott, N.
				\newblock Study of the ammonia ice cloud layer in the equatorial region of
				Jupiter from the infrared interferometric experiment on Voyager.
				\newblock {\em Icarus} {\bf 1981}, {\em 46},~233--248.
				
				\bibitem[Gautier \em{et~al.}(1981)Gautier, Conrath, Flasar, Hanel, Kunde,
				Chedin, and Scott]{gautier1981heliumjupiter}
				Gautier, D.; Conrath, B.; Flasar, M.; Hanel, R.; Kunde, V.; Chedin, A.; Scott,
				N.
				\newblock The helium abundance of Jupiter from Voyager.
				\newblock {\em J. Geophys. Res. Space Phys.} {\bf 1981}, {\em
					86},~8713--8720.
				
				\bibitem[Pirraglia \em{et~al.}(1981)Pirraglia, Conrath, Allison, and
				Gierasch]{pirraglia1981thermal}
				Pirraglia, J.; Conrath, B.; Allison, M.; Gierasch, P.
				\newblock Thermal structure and dynamics of Saturn and Jupiter.
				\newblock {\em Nature} {\bf 1981}, {\em 292},~677--679.
		
				
				\bibitem[Conrath \em{et~al.}(1984)Conrath, Gautier, Hanel, and
				Hornstein]{conrath1984helium}
				Conrath, B.; Gautier, D.; Hanel, R.; Hornstein, J.
				\newblock The helium abundance of Saturn from Voyager measurements.
				\newblock {\em Astrophys. J.} {\bf 1984}, {\em 282},~807--815.
				
				\bibitem[Hanel \em{et~al.}(1983)Hanel, Conrath, Kunde, Pearl, and
				Pirraglia]{hanel1983albedosaturn}
				Hanel, R.; Conrath, B.; Kunde, V.; Pearl, J.; Pirraglia, J.
				\newblock Albedo, internal heat flux, and energy balance of Saturn.
				\newblock {\em Icarus} {\bf 1983}, {\em 53},~262--285.
				
				\bibitem[B{\'e}zard \em{et~al.}(1984)B{\'e}zard, Gautier, and
				Conrath]{bezard1984seasonal}
				B{\'e}zard, B.; Gautier, D.; Conrath, B.
				\newblock A seasonal model of the Saturnian upper troposphere: Comparison with
				Voyager infrared measurements.
				\newblock {\em Icarus} {\bf 1984}, {\em 60},~274--288.
				
				\bibitem[Conrath and Gautier(2000)]{conrath2000saturn}
				Conrath, B.J.; Gautier, D.
				\newblock Saturn helium abundance: A reanalysis of Voyager measurements.
				\newblock {\em Icarus} {\bf 2000}, {\em 144},~124--134.
				
				\bibitem[Pearl \em{et~al.}(1990)Pearl, Conrath, Hanel, Pirraglia, and
				Coustenis]{pearl1990albedo}
				Pearl, J.; Conrath, B.; Hanel, R.; Pirraglia, J.; Coustenis, A.
				\newblock The albedo, effective temperature, and energy balance of Uranus, as
				determined from Voyager IRIS data.
				\newblock {\em Icarus} {\bf 1990}, {\em 84},~12--28.
				
				\bibitem[Conrath \em{et~al.}(1987)Conrath, Gautier, Hanel, Lindal, and
				Marten]{conrath1987helium}
				Conrath, B.; Gautier, D.; Hanel, R.; Lindal, G.; Marten, A.
				\newblock The helium abundance of Uranus from Voyager measurements.
				\newblock {\em J. Geophys. Res. Space Phys.} {\bf 1987}, {\em
					92},~15003--15010.
				
				\bibitem[Conrath \em{et~al.}(1991)Conrath, Gautier, Lindal, Samuelson, and
				Shaffer]{conrath1991helium}
				Conrath, B.; Gautier, D.; Lindal, G.; Samuelson, R.; Shaffer, W.
				\newblock The helium abundance of Neptune from Voyager measurements.
				\newblock {\em J. Geophys. Res. Space Phys.} {\bf 1991}, {\em
					96},~18907--18919.
				
				\bibitem[Smith \em{et~al.}(1989)Smith, Soderblom, Banfield, Basilevsky, Beebe,
				Bollinger, Boyce, Brahic, Briggs, Brown, et~al.]{smith1989voyager}
				Smith, B.A.; Soderblom, L.A.; Banfield, D.; Basilevsky, A.; Beebe, R.;
				Bollinger, K.; Boyce, J.; Brahic, A.; Briggs, G.; Brown, R.;  et~al.
				\newblock Voyager 2 at Neptune: Imaging science results.
				\newblock {\em Science} {\bf 1989}, {\em 246},~1422--1449.
				
				\bibitem[Conrath \em{et~al.}(1989)Conrath, Flasar, Hanel, Kunde, Maguire,
				Pearl, Pirraglia, Samuelson, Gierasch, Weir, et~al.]{conrath1989neptune}
				Conrath, B.; Flasar, F.; Hanel, R.; Kunde, V.; Maguire, W.; Pearl, J.;
				Pirraglia, J.; Samuelson, R.; Gierasch, P.; Weir, A.;  et~al.
				\newblock Infrared observations of the Neptunian system.
				\newblock {\em Science} {\bf 1989}, {\em 246},~1454--1459.
				
				\bibitem[Conrath \em{et~al.}(1990)Conrath, Gierasch, and
				Leroy]{conrath1990temperature}
				Conrath, B.J.; Gierasch, P.J.; Leroy, S.S.
				\newblock Temperature and circulation in the stratosphere of the outer planets.
				\newblock {\em Icarus} {\bf 1990}, {\em 83},~255--281.
				
				\bibitem[Smith(1978)]{smith1978ortho}
				Smith, W.H.
				\newblock On the ortho-para equilibrium of H2 in the atmospheres of the Jovian
				planets.
				\newblock {\em Icarus} {\bf 1978}, {\em 33},~210--216.
				
				\bibitem[Fletcher \em{et~al.}(2018)Fletcher, Gustafsson, and
				Orton]{fletcher2018hydrogen}
				Fletcher, L.N.; Gustafsson, M.; Orton, G.S.
				\newblock Hydrogen dimers in giant-planet infrared spectra.
				\newblock {\em Astrophys. J. Suppl. Ser.} {\bf 2018}, {\em
					235},~24.
				
				\bibitem[Conrath and Gierasch(1983)]{conrath1983evidence}
				Conrath, B.J.; Gierasch, P.J.
				\newblock Evidence for disequilibrium of ortho and para hydrogen on Jupiter
				from Voyager IRIS measurements.
				\newblock {\em Nature} {\bf 1983}, {\em 306},~571--572.
				
				\bibitem[Gierasch \em{et~al.}(1986)Gierasch, Conrath, Magalha,
				et~al.]{gierasch1986zonal}
				Gierasch, P.J.; Conrath, B.J.; Magalha, J.A.;  et~al.
				\newblock Zonal mean properties of Jupiter's upper troposphere from Voyager
				infrared observations.
				\newblock {\em Icarus} {\bf 1986}, {\em 67},~456--483.
				
				\bibitem[Bagenal \em{et~al.}(2007)Bagenal, Dowling, McKinnon, and
				McKinnon]{bagenal2007jupiter}
				Bagenal, F.; Dowling, T.E.; McKinnon, W.B.; McKinnon, W.
				\newblock {\em Jupiter: The Planet, Satellites and Magnetosphere}; 
				Cambridge University Press: Cambridge, UK,  2007; Volume~1.
				
				\bibitem[Fischer(2001)]{fischer2001mission}
				Fischer, D.
				\newblock {\em Mission Jupiter: The Spectacular Journey of the Galileo
					Spacecraft}; Springer:  Berlin/Heidelberg, Germany,  2001.
				
				\bibitem[Hunten \em{et~al.}(1986)Hunten, Colin, and
				Hansen]{hunten1986atmospheric}
				Hunten, D.; Colin, L.; Hansen, J.
				\newblock Atmospheric science on the Galileo mission.
				\newblock {\em Space Sci. Rev.} {\bf 1986}, {\em 44},~191--240.
				
				\bibitem[Russell(2012)]{russell2012galileo}
				Russell, C.T.
				\newblock {\em The Galileo Mission}; Springer Science \& Business Media:  Berlin/Heidelberg, Germany,  2012.
				
				\bibitem[Carlson \em{et~al.}(1992)Carlson, Weissman, Smythe, Mahoney,
				et~al.]{carlson1992NIMS}
				Carlson, R.; Weissman, P.; Smythe, W.; Mahoney, J. 
				\newblock Near-infrared mapping spectrometer experiment on Galileo. In
				\newblock {\em The Galileo Mission}; Springer Science+Business Media Dordrecht, The Netherlands {1992}; pp. 457--502.  %MDPI: please add publisher and publisher's location (city, country)
				
				\bibitem[Russell \em{et~al.}(1992)Russell, Brown, Chandos, Fincher, Kubel,
				Lacis, and Travis]{russell1992galileoPPR}
				Russell, E.; Brown, F.; Chandos, R.; Fincher, W.; Kubel, L.; Lacis, A.; Travis,
				L.
				\newblock Galileo photopolarimeter/radiometer experiment.
				\newblock {\em Space Sci. Rev.} {\bf 1992}, {\em 60},~531--563.
				
				\bibitem[Nixon \em{et~al.}(2001)Nixon, Irwin, Calcutt, Taylor, and
				Carlson]{nixon2001NIMSjupiter}
				Nixon, C.; Irwin, P.; Calcutt, S.; Taylor, F.; Carlson, R.
				\newblock Atmospheric composition and cloud structure in Jovian 5-$\upmu$m
				hotspots from analysis of Galileo NIMS measurements.
				\newblock {\em Icarus} {\bf 2001}, {\em 150},~48--68.
				
				\bibitem[Irwin \em{et~al.}(1997)Irwin, Calcutt, and Taylor]{irwin1997radiative}
				Irwin, P.; Calcutt, S.; Taylor, F.
				\newblock Radiative transfer models for Galileo NIMS studies of the atmosphere
				of Jupiter.
				\newblock {\em Adv. Space Res.} {\bf 1997}, {\em 19},~1149--1158.
				
				\bibitem[Dyudina \em{et~al.}(2001)Dyudina, Ingersoll, Danielson, Baines,
				Carlson, NIMS, and Teams]{dyudina2001interpretation}
				Dyudina, U.; Ingersoll, A.; Danielson, G.; Baines, K.; Carlson, R.; NIMS, T.G.;
				Teams, S.
				\newblock Interpretation of NIMS and SSI images on the Jovian cloud structure.
				\newblock {\em Icarus} {\bf 2001}, {\em 150},~219--233.
				
				\bibitem[Irwin and Dyudina(2002)]{irwin2002retrieval}
				Irwin, P.; Dyudina, U.
				\newblock The retrieval of cloud structure maps in the equatorial region of
				Jupiter using a principal component analysis of Galileo/NIMS data.
				\newblock {\em Icarus} {\bf 2002}, {\em 156},~52--63.
				
				\bibitem[Orton \em{et~al.}(1996)Orton, Spencer, Travis, Martin, and
				Tamppari]{orton1996galileo}
				Orton, G.; Spencer, J.; Travis, L.; Martin, T.; Tamppari, L.
				\newblock Galileo photopolarimeter-radiometer observations of Jupiter and the
				Galilean satellites.
				\newblock {\em Science} {\bf 1996}, {\em 274},~389--391.
				
				\bibitem[Matson \em{et~al.}(2002)Matson, Spilker, and
				Lebreton]{matson2002cassini}
				Matson, D.L.; Spilker, L.J.; Lebreton, J.P.
				\newblock The Cassini/Huygens mission to the Saturnian system.
				\newblock {\em Space Sci. Rev.} {\bf 2002}, {\em 104},~1--58.
				
				\bibitem[Dougherty \em{et~al.}(2009)Dougherty, Esposito, and
				Krimigis]{dougherty2009saturn}
				Dougherty, M.; Esposito, L.; Krimigis, S.M.
				\newblock \emph{Saturn from Cassini-huygens}; Springer:   Berlin/Heidelberg, Germany, {2009}.
				
				\bibitem[Hansen \em{et~al.}(2004)Hansen, Bolton, Matson, Spilker, and
				Lebreton]{hansen2004cassini}
				Hansen, C.J.; Bolton, S.J.; Matson, D.L.; Spilker, L.J.; Lebreton, J.P.
				\newblock The Cassini--Huygens flyby of jupiter.
				\newblock {\em Icarus} {\bf 2004}, {\em 172},~1--8.
				
				\bibitem[Brown \em{et~al.}(2004)Brown, Baines, Bellucci, Bibring, Buratti,
				Capaccioni, Cerroni, Clark, Coradini, Cruikshank, et~al.]{brown2004cassini}
				Brown, R.H.; Baines, K.H.; Bellucci, G.; Bibring, J.P.; Buratti, B.J.;
				Capaccioni, F.; Cerroni, P.; Clark, R.N.; Coradini, A.; Cruikshank, D.P.;
				et~al.
				\newblock The Cassini visual and infrared mapping spectrometer (VIMS)
				investigation. In 
				\newblock {\em The Cassini-Huygens Mission: Orbiter Remote Sensing
					Investigations}; Springer: Dordrecht, The Netherlands, {{2004}}; pp. 111--168.  
				
				\bibitem[Flasar \em{et~al.}(2004)Flasar, Kunde, Abbas, Achterberg, Ade,
				Barucci, B{\'e}zard, Bjoraker, Brasunas, Calcutt,
				et~al.]{flasar2004exploring}
				Flasar, F.M.; Kunde, V.; Abbas, M.; Achterberg, R.; Ade, P.; Barucci, A.;
				B{\'e}zard, B.; Bjoraker, G.; Brasunas, J.; Calcutt, S.;  et~al.
				\newblock Exploring the Saturn system in the thermal infrared: The composite
				infrared spectrometer.
				\newblock {\em The Cassini-Huygens Mission}; Springer: Dordrecht, The Netherlands, {2004}; pp. 169--297.
				
				\bibitem[Jennings \em{et~al.}(2017)Jennings, Flasar, Kunde, Nixon, Segura,
				Romani, Gorius, Albright, Brasunas, Carlson, et~al.]{jennings2017composite}
				Jennings, D.E.; Flasar, F.; Kunde, V.; Nixon, C.; Segura, M.; Romani, P.;
				Gorius, N.; Albright, S.; Brasunas, J.; Carlson, R.;  et~al.
				\newblock Composite infrared spectrometer (CIRS) on Cassini.
				\newblock {\em Appl. Opt.} {\bf 2017}, {\em 56},~5274--5294.
				
				\bibitem[Sromovsky and Fry(2010)]{sromovsky2010source}
				Sromovsky, L.; Fry, P.
				\newblock The source of widespread 3-$\upmu$m absorption in Jupiter’s clouds:
				Constraints from 2000 Cassini VIMS observations.
				\newblock {\em Icarus} {\bf 2010}, {\em 210},~230--257.
				
				\bibitem[Sromovsky \em{et~al.}(2018)Sromovsky, Baines, and
				Fry]{sromovsky2018vims}
				Sromovsky, L.; Baines, K.; Fry, P.
				\newblock Models of bright storm clouds and related dark ovals in Saturn’s
				Storm Alley as constrained by 2008 Cassini/VIMS spectra.
				\newblock {\em Icarus} {\bf 2018}, {\em 302},~360--385.
				
				\bibitem[Sromovsky \em{et~al.}(2021)Sromovsky, Baines, and
				Fry]{sromovsky2021evolution}
				Sromovsky, L.; Baines, K.; Fry, P.
				\newblock Evolution of Saturn’s north polar color and cloud structure between
				2012 and 2017 inferred from Cassini VIMS and ISS observations.
				\newblock {\em Icarus} {\bf 2021}, {\em 362},~114409.
				
				\bibitem[Simon-Miller \em{et~al.}(2006)Simon-Miller, Conrath, Gierasch, Orton,
				Achterberg, Flasar, and Fisher]{simon2006jupiter}
				Simon-Miller, A.A.; Conrath, B.J.; Gierasch, P.J.; Orton, G.S.; Achterberg,
				R.K.; Flasar, F.M.; Fisher, B.M.
				\newblock Jupiter's atmospheric temperatures: From Voyager IRIS to Cassini
					CIRS. %MDPI: Refs. 230 and 388 are duplicated. Please remove one of them and rearrange all the references to appear in numerical order. Please ensure that there are no duplicated references.; MTR: resolved
				\newblock {\em Icarus} {\bf 2006}, {\em 180},~98--112.
				
				\bibitem[Li \em{et~al.}(2012)Li, Baines, Smith, West, P{\'e}rez-Hoyos,
				Trammell, Simon-Miller, Conrath, Gierasch, Orton,
				et~al.]{li2012emittedcirsjupiter}
				Li, L.; Baines, K.H.; Smith, M.A.; West, R.A.; P{\'e}rez-Hoyos, S.; Trammell,
				H.J.; Simon-Miller, A.A.; Conrath, B.J.; Gierasch, P.J.; Orton, G.S.;  et~al.
				\newblock Emitted power of Jupiter based on Cassini CIRS and VIMS observations.
				\newblock {\em J. Geophys. Res. Planets} {\bf 2012}, {\em 117}.
				
				\bibitem[Matcheva \em{et~al.}(2005)Matcheva, Conrath, Gierasch, and
				Flasar]{matcheva2005cloud}
				Matcheva, K.I.; Conrath, B.J.; Gierasch, P.J.; Flasar, F.M.
				\newblock The cloud structure of the jovian atmosphere as seen by the
				Cassini/CIRS experiment.
				\newblock {\em Icarus} {\bf 2005}, {\em 179},~432--448.
				
				\bibitem[Wong \em{et~al.}(2004)Wong, Bjoraker, Smith, Flasar, and
				Nixon]{wong2004identification}
				Wong, M.H.; Bjoraker, G.L.; Smith, M.D.; Flasar, F.M.; Nixon, C.A.
				\newblock Identification of the 10-$\upmu$m ammonia ice feature on Jupiter.
				\newblock {\em Planet. Space Sci.} {\bf 2004}, {\em 52},~385--395.
				
				\bibitem[Achterberg \em{et~al.}(2006)Achterberg, Conrath, and
				Gierasch]{achterberg2006cassini}
				Achterberg, R.K.; Conrath, B.J.; Gierasch, P.J.
				\newblock Cassini CIRS retrievals of ammonia in Jupiter's upper troposphere.
				\newblock {\em Icarus} {\bf 2006}, {\em 182},~169--180.
				
				\bibitem[Irwin \em{et~al.}(2004)Irwin, Parrish, Fouchet, Calcutt, Taylor,
				Simon-Miller, and Nixon]{irwin2004retrievals}
				Irwin, P.; Parrish, P.; Fouchet, T.; Calcutt, S.; Taylor, F.; Simon-Miller, A.;
				Nixon, C.
				\newblock Retrievals of jovian tropospheric phosphine from Cassini/CIRS.
				\newblock {\em Icarus} {\bf 2004}, {\em 172},~37--49.
				
				\bibitem[Fletcher \em{et~al.}(2009)Fletcher, Orton, Teanby, and
				Irwin]{fletcher2009phosphine}
				Fletcher, L.; Orton, G.; Teanby, N.; Irwin, P.
				\newblock Phosphine on jupiter and saturn from cassini/cirs.
				\newblock {\em Icarus} {\bf 2009}, {\em 202},~543--564.
				
				\bibitem[Nixon \em{et~al.}(2007)Nixon, Achterberg, Conrath, Irwin, Teanby,
				Fouchet, Parrish, Romani, Abbas, LeClair, et~al.]{nixon2007meridional}
				Nixon, C.; Achterberg, R.; Conrath, B.; Irwin, P.; Teanby, N.; Fouchet, T.;
				Parrish, P.; Romani, P.; Abbas, M.; LeClair, A.;  et~al.
				\newblock Meridional variations of C2H2 and C2H6 in Jupiter's atmosphere from
				Cassini CIRS infrared spectra.
				\newblock {\em Icarus} {\bf 2007}, {\em 188},~47--71.
				
				\bibitem[Pierel \em{et~al.}(2017)Pierel, Nixon, Lellouch, Fletcher, Bjoraker,
				Achterberg, Bezard, Hesman, Irwin, and Flasar]{pierel2017d2hratios}
				Pierel, J.; Nixon, C.; Lellouch, E.; Fletcher, L.; Bjoraker, G.; Achterberg,
				R.; Bezard, B.; Hesman, B.; Irwin, P.; Flasar, F.
				\newblock D/H ratios on Saturn and Jupiter from Cassini CIRS.
				\newblock {\em Astron. J.} {\bf 2017}, {\em 154},~178.
				
				\bibitem[Fouchet \em{et~al.}(2004)Fouchet, Orton, Irwin, Calcutt, and
				Nixon]{fouchet2004upper}
				Fouchet, T.; Orton, G.; Irwin, P.G.; Calcutt, S.B.; Nixon, C.A.
				\newblock Upper limits on hydrogen halides in Jupiter from Cassini/CIRS
				observations.
				\newblock {\em Icarus} {\bf 2004}, {\em 170},~237--241.
				
				\bibitem[Nixon \em{et~al.}(2010)Nixon, Achterberg, Romani, Allen, Zhang,
				Teanby, Irwin, and Flasar]{nixon2010abundances}
				Nixon, C.A.; Achterberg, R.K.; Romani, P.N.; Allen, M.; Zhang, X.; Teanby,
				N.A.; Irwin, P.G.; Flasar, F.M.
				\newblock Abundances of Jupiter's trace hydrocarbons from Voyager and Cassini.
				\newblock {\em Planet.  Space Sci.} {\bf 2010}, {\em 58},~1667--1680.
				
				\bibitem[Sinclair \em{et~al.}(2019)Sinclair, Moses, Hue, Greathouse, Orton,
				Fletcher, and Irwin]{sinclair2019jupiter}
				Sinclair, J.; Moses, J.; Hue, V.; Greathouse, T.; Orton, G.; Fletcher, L.;
				Irwin, P.
				\newblock Jupiter's auroral-related stratospheric heating and chemistry III:
				Abundances of C$_2$H$_4$, CH$_3$C$_2$H, C$_4$H$_2$ and C$_6$H$_6$ from Voyager-IRIS and Cassini-CIRS.
				\newblock {\em Icarus} {\bf 2019}, {\em 328},~176--193.
				
				\bibitem[Fletcher \em{et~al.}(2007)Fletcher, Irwin, Teanby, Orton, Parrish,
				de~Kok, Howett, Calcutt, Bowles, and Taylor]{fletcher2007characterising}
				Fletcher, L.; Irwin, P.; Teanby, N.; Orton, G.; Parrish, P.; de~Kok, R.;
				Howett, C.; Calcutt, S.; Bowles, N.; Taylor, F.
				\newblock Characterising Saturn's vertical temperature structure from
				Cassini/CIRS.
				\newblock {\em Icarus} {\bf 2007}, {\em 189},~457--478.
				
				\bibitem[Fletcher \em{et~al.}(2010)Fletcher, Achterberg, Greathouse, Orton,
				Conrath, Simon-Miller, Teanby, Guerlet, Irwin, and
				Flasar]{fletcher2010seasonal}
				Fletcher, L.N.; Achterberg, R.K.; Greathouse, T.K.; Orton, G.S.; Conrath, B.J.;
				Simon-Miller, A.A.; Teanby, N.; Guerlet, S.; Irwin, P.G.; Flasar, F.
				\newblock Seasonal change on Saturn from Cassini/CIRS observations, 2004--2009.
				\newblock {\em Icarus} {\bf 2010}, {\em 208},~337--352.
				
				\bibitem[Guerlet \em{et~al.}(2011)Guerlet, Fouchet, B{\'e}zard, Flasar, and
				Simon-Miller]{guerlet2011evolution}
				Guerlet, S.; Fouchet, T.; B{\'e}zard, B.; Flasar, F.; Simon-Miller, A.
				\newblock Evolution of the equatorial oscillation in Saturn's stratosphere
				between 2005 and 2010 from Cassini/CIRS limb data analysis.
				\newblock {\em Geophys. Res. Lett.} {\bf 2011}, {\em 38}. https://doi.org/10.1029/2011GL047192.
				
				\bibitem[Sinclair \em{et~al.}(2013)Sinclair, Irwin, Fletcher, Moses,
				Greathouse, Friedson, Hesman, Hurley, and Merlet]{sinclair2013seasonalchem}
				Sinclair, J.; Irwin, P.; Fletcher, L.; Moses, J.; Greathouse, T.; Friedson, A.;
				Hesman, B.; Hurley, J.; Merlet, C.
				\newblock Seasonal variations of temperature, acetylene and ethane in
				Saturn’s atmosphere from 2005 to 2010, as observed by Cassini-CIRS.
				\newblock {\em Icarus} {\bf 2013}, {\em 225},~257--271.
				
				\bibitem[Sylvestre \em{et~al.}(2015)Sylvestre, Guerlet, Fouchet, Spiga, Flasar,
				Hesman, and Bjoraker]{sylvestre2015seasonal}
				Sylvestre, M.; Guerlet, S.; Fouchet, T.; Spiga, A.; Flasar, F.; Hesman, B.;
				Bjoraker, G.
				\newblock Seasonal changes in Saturn’s stratosphere inferred from
				Cassini/CIRS limb observations.
				\newblock {\em Icarus} {\bf 2015}, {\em 258},~224--238.
				
				\bibitem[Fletcher \em{et~al.}(2016)Fletcher, Irwin, Achterberg, Orton, and
				Flasar]{fletcher2016seasonal}
				Fletcher, L.N.; Irwin, P.G.; Achterberg, R.K.; Orton, G.S.; Flasar, F.M.
				\newblock Seasonal variability of Saturn’s tropospheric temperatures, winds
				and para-H2 from Cassini far-IR spectroscopy.
				\newblock {\em Icarus} {\bf 2016}, {\em 264},~137--159.
				
				
				\bibitem[Guerlet \em{et~al.}(2010)Guerlet, Fouchet, B{\'e}zard, Moses,
				Fletcher, Simon-Miller, and Flasar]{guerlet2010meridional}
				Guerlet, S.; Fouchet, T.; B{\'e}zard, B.; Moses, J.I.; Fletcher, L.N.;
				Simon-Miller, A.A.; Flasar, F.M.
				\newblock Meridional distribution of CH$_3$C$_2$H and C$_4$H$_2$ in Saturn’s stratosphere
				from CIRS/Cassini limb and nadir observations.
				\newblock {\em Icarus} {\bf 2010}, {\em 209},~682--695.
				
				\bibitem[Guerlet \em{et~al.}(2009)Guerlet, Fouchet, B{\'e}zard, Simon-Miller,
				and Flasar]{guerlet2009vertical}
				Guerlet, S.; Fouchet, T.; B{\'e}zard, B.; Simon-Miller, A.A.; Flasar, F.M.
				\newblock Vertical and meridional distribution of ethane, acetylene and propane
				in Saturn’s stratosphere from CIRS/Cassini limb observations.
				\newblock {\em Icarus} {\bf 2009}, {\em 203},~214--232.
				
				\bibitem[Teanby \em{et~al.}(2006)Teanby, Fletcher, Irwin, Fouchet, and
				Orton]{teanby2006new}
				Teanby, N.; Fletcher, L.; Irwin, P.; Fouchet, T.; Orton, G.
				\newblock New upper limits for hydrogen halides on Saturn derived from
				Cassini-CIRS data.
				\newblock {\em Icarus} {\bf 2006}, {\em 185},~466--475.
				
				\bibitem[Howett \em{et~al.}(2007)Howett, Irwin, Teanby, Simon-Miller, Calcutt,
				Fletcher, and de~Kok]{howett2007meridional}
				Howett, C.; Irwin, P.; Teanby, N.; Simon-Miller, A.; Calcutt, S.; Fletcher, L.;
				de~Kok, R.
				\newblock Meridional variations in stratospheric acetylene and ethane in the
				southern hemisphere of the saturnian atmosphere as determined from
				Cassini/CIRS measurements.
				\newblock {\em Icarus} {\bf 2007}, {\em 190},~556--572.
				
				\bibitem[Hesman \em{et~al.}(2012)Hesman, Bjoraker, Sada, Achterberg, Jennings,
				Romani, Lunsford, Fletcher, Boyle, Simon-Miller, et~al.]{hesman2012elusive}
				Hesman, B.; Bjoraker, G.; Sada, P.; Achterberg, R.; Jennings, D.; Romani, P.;
				Lunsford, A.; Fletcher, L.; Boyle, R.; Simon-Miller, A.;  et~al.
				\newblock Elusive ethylene detected in Saturn's northern storm region.
				\newblock {\em Astrophys. J.} {\bf 2012}, {\em 760},~24.
				
				\bibitem[Hurley \em{et~al.}(2012)Hurley, Fletcher, Irwin, Calcutt, Sinclair,
				and Merlet]{hurley2012latitudinal}
				Hurley, J.; Fletcher, L.; Irwin, P.; Calcutt, S.; Sinclair, J.; Merlet, C.
				\newblock Latitudinal variation of upper tropospheric NH3 on Saturn derived
				from Cassini/CIRS far-infrared measurements.
				\newblock {\em Planet.  Space Sci.} {\bf 2012}, {\em 73},~347--363.
				
				\bibitem[Abbas \em{et~al.}(2013)Abbas, LeClair, Woodard, Young, Stanbro,
				Flasar, Kunde, Achterberg, Bjoraker, Brasunas,
				et~al.]{abbas2013co2distribution}
				Abbas, M.; LeClair, A.; Woodard, E.; Young, M.; Stanbro, M.; Flasar, F.; Kunde,
				V.; Achterberg, R.; Bjoraker, G.; Brasunas, J.;  et~al.
				\newblock Distribution of CO$_2$ in Saturn's atmosphere from Cassini/CIRS infrared
				observations.
				\newblock {\em Astrophys. J.} {\bf 2013}, {\em 776},~73.
				
				\bibitem[Koskinen \em{et~al.}(2016)Koskinen, Moses, West, Guerlet, and
				Jouchoux]{koskinen2016detection}
				Koskinen, T.; Moses, J.; West, R.; Guerlet, S.; Jouchoux, A.
				\newblock The detection of benzene in Saturn's upper atmosphere.
				\newblock {\em Geophys. Res. Lett.} {\bf 2016}, {\em 43},~7895--7901.
				
				\bibitem[Koskinen and Guerlet(2018)]{koskinen2018atmospheric}
				Koskinen, T.; Guerlet, S.
				\newblock Atmospheric structure and helium abundance on Saturn from
				Cassini/UVIS and CIRS observations.
				\newblock {\em Icarus} {\bf 2018}, {\em 307},~161--171.
				
				\bibitem[Ingersoll(2020)]{ingersoll2020cassini}
				Ingersoll, A.P.
				\newblock Cassini exploration of the planet Saturn: a comprehensive review.
				\newblock {\em Space Sci. Rev.} {\bf 2020}, {\em 216},~1--51.
				
				\bibitem[Leech \em{et~al.}(2003)Leech, Kester, Shipman, Beintema, Feuchtgruber,
				Heras, Huygen, Lahuis, Lutz, Morris, et~al.]{leech2003iso}
				Leech, K.; Kester, D.; Shipman, R.; Beintema, D.; Feuchtgruber, H.; Heras, A.;
				Huygen, R.; Lahuis, F.; Lutz, D.; Morris, P.;  et~al.
				\newblock \emph{The ISO Handbook, Volume V-SWS-The Short Wavelength Spectrometer}; 
				\newblock {ESA Special Publication}: {2003}.  %MDPI: please add publisher's location (city, country): MTR: Appears to be digital
				
				\bibitem[Sloan \em{et~al.}(2003)Sloan, Kraemer, Price, and
				Shipman]{sloan2003uniform}
				Sloan, G.; Kraemer, K.E.; Price, S.D.; Shipman, R.F.
				\newblock A uniform database of 2.4--45.4 micron spectra from the Infrared Space
				Observatory Short Wavelength Spectrometer.
				\newblock {\em Astrophys. J. Suppl. Ser.} {\bf 2003}, {\em
					147},~379.
				
				\bibitem[Encrenaz \em{et~al.}(1996)Encrenaz, de~Graauw, Schaeidt, Lellouch,
				Feuchtgruber, Beintema, B{\'e}zard, Drossart, Griffin, Heras,
				et~al.]{encrenaz1996first}
				Encrenaz, T.; de~Graauw, T.; Schaeidt, S.; Lellouch, E.; Feuchtgruber, H.;
				Beintema, D.; B{\'e}zard, B.; Drossart, P.; Griffin, M.; Heras, A.;  et~al.
				\newblock First results of ISO-SWS observations of Jupiter.
				\newblock {\em Astron. Astrophys.} {\bf 1996}, {\em 315},~L397--L400.
				
				\bibitem[B{\'e}zard \em{et~al.}(1998)B{\'e}zard, Feuchtgruber, Moses, and
				Encrenaz]{bezard1998detection}
				B{\'e}zard, B.; Feuchtgruber, H.; Moses, J.; Encrenaz, T.
				\newblock Detection of methyl radicals (CH$_3$) on Saturn.
				\newblock {\em Astron. Astrophys.} {\bf 1998}, {\em 334},~L41--L44.
				
				\bibitem[De~Graauw \em{et~al.}(1997)De~Graauw, Feuchtgruber, Bezard, Drossart,
				Encrenaz, Beintema, Griffin, Heras, Kessler, Leech, et~al.]{de1997first}
				De~Graauw, T.; Feuchtgruber, H.; Bezard, B.; Drossart, P.; Encrenaz, T.;
				Beintema, D.; Griffin, M.; Heras, A.; Kessler, M.; Leech, K.;  et~al.
				\newblock First results of ISO-SWS observations of Saturn: Detection of
				CO$_2$, CH$_3$C$_2$H, C$_4$H$_2$ and tropospheric H$_2$O.
				\newblock {\em Astron. Astrophys.} {\bf 1997}, {\em 321},~L13--L16.
				
				\bibitem[Lellouch \em{et~al.}(1997)Lellouch, Feuchtgruber, de~Graauw, Bezard,
				Encrenaz, and Griffin]{lellouch1997h_2o}
				Lellouch, E.; Feuchtgruber, H.; de~Graauw, T.; Bezard, B.; Encrenaz, T.;
				Griffin, M.
				\newblock H$_2$O and CO$_2$ in the Upper Atmospheres of the Giant Planets. In
				\newblock  \emph{AAS/Division for Planetary Sciences Meeting Abstracts\# 29}; Washington, DC USA 1997; 
				Volume~29, p. 992. %MDPI: please add publisher and publisher's location (city, country)
				
				\bibitem[B{\'e}zard \em{et~al.}(1999)B{\'e}zard, Romani, Feuchtgruber, and
				Encrenaz]{bezard1999detection}
				B{\'e}zard, B.; Romani, P.; Feuchtgruber, H.; Encrenaz, T.
				\newblock Detection of the methyl radical on Neptune.
				\newblock {\em Astrophys. J.} {\bf 1999}, {\em 515},~868.
				
				\bibitem[Encrenaz \em{et~al.}(1999)Encrenaz, Drossart, Feuchtgruber, Lellouch,
				B{\'e}zard, Fouchet, and Atreya]{encrenaz1999atmospheric}
				Encrenaz, T.; Drossart, P.; Feuchtgruber, H.; Lellouch, E.; B{\'e}zard, B.;
				Fouchet, T.; Atreya, S.
				\newblock The atmospheric composition and structure of Jupiter and Saturn from
				ISO observations: A preliminary review.
				\newblock {\em Planet.  Space Sci.} {\bf 1999}, {\em 47},~1225--1242.
				
				\bibitem[Werner \em{et~al.}(2004)Werner, Roellig, Low, Rieke, Rieke, Hoffmann,
				Young, Houck, Brandl, Fazio, et~al.]{werner2004spitzer}
				Werner, M.W.; Roellig, T.; Low, F.; Rieke, G.H.; Rieke, M.; Hoffmann, W.;
				Young, E.; Houck, J.; Brandl, B.; Fazio, G.;  et~al.
				\newblock The Spitzer space telescope mission.
				\newblock {\em Astrophys. J. Suppl. Ser.} {\bf 2004}, {\em
					154},~1.
				
				\bibitem[Houck \em{et~al.}(2004)Houck, Roellig, Van~Cleve, Forrest, Herter,
				Lawrence, Matthews, Reitsema, Soifer, Watson, et~al.]{houck2004infrared}
				Houck, J.R.; Roellig, T.L.; Van~Cleve, J.; Forrest, W.J.; Herter, T.; Lawrence,
				C.R.; Matthews, K.; Reitsema, H.J.; Soifer, B.T.; Watson, D.M.;  et~al.
				\newblock The infrared spectrograph*(IRS) on the Spitzer space telescope.
				\newblock {\em Astrophys. J. Suppl. Ser.} {\bf 2004}, {\em
					154},~18.
				
				\bibitem[Meadows \em{et~al.}(2008)Meadows, Orton, Line, Liang, Yung, Van~Cleve,
				and Burgdorf]{meadows2008first}
				Meadows, V.S.; Orton, G.; Line, M.; Liang, M.C.; Yung, Y.L.; Van~Cleve, J.;
				Burgdorf, M.J.
				\newblock First Spitzer observations of Neptune: Detection of new hydrocarbons.
				\newblock {\em Icarus} {\bf 2008}, {\em 197},~585--589.
				
				\bibitem[Rowe-Gurney \em{et~al.}(2021{\natexlab{a}})Rowe-Gurney, Fletcher,
				Orton, Roman, Sinclair, Moses, and Irwin]{rowe2021neptune}
				Rowe-Gurney, N.; Fletcher, L.; Orton, G.; Roman, M.; Sinclair, J.; Moses, J.;
				Irwin, P.
				\newblock  {Neptune's Atmospheric Structure from the Spitzer Infrared
					Spectrometer}. Technical Report. In Proceedings of the  15th Europlanet Science Congress 2021,  Online, 13--24 September 2021; 
				
				\bibitem[Rowe-Gurney \em{et~al.}(2021{\natexlab{b}})Rowe-Gurney, Fletcher,
				Orton, Roman, Mainzer, Moses, De~Pater, and Irwin]{rowe2021longitudinal}
				Rowe-Gurney, N.; Fletcher, L.N.; Orton, G.S.; Roman, M.T.; Mainzer, A.; Moses,
				J.I.; De~Pater, I.; Irwin, P.G.
				\newblock Longitudinal variations in the stratosphere of Uranus from the
				Spitzer infrared spectrometer.
				\newblock {\em Icarus} {\bf 2021}, {\em 365},~114506.
				
				\bibitem[{Gezari} \em{et~al.}(1989){Gezari}, {Mumma}, {Espenak}, {Deming},
				{Bjoraker}, {Woods}, and {Folz}]{Gezari1989Nature}
				{Gezari}, D.Y.; {Mumma}, M.J.; {Espenak}, F.; {Deming}, D.; {Bjoraker}, G.;
				{Woods}, L.; {Folz}, W.
				\newblock {New features in Saturn's atmosphere revealed by high-resolution
					thermal infrared images}.
				\newblock {\em Nature} {\bf 1989}, {\em 342},~777--780.
				\newblock
				{\changeurlcolor{black}\href{https://doi.org/10.1038/342777a0}{\detokenize{https://doi.org/10.1038/342777a0}}}.
				
				\bibitem[Hoffmann \em{et~al.}(1993)Hoffmann, Fazio, Shivanandan, Hora, and
				Deutsch]{hoffmann1993mirac}
				Hoffmann, W.F.; Fazio, G.G.; Shivanandan, K.; Hora, J.L.; Deutsch, L.K.
				\newblock MIRAC: A mid-infrared array camera for astronomy. In
				\newblock  \emph{Infrared Detectors and Instrumentation}; SPIE: Bellingham, WA USA  %MDPI: please add location (city, country) of publisher
				1993; Volume 1946, pp.
				449--460.
				
				\bibitem[{Hoffmann} \em{et~al.}(1994){Hoffmann}, {Fazio}, {Shivanandan},
				{Hora}, and {Deutsch}]{Hoffman1994In}
				{Hoffmann}, W.F.; {Fazio}, G.G.; {Shivanandan}, K.; {Hora}, J.L.; {Deutsch},
				L.K.
				\newblock {Astronomical observations with the Mid-Infrared Array Camera,
					MIRAC}.
				\newblock {\em Infrared Phys. Technol.} {\bf 1994}, {\em 35},~175--194.
				\newblock
				{\changeurlcolor{black}\href{https://doi.org/10.1016/1350-4495(94)90079-5}{\detokenize{https://doi.org/10.1016/1350-4495(94)90079-5}}}.
				
				\bibitem[Jones and Puetter(1993)]{jones1993keck}
				Jones, B.; Puetter, R.C.
				\newblock Keck long-wavelength spectrometer. In 
				\newblock  \emph{Infrared Detectors and Instrumentation}; International Society for
					Optics and Photonics: Bellingham, WA USA 1993; Volume 1946, pp. 610--621. %MDPI: please add publisher's location (city, country)
				
				\bibitem[Glasse \em{et~al.}(1997)Glasse, Ettedgui-Atad, and
				Harris]{glasse1997michelle}
				Glasse, A.C.; Ettedgui-Atad, E.; Harris, J.W.
				\newblock Michelle midinfrared spectrometer and imager. In 
				\newblock  \emph{Optical Telescopes of Today and Tomorrow}; International Society for
					Optics and Photonics: Bellingham, WA USA 1997; Volume 2871, pp. 1197--1203. %MDPI: please add publisher's location (city, country)
				
				\bibitem[Lagage \em{et~al.}(2004)Lagage, Pel, Authier, Belorgey, Claret,
				Doucet, Dubreuil, Durand, Elswijk, Girardot, et~al.]{lagage2004visir}
				Lagage, P.; Pel, J.; Authier, M.; Belorgey, J.; Claret, A.; Doucet, C.;
				Dubreuil, D.; Durand, G.; Elswijk, E.; Girardot, P.;  et~al.
				\newblock Successful Commissioning OF.
				\newblock {\em  Messenger} {\bf 2004}, {\em 117},~12.
				
				\bibitem[De~Buizer and Fisher(2005)]{deBuiz2005Trecs}
				De~Buizer, J.M.; Fisher, R.S.
				\newblock T-ReCS and Michelle: The mid-infrared spectroscopic capabilities of
				the Gemini observatory. In {\em High Resolution Infrared Spectroscopy in
					Astronomy}; Springer:  Berlin/Heidelberg, Germany,  2005; pp. 84--87.
				
				\bibitem[Kataza \em{et~al.}(2000)Kataza, Okamoto, Takubo, Onaka, Sako,
				Nakamura, Miyata, and Yamashita]{kataza2000comics}
				Kataza, H.; Okamoto, Y.; Takubo, S.; Onaka, T.; Sako, S.; Nakamura, K.; Miyata,
				T.; Yamashita, T.
				\newblock COMICS: the cooled mid-infrared camera and spectrometer for the
				Subaru telescope. In
				\newblock  \emph{Optical and IR Telescope Instrumentation and Detectors};  
				International Society for Optics and Photonics:Bellingham, WA USA  2000; Volume 4008, pp.
				1144--1152. %MDPI: please add publisher's location (city, country)
				
				\bibitem[Fletcher \em{et~al.}(2011)Fletcher, Hesman, Irwin, Baines, Momary,
				Sanchez-Lavega, Flasar, Read, Orton, Simon-Miller,
				et~al.]{fletcher2011thermalsaturn}
				Fletcher, L.N.; Hesman, B.E.; Irwin, P.G.; Baines, K.H.; Momary, T.W.;
				Sanchez-Lavega, A.; Flasar, F.M.; Read, P.L.; Orton, G.S.; Simon-Miller, A.;
				et~al.
				\newblock Thermal structure and dynamics of Saturn’s northern springtime
				disturbance.
				\newblock {\em Science} {\bf 2011}, {\em 332},~1413--1417.
				
				
				\bibitem[Lacy \em{et~al.}(2002)Lacy, Richter, Greathouse, Jaffe, and
				Zhu]{lacy2002texes}
				Lacy, J.; Richter, M.; Greathouse, T.; Jaffe, D.; Zhu, Q.
				\newblock TEXES: A sensitive high-resolution grating spectrograph for the
				mid-infrared.
				\newblock {\em Publ. Astron. Soc. Pac.} {\bf
					2002}, {\em 114},~153.
				
				\bibitem[Fletcher \em{et~al.}(2014)Fletcher, Greathouse, Orton, Irwin, Mousis,
				Sinclair, and Giles]{fletcher2014origin}
				Fletcher, L.N.; Greathouse, T.K.; Orton, G.S.; Irwin, P.G.; Mousis, O.;
				Sinclair, J.A.; Giles, R.S.
				\newblock The origin of nitrogen on Jupiter and Saturn from the 15N/14N ratio.
				\newblock {\em Icarus} {\bf 2014}, {\em 238},~170--190.
				
				\bibitem[Sinclair \em{et~al.}(2018)Sinclair, Orton, Greathouse, Fletcher,
				Moses, Hue, and Irwin]{sinclair2018jupiter}
				Sinclair, J.; Orton, G.; Greathouse, T.; Fletcher, L.N.; Moses, J.; Hue, V.;
				Irwin, P.
				\newblock Jupiter’s auroral-related stratospheric heating and chemistry II:
				analysis of IRTF-TEXES spectra measured in December 2014.
				\newblock {\em Icarus} {\bf 2018}, {\em 300},~305--326.
				
				\bibitem[Fletcher \em{et~al.}(2018)Fletcher, Melin, Adriani, Simon,
				Sanchez-Lavega, Donnelly, Antu{\~n}ano, Orton, Hueso, Kraaikamp,
				et~al.]{fletcher2018jupiter}
				Fletcher, L.N.; Melin, H.; Adriani, A.; Simon, A.; Sanchez-Lavega, A.;
				Donnelly, P.; Antu{\~n}ano, A.; Orton, G.; Hueso, R.; Kraaikamp, E.;  et~al.
				\newblock Jupiter’s Mesoscale Waves Observed at 5 $\upmu$m by Ground-based
				Observations and Juno JIRAM.
				\newblock {\em Astron. J.} {\bf 2018}, {\em 156},~67.
				
				\bibitem[Melin \em{et~al.}(2018)Melin, Fletcher, Donnelly, Greathouse, Lacy,
				Orton, Giles, Sinclair, and Irwin]{melin2018assessing}
				Melin, H.; Fletcher, L.; Donnelly, P.; Greathouse, T.; Lacy, J.; Orton, G.;
				Giles, R.; Sinclair, J.; Irwin, P.
				\newblock Assessing the long-term variability of acetylene and ethane in the
				stratosphere of Jupiter.
				\newblock {\em Icarus} {\bf 2018}, {\em 305},~301--313.
				
				\bibitem[Blain \em{et~al.}(2018)Blain, Fouchet, Greathouse, Encrenaz, Charnay,
				B{\'e}zard, Li, Lellouch, Orton, Fletcher, et~al.]{blain2018mapping}
				Blain, D.; Fouchet, T.; Greathouse, T.; Encrenaz, T.; Charnay, B.; B{\'e}zard,
				B.; Li, C.; Lellouch, E.; Orton, G.; Fletcher, L.N.;  et~al.
				\newblock Mapping of Jupiter’s tropospheric NH3 abundance using ground-based
				IRTF/TEXES observations at 5 $\upmu$m.
				\newblock {\em Icarus} {\bf 2018}, {\em 314},~106--120.
				
				\bibitem[Sinclair \em{et~al.}(2020)Sinclair, Greathouse, Giles, Antu{\~n}ano,
				Moses, Fouchet, B{\'e}zard, Tao, Mart{\'\i}n-Torres, Clark,
				et~al.]{sinclair2020spatial}
				Sinclair, J.A.; Greathouse, T.K.; Giles, R.S.; Antu{\~n}ano, A.; Moses, J.I.;
				Fouchet, T.; B{\'e}zard, B.; Tao, C.; Mart{\'\i}n-Torres, J.; Clark, G.B.;
				et~al.
				\newblock Spatial Variations in the Altitude of the CH4 Homopause at
				Jupiter’s Mid-to-high Latitudes, as Constrained from IRTF-TEXES Spectra.
				\newblock {\em  Planet. Sci. J.} {\bf 2020}, {\em 1},~85.
				
				\bibitem[Greathouse \em{et~al.}(2005)Greathouse, Lacy, B{\'e}zard, Moses,
				Griffith, and Richter]{greathouse2005meridional}
				Greathouse, T.K.; Lacy, J.H.; B{\'e}zard, B.; Moses, J.I.; Griffith, C.A.;
				Richter, M.J.
				\newblock Meridional variations of temperature, C2H2 and C2H6 abundances in
				Saturn's stratosphere at southern summer solstice.
				\newblock {\em Icarus} {\bf 2005}, {\em 177},~18--31.
				
				\bibitem[Moses and Greathouse(2005)]{moses2005latitudinal}
				Moses, J.; Greathouse, T.
				\newblock Latitudinal and seasonal models of stratospheric photochemistry on
				Saturn: Comparison with infrared data from IRTF/TEXES.
				\newblock {\em J. Geophys. Res. Planets} {\bf 2005}, {\em 110}.  https://doi.org/10.1029/2005JE002450.
				
				\bibitem[Greathouse \em{et~al.}(2006)Greathouse, Lacy, B{\'e}zard, Moses,
				Richter, and Knez]{greathouse2006first}
				Greathouse, T.K.; Lacy, J.H.; B{\'e}zard, B.; Moses, J.I.; Richter, M.J.; Knez,
				C.
				\newblock The first detection of propane on Saturn.
				\newblock {\em Icarus} {\bf 2006}, {\em 181},~266--271.
				
				\bibitem[Fouchet \em{et~al.}(2016)Fouchet, Greathouse, Spiga, Fletcher,
				Guerlet, Leconte, and Orton]{fouchet2016stratospheric}
				Fouchet, T.; Greathouse, T.K.; Spiga, A.; Fletcher, L.N.; Guerlet, S.; Leconte,
				J.; Orton, G.S.
				\newblock Stratospheric aftermath of the 2010 Storm on Saturn as observed by
				the TEXES instrument. I. Temperature structure.
				\newblock {\em Icarus} {\bf 2016}, {\em 277},~196--214.
				
				\bibitem[Trafton \em{et~al.}(2012)Trafton, Orton, Greathouse, Lacy, and
				Encrenaz]{trafton2012mid}
				Trafton, L.M.; Orton, G.; Greathouse, T.; Lacy, J.; Encrenaz, T.
				\newblock Mid-IR Observations of Uranus’ H2 Quadrupole Emission Near Equinox. In 
				\newblock  \emph{AAS/Division for Planetary Sciences Meeting Abstracts\# 44}; AAS: Washington, DC USA, 2012; 
				Volume~44, p. 412.21. %MDPI: please add publisher and publisher's location (city, country)
				
				\bibitem[Orton \em{et~al.}(2019)Orton, Trafron, Fletcher, Encrenaz, Roman,
				Greathouse, Lacy, Sinclair, Moses, Leyrat, et~al.]{orton2019spatial}
				Orton, G.; Trafron, L.; Fletcher, L.; Encrenaz, T.; Roman, M.; Greathouse, T.;
				Lacy, J.; Sinclair, J.; Moses, J.; Leyrat, C.;  et~al.
				\newblock Spatial Variability in the Stratosphere of Uranus.
				\newblock  \emph{Geophys. Res. Abstr.}  \textbf{2019}, \emph{21}, 1.
				
				\bibitem[Kamizuka \em{et~al.}(2012)Kamizuka, Miyata, Sako, Nakamura, Asano,
				Uchiyama, Okada, Onaka, Sakon, Kataza, et~al.]{kamizuka2012development}
				Kamizuka, T.; Miyata, T.; Sako, S.; Nakamura, T.; Asano, K.; Uchiyama, M.;
				Okada, K.; Onaka, T.; Sakon, I.; Kataza, H.;  et~al.
				\newblock Development of MIMIZUKU: A mid-infrared multi-field imager for 6.5-m
				TAO telescope. In
				\newblock  \emph{Ground-based and Airborne Instrumentation for Astronomy IV}; SPIE: Bellingham, WA USA %MDPI: please add location (city, country) of publisher 
				2012; Volume 8446, pp. 1982--1992.
				
				\bibitem[Miyata \em{et~al.}(2022)Miyata, Yoshii, Doi, Kohno, Tanaka, Motohara,
				Minezaki, Sako, Morokuma, Tanabe, et~al.]{miyata2022university}
				Miyata, T.; Yoshii, Y.; Doi, M.; Kohno, K.; Tanaka, M.; Motohara, K.; Minezaki,
				T.; Sako, S.; Morokuma, T.; Tanabe, T.;  et~al.
				\newblock The University of Tokyo Atacama Observatory 6.5 m telescope: project
				status 2022. In
				\newblock  \emph{Ground-based and Airborne Telescopes IX}; SPIE: Bellingham, WA USA %MDPI: please add location (city, country) of publisher
				2022; Volume 12182,
				pp. 385--393.
				
				\bibitem[Kamizuka(2022)]{kamizuka2022challenging}
				Kamizuka, T.
				\newblock Challenging the difficulties in ground-based MIR observations: The
				case of TAO/MIMIZUKU.
				\newblock {IR2022: An Infrared Bright Future for Ground-based IR
					Observatories in the Era of JWST.} {2022}; p.~19. Available online: \url{https://zenodo.org/communities/ir2022} (accessed on 27, December 2022). %MDPI: please add access date (day month year): added
				
				\bibitem[Kamizuka \em{et~al.}(2020)Kamizuka, Miyata, Sako, Ohsawa, Asano,
				Uchiyama, Mori, Yoshida, Tachibana, Michifuji,
				et~al.]{kamizuka2020university}
				Kamizuka, T.; Miyata, T.; Sako, S.; Ohsawa, R.; Asano, K.; Uchiyama, M.S.;
				Mori, T.; Yoshida, Y.; Tachibana, K.; Michifuji, T.;  et~al.
				\newblock The University of Tokyo Atacama Observatory 6.5 m telescope: On-sky
				performance evaluations of the mid-infrared instrument MIMIZUKU on the Subaru
				telescope. In
				\newblock  \emph{Ground-Based and Airborne Instrumentation for Astronomy VIII}; SPIE: Bellingham, WA USA %MDPI: please add location (city, country) of publisher
				2020; Volume 11447, pp. 1296--1314.
				
				\bibitem[Brandl \em{et~al.}(2012)Brandl, Lenzen, Pantin, Glasse, Blommaert,
				Meyer, Guedel, Venema, Molster, Stuik, et~al.]{brandl2012metis}
				Brandl, B.R.; Lenzen, R.; Pantin, E.; Glasse, A.; Blommaert, J.; Meyer, M.;
				Guedel, M.; Venema, L.; Molster, F.; Stuik, R.;  et~al.
				\newblock METIS: the thermal infrared instrument for the E-ELT. In
				\newblock  \emph{Ground-Based and Airborne Instrumentation for Astronomy IV}; SPIE: Bellingham, WA USA  %MDPI: please add location (city, country) of publisher 
				2012; Volume 8446, pp. 554--566.
				
				\bibitem[Brandl \em{et~al.}(2022)Brandl, Bettonvil, van Boekel, Glauser, Quanz,
				Absil, Feldt, Garcia, Glasse, Guedel, et~al.]{brandl2022status}
				Brandl, B.R.; Bettonvil, F.; van Boekel, R.; Glauser, A.; Quanz, S.P.; Absil,
				O.; Feldt, M.; Garcia, P.J.; Glasse, A.; Guedel, M.;  et~al.
				\newblock Status update on the development of METIS, the mid-infrared ELT
				imager and spectrograph. In
				\newblock  \emph{Ground-Based and Airborne Instrumentation for Astronomy IX}; SPIE: Bellingham, WA USA  %MDPI: please add location (city, country) of publisher
				2022; Volume 12184, pp. 690--705.
				
				\bibitem[Antunano \em{et~al.}(2021)Antunano, Fletcher, Orton, Melin, Donnelly,
				Roman, Sinclair, and Kasaba]{antunano2021cycles}
				Antunano, A.; Fletcher, L.N.; Orton, G.S.; Melin, H.; Donnelly, P.T.; Roman,
				M.T.; Sinclair, J.A.; Kasaba, Y.
				\newblock Cycles of Variability in Jupiter's Atmosphere from Ground-Based
				Mid-Infrared Observations. Technical Report. In Proceedings of the European Planetary Science Congress 2021, Online, 13--24 September 2021.
				
				
				
				
				\bibitem[Drossart(1998)]{drossart1998saturn}
				Drossart, P.
				\newblock Saturn tropospheric water measured with ISO/SWS. In
				\newblock  \emph{AAS/Division for Planetary Sciences Meeting Abstracts\# 30}; AAS: Washington, DC USA, 1998;
				Volume~30, p. 1060. %MDPI: please add publisher and publisher's location (city, country)
				
				\bibitem[{\"O}berg \em{et~al.}(2011){\"O}berg, Murray-Clay, and
				Bergin]{oberg2011effects}
				{\"O}berg, K.I.; Murray-Clay, R.; Bergin, E.A.
				\newblock The effects of snowlines on C/O in planetary atmospheres.
				\newblock {\em Astrophys. J. Lett.} {\bf 2011}, {\em 743},~L16.
				
				\bibitem[Mousis \em{et~al.}(2012)Mousis, Lunine, Madhusudhan, and
				Johnson]{mousis2012nebular}
				Mousis, O.; Lunine, J.I.; Madhusudhan, N.; Johnson, T.V.
				\newblock Nebular water depletion as the cause of Jupiter's low oxygen
				abundance.
				\newblock {\em Astrophys. J. Lett.} {\bf 2012}, {\em 751},~L7.
				
				\bibitem[Lunine and Hunten(1989)]{lunine1989abundance}
				Lunine, J.I.; Hunten, D.M.
				\newblock Abundance of condensable species at planetary cold traps: The role of
				moist convection.
				\newblock {\em Planet.  Space Sci.} {\bf 1989}, {\em 37},~151--166.
				
				\bibitem[Taylor \em{et~al.}(2004)Taylor, Atreya, Encrenaz, Hunten, Irwin, and
				Owen]{taylor2004composition}
				Taylor, F.; Atreya, S.; Encrenaz, T.; Hunten, D.; Irwin, P.; Owen, T.
				\newblock The composition of the atmosphere of Jupiter. In
				\newblock {\em Jupiter: The Planet, Satellites and Magnetosphere}; Cambridge University Press: Cambridge, UK, {2004}; 
				pp. 59--78. %MDPI: please add publisher and publisher's location (city, country)
				
				\bibitem[Giles \em{et~al.}(2017)Giles, Fletcher, and
				Irwin]{giles2017latitudinal}
				Giles, R.S.; Fletcher, L.N.; Irwin, P.G.
				\newblock Latitudinal variability in Jupiter’s tropospheric disequilibrium
				species: GeH$_4$, AsH$_3$ and PH$_3$.
				\newblock {\em Icarus} {\bf 2017}, {\em 289},~254--269.
				
				\bibitem[Moreno \em{et~al.}(2009)Moreno, Marten, and
				Lellouch]{moreno2009search}
				Moreno, R.; Marten, A.; Lellouch, E.
				\newblock Search for PH3 in the atmospheres of Uranus and Neptune at millimeter
				wavelength. In
				\newblock  \emph{AAS/Division for Planetary Sciences Meeting Abstracts\# 41}; AAS: Washington, DC USA, 2009,
				Volume~41, p. 28.02. %MDPI: please add publisher and publisher's location (city, country)
				
				\bibitem[Lecluse \em{et~al.}(1996)Lecluse, Robert, Gautier, and
				Guiraud]{lecluse1996deuterium}
				Lecluse, C.; Robert, F.; Gautier, D.; Guiraud, M.
				\newblock Deuterium enrichment in giant planets.
				\newblock {\em Planet.  Space Sci.} {\bf 1996}, {\em 44},~1579--1592.
				
				\bibitem[Griffin \em{et~al.}(1996)Griffin, Naylor, Davis, Ade, Oldham,
				Swinyard, Gautier, Lellouch, Orton, Encrenaz, et~al.]{griffin1996first}
				Griffin, M.; Naylor, D.; Davis, G.; Ade, P.A.; Oldham, P.; Swinyard, B.;
				Gautier, D.; Lellouch, E.; Orton, G.; Encrenaz, T.;  et~al.
				\newblock First detection of the 56-mum rotational line of HD in Saturn's
				atmosphere.
				\newblock {\em Astron. Astrophys.} {\bf 1996}, {\em 315},~L389--L392.
				
				\bibitem[Feuchtgruber \em{et~al.}(1999)Feuchtgruber, Lellouch, B{\'e}zard,
				Encrenaz, de~Graauw, and Davis]{feuchtgruber1999detection}
				Feuchtgruber, H.; Lellouch, E.; B{\'e}zard, B.; Encrenaz, T.; de~Graauw, T.;
				Davis, G.
				\newblock Detection of HD in the atmospheres of Uranus and Neptune: a new
				determination of the D/H ratio.
				\newblock {\em Astron. Astrophys.} {\bf 1999}, {\em 341},~L17--L21.
				
				\bibitem[Lellouch \em{et~al.}(2001)Lellouch, B{\'e}zard, Fouchet, Feuchtgruber,
				Encrenaz, and de~Graauw]{lellouch2001deuterium}
				Lellouch, E.; B{\'e}zard, B.; Fouchet, T.; Feuchtgruber, H.; Encrenaz, T.;
				de~Graauw, T.
				\newblock The deuterium abundance in Jupiter and Saturn from ISO-SWS
				observations.
				\newblock {\em Astron. Astrophys.} {\bf 2001}, {\em 370},~610--622.
				
				\bibitem[Moses \em{et~al.}(1992)Moses, Allen, and Yung]{moses1992hydrocarbon}
				Moses, J.I.; Allen, M.; Yung, Y.L.
				\newblock Hydrocarbon nucleation and aerosol formation in Neptune's atmosphere.
				\newblock {\em Icarus} {\bf 1992}, {\em 99},~318--346.
				
				\bibitem[Moses \em{et~al.}(2018)Moses, Fletcher, Greathouse, Orton, and
				Hue]{moses2018seasonal}
				Moses, J.I.; Fletcher, L.N.; Greathouse, T.K.; Orton, G.S.; Hue, V.
				\newblock Seasonal stratospheric photochemistry on Uranus and Neptune.
				\newblock {\em Icarus} {\bf 2018}, {\em 307},~124--145.
				
				\bibitem[Lellouch \em{et~al.}(2015)Lellouch, Moreno, Orton, Feuchtgruber,
				Cavali{\'e}, Moses, Hartogh, Jarchow, and Sagawa]{lellouch2015new}
				Lellouch, E.; Moreno, R.; Orton, G.; Feuchtgruber, H.; Cavali{\'e}, T.; Moses,
				J.; Hartogh, P.; Jarchow, C.; Sagawa, H.
				\newblock New constraints on the CH4 vertical profile in Uranus and Neptune
				from Herschel observations.
				\newblock {\em Astron. Astrophys.} {\bf 2015}, {\em 579},~A121.
				
				\bibitem[Baines and Smith(1990)]{baines_UV_neptune_1990}
				Baines, K.H.; Smith, W.H.
				\newblock The atmospheric structure and dynamical properties of Neptune derived
				from ground-based and IUE spectrophotometry.
				\newblock {\em Icarus} {\bf 1990}, {\em 85},~65--108.
				
				\bibitem[Baines and Hammel(1994)]{baines1994clouds}
				Baines, K.H.; Hammel, H.B.
				\newblock Clouds, hazes, and the stratospheric methane abundance in Neptune.
				\newblock {\em Icarus} {\bf 1994}, {\em 109},~20--39.
				
				\bibitem[Stoker(1986)]{stoker1986moist}
				Stoker, C.R.
				\newblock Moist convection: A mechanism for producing the vertical structure of
				the Jovian equatorial plumes.
				\newblock {\em Icarus} {\bf 1986}, {\em 67},~106--125.
				
				\bibitem[Orton \em{et~al.}(2007)Orton, Encrenaz, Leyrat, Puetter, and
				Friedson]{orton2007evidencehotspot}
				Orton, G.S.; Encrenaz, T.; Leyrat, C.; Puetter, R.; Friedson, A.J.
				\newblock Evidence for methane escape and strong seasonal and dynamical
					perturbations of Neptune's atmospheric temperatures. %MDPI: Refs. 319 and 352 are duplicated. Please remove one of them and rearrange all the references to appear in numerical order. Please ensure that there are no duplicated references.
				\newblock {\em Astron. Astrophys.} {\bf 2007}, {\em 473},~L5--L8.
				
				\bibitem[Appleby(1986)]{appleby1986radiative}
				Appleby, J.F.
				\newblock Radiative-convective equilibrium models of Uranus and Neptune.
				\newblock {\em Icarus} {\bf 1986}, {\em 65},~383--405.
				
				\bibitem[Friedson and Ingersoll(1987)]{friedson1987seasonal}
				Friedson, J.; Ingersoll, A.P.
				\newblock Seasonal meridional energy balance and thermal structure of the
				atmosphere of Uranus: A radiative-convective-dynamical model.
				\newblock {\em Icarus} {\bf 1987}, {\em 69},~135--156.
				
				\bibitem[Marley and McKay(1999)]{marley1999thermal}
				Marley, M.S.; McKay, C.P.
				\newblock Thermal structure of Uranus' atmosphere.
				\newblock {\em Icarus} {\bf 1999}, {\em 138},~268--286.
				
				\bibitem[Li \em{et~al.}(2018)Li, Le, Zhang, and Yung]{li2018high}
				Li, C.; Le, T.; Zhang, X.; Yung, Y.L.
				\newblock A high-performance atmospheric radiation package: With applications
				to the radiative energy budgets of giant planets.
				\newblock {\em J. Quant. Spectrosc. Radiat. Transf.}
				{\bf 2018}, {\em 217},~353--362.
				
				\bibitem[Melin \em{et~al.}(2020)Melin, Fletcher, Irwin, and
				Edgington]{melin2020jupiter}
				Melin, H.; Fletcher, L.N.; Irwin, P.G.; Edgington, S.G.
				\newblock Jupiter in the Ultraviolet: Acetylene and Ethane Abundances in the
				Stratosphere of Jupiter from Cassini Observations between 0.15 and 0.19
				$\upmu$m.
				\newblock {\em Astron. J.} {\bf 2020}, {\em 159},~291.
				
				\bibitem[Lellouch \em{et~al.}(2002)Lellouch, B{\'e}zard, Moses, Davis,
				Drossart, Feuchtgruber, Bergin, Moreno, and
				Encrenaz]{lellouch2002originh2oco2}
				Lellouch, E.; B{\'e}zard, B.; Moses, J.; Davis, G.; Drossart, P.; Feuchtgruber,
				H.; Bergin, E.; Moreno, R.; Encrenaz, T.
				\newblock The origin of water vapor and carbon dioxide in Jupiter's
				stratosphere.
				\newblock {\em Icarus} {\bf 2002}, {\em 159},~112--131.
				
				\bibitem[Encrenaz \em{et~al.}(1997)Encrenaz, Lellouch, Feuchtgruber, Altieri,
				B{\'e}zard, Davis, de~Graauw, Drossart, Griffin, Kessler,
				et~al.]{encrenaz1997giant}
				Encrenaz, T.; Lellouch, E.; Feuchtgruber, H.; Altieri, B.; B{\'e}zard, B.;
				Davis, M.; de~Graauw, T.; Drossart, P.; Griffin, M.; Kessler, M.;  et~al.
				\newblock The giant planets as seen by ISO. In
				\newblock The First ISO Workshop on Analytical Spectroscopy; ESA Publications Division:Noordwijk, The Netherlands,  1997; Volume 419,
				p. 125. %MDPI: please add publisher and publisher's location (city, country)
				
				\bibitem[Moses and Poppe(2017)]{moses2017dust}
				Moses, J.I.; Poppe, A.R.
				\newblock Dust ablation on the giant planets: consequences for stratospheric
				photochemistry.
				\newblock {\em Icarus} {\bf 2017}, {\em 297},~33--58.
				
				\bibitem[Seager and Deming(2010)]{seager2010exoplanet}
				Seager, S.; Deming, D.
				\newblock Exoplanet atmospheres.
				\newblock {\em Annu. Rev. Astron. Astrophys.} {\bf 2010}, {\em
					48},~631--672.
				
				\bibitem[Barstow and Heng(2020)]{barstow2020outstanding}
				Barstow, J.K.; Heng, K.
				\newblock Outstanding challenges of exoplanet atmospheric retrievals.
				\newblock {\em Space Sci. Rev.} {\bf 2020}, {\em 216},~1--25.
				
				
				
				
				\bibitem[Fletcher \em{et~al.}(2020)Fletcher, Kaspi, Guillot, and
				Showman]{fletcher2020beltzones}
				Fletcher, L.N.; Kaspi, Y.; Guillot, T.; Showman, A.P.
				\newblock How well do we understand the belt/zone circulation of giant planet
				atmospheres?
				\newblock {\em Space Sci. Rev.} {\bf 2020}, {\em 216},~1--33.
				
				\bibitem[Dowling(2021)]{dowling2021emoticons}
				Dowling, T.
				\newblock Emoticons for Teaching Jupiter's Belts, Zones, and Spots. In 
				\newblock  \emph{AAS/Division for Planetary Sciences Meeting Abstracts}; AAS: Washington, DC USA, 2021,
				Volume~53, p. 410.10. %MDPI: please add publisher and publisher's location (city, country)
				
				\bibitem[Conrath and Gierasch(1984)]{conrath1984global}
				Conrath, B.J.; Gierasch, P.J.
				\newblock Global variation of the para hydrogen fraction in Jupiter's
				atmosphere and implications for dynamics on the outer planets.
				\newblock {\em Icarus} {\bf 1984}, {\em 57},~184--204.
				
				\bibitem[de~Pater \em{et~al.}(2021)de~Pater, Fletcher, Reach, Goullaud, Orton,
				Wong, and Gehrz]{de2021sofia}
				de~Pater, I.; Fletcher, L.N.; Reach, W.T.; Goullaud, C.; Orton, G.S.; Wong,
				M.H.; Gehrz, R.D.
				\newblock SOFIA Observations of Variability in Jupiter's Para-H2 Distribution
				and Subsurface Emission Characteristics of the Galilean Satellites.
				\newblock {\em  Planet. Sci. J.} {\bf 2021}, {\em 2},~226.
				
				
				\bibitem[Porco \em{et~al.}(2003)Porco, West, McEwen, Del~Genio, Ingersoll,
				Thomas, Squyres, Dones, Murray, Johnson, et~al.]{porco2003cassini}
				Porco, C.C.; West, R.A.; McEwen, A.; Del~Genio, A.D.; Ingersoll, A.P.; Thomas,
				P.; Squyres, S.; Dones, L.; Murray, C.D.; Johnson, T.V.;  et~al.
				\newblock Cassini imaging of Jupiter's atmosphere, satellites, and rings.
				\newblock {\em Science} {\bf 2003}, {\em 299},~1541--1547.
				
				\bibitem[Garc{\'\i}a-Melendo \em{et~al.}(2011)Garc{\'\i}a-Melendo,
				P{\'e}rez-Hoyos, S{\'a}nchez-Lavega, and Hueso]{garcia2011saturn}
				Garc{\'\i}a-Melendo, E.; P{\'e}rez-Hoyos, S.; S{\'a}nchez-Lavega, A.; Hueso, R.
				\newblock Saturn’s zonal wind profile in 2004--2009 from Cassini ISS images
				and its long-term variability.
				\newblock {\em Icarus} {\bf 2011}, {\em 215},~62--74.
				
				
				
				\bibitem[{Tollefson} \em{et~al.}(2017){Tollefson}, {Wong}, {Pater}, {Simon},
				{Orton}, {Rogers}, {Atreya}, {Cosentino}, {Januszewski},
				{Morales-Juber{\'\i}as}, and {Marcus}]{Tollefson2017jupiterzonal}
				{Tollefson}, J.; {Wong}, M.H.; {Pater}, I.d.; {Simon}, A.A.; {Orton}, G.S.;
				{Rogers}, J.H.; {Atreya}, S.K.; {Cosentino}, R.G.; {Januszewski}, W.;
				{Morales-Juber{\'\i}as}, R.;  et~al.
				\newblock {Changes in Jupiter's Zonal Wind Profile preceding and during the
					Juno mission}.
				\newblock {\em Icarus} {\bf 2017}, {\em 296},~163--178.
				\newblock
				{\changeurlcolor{black}\href{https://doi.org/10.1016/j.icarus.2017.06.007}{\detokenize{https://doi.org/10.1016/j.icarus.2017.06.007}}}.
				
				\bibitem[Sromovsky \em{et~al.}(2001)Sromovsky, Fry, Dowling, Baines, and
				Limaye]{sromovsky2001neptune}
				Sromovsky, L.; Fry, P.; Dowling, T.; Baines, K.; Limaye, S.
				\newblock Neptune's atmospheric circulation and cloud morphology: Changes
				revealed by 1998 HST imaging.
				\newblock {\em Icarus} {\bf 2001}, {\em 150},~244--260.
				
				\bibitem[Sromovsky and Fry(2005)]{sromovsky2005dynamics}
				Sromovsky, L.; Fry, P.
				\newblock Dynamics of cloud features on Uranus.
				\newblock {\em Icarus} {\bf 2005}, {\em 179},~459--484.
				
				
				
				
				
				
				
				\bibitem[Liang \em{et~al.}(2005)Liang, Shia, Lee, Allen, Friedson, and
				Yung]{liang2005meridional}
				Liang, M.C.; Shia, R.L.; Lee, A.Y.T.; Allen, M.; Friedson, A.J.; Yung, Y.L.
				\newblock Meridional transport in the stratosphere of Jupiter.
				\newblock {\em Astrophys. J.} {\bf 2005}, {\em 635},~L177.
				
				\bibitem[Zhang \em{et~al.}(2013)Zhang, West, Banfield, and
				Yung]{zhang2013stratospheric}
				Zhang, X.; West, R.; Banfield, D.; Yung, Y.
				\newblock Stratospheric aerosols on Jupiter from Cassini observations.
				\newblock {\em Icarus} {\bf 2013}, {\em 226},~159--171.
				
				\bibitem[Orsolini and Leovy(1993)]{orsolini1993model}
				Orsolini, Y.; Leovy, C.
				\newblock A model of large-scale instabilities in the jovian troposphere: 1.
				Linear model.
				\newblock {\em Icarus} {\bf 1993}, {\em 106},~392--405.
				
				\bibitem[Ingersoll \em{et~al.}(2000)Ingersoll, Gierasch, Banfield, Vasavada,
				and Team]{ingersoll2000moist}
				Ingersoll, A.; Gierasch, P.; Banfield, D.; Vasavada, A.; Team, G.I.
				\newblock Moist convection as an energy source for the large-scale motions in
				Jupiter's atmosphere.
				\newblock {\em Nature} {\bf 2000}, {\em 403},~630--632.
				
				\bibitem[Ingersoll \em{et~al.}(2021)Ingersoll, Atreya, Bolton, Brueshaber,
				Fletcher, Levin, Li, Li, Lunine, Orton, et~al.]{ingersoll2021jupiter}
				Ingersoll, A.P.; Atreya, S.; Bolton, S.J.; Brueshaber, S.; Fletcher, L.N.;
				Levin, S.M.; Li, C.; Li, L.; Lunine, J.I.; Orton, G.S.;  et~al.
				\newblock Jupiter's overturning circulation: Breaking waves take the place of
				solid boundaries.
				\newblock {\em Geophys. Res. Lett.} {\bf 2021}, {\em
					48},~e2021GL095756.
				
				\bibitem[Duer \em{et~al.}(2021)Duer, Gavriel, Galanti, Kaspi, Fletcher,
				Guillot, Bolton, Levin, Atreya, Grassi, et~al.]{duer2021evidence}
				Duer, K.; Gavriel, N.; Galanti, E.; Kaspi, Y.; Fletcher, L.N.; Guillot, T.;
				Bolton, S.J.; Levin, S.M.; Atreya, S.K.; Grassi, D.;  et~al.
				\newblock Evidence for multiple Ferrel-like cells on Jupiter.
				\newblock {\em Geophys. Res. Lett.} {\bf 2021}, {\em
					48},~e2021GL095651.
				
				\bibitem[Fletcher \em{et~al.}(2021)Fletcher, Oyafuso, Allison, Ingersoll, Li,
				Kaspi, Galanti, Wong, Orton, Duer, et~al.]{fletcher2021jupiter}
				Fletcher, L.N.; Oyafuso, F.A.; Allison, M.; Ingersoll, A.; Li, L.; Kaspi, Y.;
				Galanti, E.; Wong, M.H.; Orton, G.S.; Duer, K.;  et~al.
				\newblock Jupiter's temperate belt/zone contrasts revealed at depth by Juno
				microwave observations.
				\newblock {\em J. Geophys. Res. Planets} {\bf 2021}, {\em
					126},~e2021JE006858.
				
				\bibitem[Achterberg \em{et~al.}(2018)Achterberg, Flasar, Bjoraker, Hesman,
				Gorius, Mamoutkine, Fletcher, Segura, Edgington, and
				Brooks]{achterberg2018thermal}
				Achterberg, R.; Flasar, F.; Bjoraker, G.; Hesman, B.; Gorius, N.; Mamoutkine,
				A.; Fletcher, L.; Segura, M.; Edgington, S.; Brooks, S.
				\newblock Thermal Emission From Saturn's Polar Cyclones.
				\newblock {\em Geophys. Res. Lett.} {\bf 2018}, {\em 45},~5312--5319.
				
				
				
				\bibitem[Fletcher \em{et~al.}(2017)Fletcher, Guerlet, Orton, Cosentino,
				Fouchet, Irwin, Li, Flasar, Gorius, and
				Morales-Juber{\'\i}as]{fletcher2017saturndisruption}
				Fletcher, L.N.; Guerlet, S.; Orton, G.S.; Cosentino, R.G.; Fouchet, T.; Irwin,
				P.G.; Li, L.; Flasar, F.M.; Gorius, N.; Morales-Juber{\'\i}as, R.
				\newblock Disruption of Saturn’s quasi-periodic equatorial oscillation by the
				great northern storm.
				\newblock {\em Nat. Astron.} {\bf 2017}, {\em 1},~765--770.
				
				\bibitem[Flasar \em{et~al.}(1987)Flasar, Conrath, Gierasch, and
				Pirraglia]{flasar1987voyager}
				Flasar, F.; Conrath, B.; Gierasch, P.; Pirraglia, J.
				\newblock Voyager infrared observations of Uranus' atmosphere: Thermal
				structure and dynamics.
				\newblock {\em J. Geophys. Res. Space Phys.} {\bf 1987}, {\em
					92},~15011--15018.
				
				
				
				
				\bibitem[S{\'a}nchez-Lavega \em{et~al.}(2019)S{\'a}nchez-Lavega, Sromovsky,
				Showman, Del~Genio, Young, Hueso, Garcia-Melendo, Kaspi, Orton,
				Barrado-Izagirre, et~al.]{sanchez2019gas}
				S{\'a}nchez-Lavega, A.; Sromovsky, L.; Showman, A.P.; Del~Genio, A.; Young, R.;
				Hueso, R.; Garcia-Melendo, E.; Kaspi, Y.; Orton, G.S.; Barrado-Izagirre, N.;
				et~al.
				\newblock Gas giants.
				\newblock Technical report, Cambridge University Press: Cambridge, UK,  2019.
				
				
				
				
				\bibitem[Hammel \em{et~al.}(2005)Hammel, de~Pater, Gibbard, Lockwood, and
				Rages]{hammel2005new}
				Hammel, H.; de~Pater, I.; Gibbard, S.; Lockwood, G.; Rages, K.
				\newblock New cloud activity on Uranus in 2004: First detection of a southern
				feature at 2.2 $\upmu$m.
				\newblock {\em Icarus} {\bf 2005}, {\em 175},~284--288.
				
				\bibitem[Hammel \em{et~al.}(2006)Hammel, Lynch, Russell, Sitko, Bernstein, and
				Hewagama]{hammel2006mid}
				Hammel, H.; Lynch, D.; Russell, R.; Sitko, M.; Bernstein, L.; Hewagama, T.
				\newblock Mid-infrared ethane emission on Neptune and Uranus.
				\newblock {\em Astrophys. J.} {\bf 2006}, {\em 644},~1326.
				
				\bibitem[Orton \em{et~al.}(2012)Orton, Fletcher, Liu, Schneider,
				Yanamandra-Fisher, de~Pater, Edwards, Geballe, Hammel, Fujiyoshi,
				et~al.]{orton2012recovery}
				Orton, G.S.; Fletcher, L.N.; Liu, J.; Schneider, T.; Yanamandra-Fisher, P.A.;
				de~Pater, I.; Edwards, M.; Geballe, T.R.; Hammel, H.B.; Fujiyoshi, T.;
				et~al.
				\newblock Recovery and characterization of Neptune's near-polar stratospheric
				hot spot.
				\newblock {\em Planet.  Space Sci.} {\bf 2012}, {\em 61},~161--167.
				
				\bibitem[Orton \em{et~al.}(2018)Orton, Moses, Encrenaz, Fletcher, Greathouse,
				Leyrat, Sinclair, Trafton, Lacy, and Pantin]{orton2018neii}
				Orton, G.; Moses, J.; Encrenaz, T.; Fletcher, L.; Greathouse, T.; Leyrat, C.;
				Sinclair, J.; Trafton, L.; Lacy, J.; Pantin, E.
				\newblock Spatial Variability in the Stratosphere of Uranus. In Proceedings of the 
				\newblock  42nd COSPAR Scientific Assembly, Pasadena, CA, USA, 14--22 July,  2018; Volume~42.
				
				\bibitem[Peek and Moore(1981)]{peek1981planet}
				Peek, B.M.; Moore, P.
				\newblock \emph{The Planet Jupiter: The Observer's Handbook};  Faber and Faber: London, UK; Boston, MA, USA, 
				{1981}.
				
				\bibitem[Rogers(1995)]{rogers1995giant}
				Rogers, J.H.
				\newblock {\em The Giant Planet Jupiter}; Cambridge University Press: Cambridge, UK,
				1995; Volume~6.
				
				\bibitem[Simon-Miller and Gierasch(2010)]{simon2010long}
				Simon-Miller, A.A.; Gierasch, P.J.
				\newblock On the long-term variability of Jupiter’s winds and brightness as
				observed from Hubble.
				\newblock {\em Icarus} {\bf 2010}, {\em 210},~258--269.
				
				\bibitem[Karkoschka and Tomasko(2005)]{karkoschka2005saturn}
				Karkoschka, E.; Tomasko, M.
				\newblock Saturn's vertical and latitudinal cloud structure 1991--2004 from HST
				imaging in 30 filters.
				\newblock {\em Icarus} {\bf 2005}, {\em 179},~195--221.
				
				\bibitem[Lockwood and Thompson(2002)]{lockwood2002photometric}
				Lockwood, G.; Thompson, D.
				\newblock Photometric variability of Neptune, 1972--2000.
				\newblock {\em Icarus} {\bf 2002}, {\em 156},~37--51.
				
				\bibitem[Lockwood and Jerzykiewicz(2006)]{lockwood2006photometric}
				Lockwood, G.; Jerzykiewicz, M.
				\newblock Photometric variability of Uranus and Neptune, 1950--2004.
				\newblock {\em Icarus} {\bf 2006}, {\em 180},~442--452.
				
				\bibitem[Sromovsky \em{et~al.}(2009)Sromovsky, Fry, Hammel, Ahue, de~Pater,
				Rages, Showalter, and van Dam]{sromovsky2009uranus}
				Sromovsky, L.A.; Fry, P.; Hammel, H.; Ahue, W.; de~Pater, I.; Rages, K.;
				Showalter, M.; van Dam, M.
				\newblock Uranus at equinox: Cloud morphology and dynamics.
				\newblock {\em Icarus} {\bf 2009}, {\em 203},~265--286.
				
				\bibitem[Roman \em{et~al.}(2018)Roman, Banfield, and
				Gierasch]{roman2018aerosols}
				Roman, M.T.; Banfield, D.; Gierasch, P.J.
				\newblock Aerosols and methane in the ice giant atmospheres inferred from
				spatially resolved, near-infrared spectra: I. Uranus, 2001--2007.
				\newblock {\em Icarus} {\bf 2018}, {\em 310},~54--76.
				
				\bibitem[Sromovsky \em{et~al.}(2019)Sromovsky, Karkoschka, Fry, de~Pater, and
				Hammel]{sromovsky2019methane}
				Sromovsky, L.A.; Karkoschka, E.; Fry, P.M.; de~Pater, I.; Hammel, H.B.
				\newblock The methane distribution and polar brightening on Uranus based on
				HST/STIS, Keck/NIRC2, and IRTF/SpeX observations through 2015.
				\newblock {\em Icarus} {\bf 2019}, {\em 317},~266--306.
				
				\bibitem[Lockwood(2019)]{lockwood2019final}
				Lockwood, G.
				\newblock Final compilation of photometry of Uranus and Neptune, 1972--2016.
				\newblock {\em Icarus} {\bf 2019}, {\em 324},~77--85.
				
				\bibitem[Karkoschka(2011)]{karkoschka2011neptune}
				Karkoschka, E.
				\newblock Neptune’s cloud and haze variations 1994--2008 from 500 HST--WFPC2
				images.
				\newblock {\em Icarus} {\bf 2011}, {\em 215},~759--773.
				
				\bibitem[Hueso \em{et~al.}(2017)Hueso, De~Pater, Simon, S{\'a}nchez-Lavega,
				Delcroix, Wong, Tollefson, Baranec, de~Kleer, Luszcz-Cook,
				et~al.]{huesoneptunelonglived2017}
				Hueso, R.; De~Pater, I.; Simon, A.; S{\'a}nchez-Lavega, A.; Delcroix, M.; Wong,
				M.; Tollefson, J.; Baranec, C.; de~Kleer, K.; Luszcz-Cook, S.;  et~al.
				\newblock Neptune long-lived atmospheric features in 2013--2015 from small
				(28-cm) to large (10-m) telescopes.
				\newblock {\em Icarus} {\bf 2017}, {\em 295},~89--109.
				
				\bibitem[Wong \em{et~al.}(2018)Wong, Tollefson, Hsu, de~Pater, Simon, Hueso,
				S{\'a}nchez-Lavega, Sromovsky, Fry, Luszcz-Cook, et~al.]{wong2018new}
				Wong, M.H.; Tollefson, J.; Hsu, A.I.; de~Pater, I.; Simon, A.A.; Hueso, R.;
				S{\'a}nchez-Lavega, A.; Sromovsky, L.; Fry, P.; Luszcz-Cook, S.;  et~al.
				\newblock A new dark vortex on Neptune.
				\newblock {\em Astron. J.} {\bf 2018}, {\em 155},~117.
				
				\bibitem[Molter \em{et~al.}(2019)Molter, de~Pater, Luszcz-Cook, Hueso,
				Tollefson, Alvarez, S{\'a}nchez-Lavega, Wong, Hsu, Sromovsky,
				et~al.]{molter2019neptune}
				Molter, E.; de~Pater, I.; Luszcz-Cook, S.; Hueso, R.; Tollefson, J.; Alvarez,
				C.; S{\'a}nchez-Lavega, A.; Wong, M.H.; Hsu, A.I.; Sromovsky, L.A.;  et~al.
				\newblock Analysis of Neptune’s 2017 bright equatorial storm.
				\newblock {\em Icarus} {\bf 2019}, {\em 321},~324--345.
				
				\bibitem[Hueso and S{\'a}nchez-Lavega(2019)]{hueso2019atmospheric}
				Hueso, R.; S{\'a}nchez-Lavega, A.
				\newblock Atmospheric dynamics and vertical structure of Uranus and Neptune’s
				weather layers.
				\newblock {\em Space Sci. Rev.} {\bf 2019}, {\em 215},~1--33.
				
				\bibitem[Simon \em{et~al.}(2022)Simon, Wong, Sromovsky, Fletcher, and
				Fry]{simon2022giantvisreview}
				Simon, A.A.; Wong, M.H.; Sromovsky, L.A.; Fletcher, L.N.; Fry, P.M.
				\newblock Giant Planet Atmospheres: Dynamics and Variability from UV to Near-IR
				Hubble and Adaptive Optics Imaging.
				\newblock {\em Remote Sens.} {\bf 2022}, {\em 14},~1518.
				
				\bibitem[Mitchell(1976)]{mitchell1976overviewvariability}
				Mitchell, J.M.
				\newblock An overview of climatic variability and its causal mechanisms.
				\newblock {\em Quat. Res.} {\bf 1976}, {\em 6},~481--493.
				
				\bibitem[Wallace(1983)]{wallace1983seasonal}
				Wallace, L.
				\newblock The seasonal variation of the thermal structure of the atmosphere of
				Uranus.
				\newblock {\em Icarus} {\bf 1983}, {\em 54},~110--132.
				
				\bibitem[Greathouse \em{et~al.}(2008)Greathouse, Strong, Moses, Orton,
				Fletcher, and Dowling]{greathouse2008general}
				Greathouse, T.; Strong, S.; Moses, J.; Orton, G.; Fletcher, L.; Dowling, T.
				\newblock A General Radiative Seasonal Climate Model Applied to Saturn, Uranus,
				and Neptune. In 
				\newblock  \emph{AGU Fall Meeting Abstracts}; Washington, DC, USA, 2008; Volume 2008, p. P21B-06. %MDPI: please add publisher and publisher's location (city, country)
				
				\bibitem[Seidelmann(1992)]{seidelmann1992explanatory}
				Seidelmann, P.K.
				\newblock {\em Explanatory Supplement to the Astronomical Almanac}; University
					Science Books: Melville, NY USA, 1992.  %MDPI: please add publisher's location (city, country)
				
				\bibitem[Meeus(1997)]{meeus1997equinoxes}
				Meeus, J.
				\newblock Equinoxes and solstices on Uranus and Neptune.
				\newblock {\em J. Br. Astron. Assoc.} {\bf 1997}, {\em 107},~332.
				
				\bibitem[Fletcher \em{et~al.}(2015)Fletcher, Irwin, Sinclair, Orton, Giles,
				Hurley, Gorius, Achterberg, Hesman, and Bjoraker]{fletcher2015seasonal}
				Fletcher, L.N.; Irwin, P.; Sinclair, J.; Orton, G.; Giles, R.; Hurley, J.;
				Gorius, N.; Achterberg, R.; Hesman, B.; Bjoraker, G.
				\newblock Seasonal evolution of Saturn’s polar temperatures and composition.
				\newblock {\em Icarus} {\bf 2015}, {\em 250},~131--153.
				
				\bibitem[Kopp(2019)]{Kopp2019solar}
				Kopp, G.
				\newblock SORCE Level 3 Total Solar Irradiance Daily Means, Version 018. 2019.  Available online: \url{https://disc.gsfc.nasa.gov/datasets/SOR3TSID_019/summary} (accessed on 15 October 2022).
				%MDPI: Please add access date (day month year)
				%MDPI: We replace doi to URL, please confirm; MTR: that's fine
				
				\bibitem[Orton \em{et~al.}(2007)Orton, Gustafsson, Burgdorf, and
				Meadows]{orton2007revisedh2h2}
				Orton, G.S.; Gustafsson, M.; Burgdorf, M.; Meadows, V.
				\newblock Revised ab initio models for H$_2$--H$_2$ collision-induced absorption at
				low temperatures.
				\newblock {\em Icarus} {\bf 2007}, {\em 189},~544--549.
				
				\bibitem[Cavali{\'e} \em{et~al.}(2015)Cavali{\'e}, Dobrijevic, Fletcher,
				Loison, Hickson, Hue, and Hartogh]{cavalie2015photochemical}
				Cavali{\'e}, T.; Dobrijevic, M.; Fletcher, L.N.; Loison, J.C.; Hickson, K.;
				Hue, V.; Hartogh, P.
				\newblock Photochemical response to the variation of temperature in the 2011-
				2012 stratospheric vortex of Saturn.
				\newblock {\em Astron. Astrophys.} {\bf 2015}, {\em 580},~A55.
				
				\bibitem[Hue \em{et~al.}(2018)Hue, Hersant, Cavali{\'e}, Dobrijevic, and
				Sinclair]{hue2018photochemistry}
				Hue, V.; Hersant, F.; Cavali{\'e}, T.; Dobrijevic, M.; Sinclair, J.
				\newblock Photochemistry, mixing and transport in Jupiter’s stratosphere
				constrained by Cassini.
				\newblock {\em Icarus} {\bf 2018}, {\em 307},~106--123.
				
				\bibitem[Orton \em{et~al.}(1994)Orton, Friedson, Yanamandra-Fisher, Caldwell,
				Hammel, Baines, Bergstralh, Martin, West, Veeder~Jr,
				et~al.]{orton1994spatial}
				Orton, G.S.; Friedson, A.J.; Yanamandra-Fisher, P.A.; Caldwell, J.; Hammel,
				H.B.; Baines, K.H.; Bergstralh, J.T.; Martin, T.Z.; West, R.A.; Veeder~Jr,
				G.J.;  et~al.
				\newblock Spatial organization and time dependence of Jupiter's tropospheric
				temperatures, 1980-1993.
				\newblock {\em Science} {\bf 1994}, {\em 265},~625--631.
				
				\bibitem[Antu{\~n}ano \em{et~al.}(2019)Antu{\~n}ano, Fletcher, Orton, Melin,
				Milan, Rogers, Greathouse, Harrington, Donnelly, and
				Giles]{antunano2019jupiter}
				Antu{\~n}ano, A.; Fletcher, L.N.; Orton, G.S.; Melin, H.; Milan, S.; Rogers,
				J.; Greathouse, T.; Harrington, J.; Donnelly, P.T.; Giles, R.
				\newblock Jupiter’s atmospheric variability from long-term ground-based
				observations at 5 $\upmu$m.
				\newblock {\em Astron. J.} {\bf 2019}, {\em 158},~130.
				
				\bibitem[Friedson(1999)]{friedson1999_QBO}
				Friedson, A.J.
				\newblock New observations and modelling of a QBO-like oscillation in Jupiter's
				stratosphere.
				\newblock {\em Icarus} {\bf 1999}, {\em 137},~34--55.
				
				\bibitem[Antu{\~n}ano \em{et~al.}(2021)Antu{\~n}ano, Cosentino, Fletcher,
				Simon, Greathouse, and Orton]{antunano2021fluctuations}
				Antu{\~n}ano, A.; Cosentino, R.G.; Fletcher, L.N.; Simon, A.A.; Greathouse,
				T.K.; Orton, G.S.
				\newblock Fluctuations in Jupiter’s equatorial stratospheric oscillation.
				\newblock {\em Nat. Astron.} {\bf 2021}, {\em 5},~71--77.
				
				\bibitem[Ortiz \em{et~al.}(1998)Ortiz, Orton, Friedson, Stewart, Fisher, and
				Spencer]{ortiz1998evolution}
				Ortiz, J.; Orton, G.; Friedson, A.; Stewart, S.; Fisher, B.; Spencer, J.
				\newblock Evolution and persistence of 5-$\upmu$m hot spots at the Galileo probe
				entry latitude.
				\newblock {\em J. Geophys. Res. Planets} {\bf 1998}, {\em
					103},~23051--23069.
				
				\bibitem[Orton \em{et~al.}(2022)Orton, Antunano, Fletcher, Sinclair, Momary,
				Fujiyoshi, Yanamandra-Fisher, Donnelly, Greco, Payne,
				et~al.]{orton2022unexpected}
				Orton, G.S.; Antunano, A.; Fletcher, L.N.; Sinclair, J.A.; Momary, T.W.;
				Fujiyoshi, T.; Yanamandra-Fisher, P.; Donnelly, P.T.; Greco, J.J.; Payne,
				A.V.;  et~al.
				\newblock Unexpected Long-Term Variability in Jupiter's Tropospheric
				Temperatures.
				\newblock {\em arXiv} {\bf 2022},    arXiv:2211.04398.
				
				\bibitem[Leovy \em{et~al.}(1991)Leovy, Friedson, and
				Orton]{leovy1991quasiquadrennial}
				Leovy, C.B.; Friedson, A.J.; Orton, G.S.
				\newblock The quasiquadrennial oscillation of Jupiter's equatorial
				stratosphere.
				\newblock {\em Nature} {\bf 1991}, {\em 354},~380--382.
				
				
				\bibitem[Greathouse \em{et~al.}(2016)Greathouse, Orton, Cosentino,
				Morales-Juberias, Fletcher, Giles, Melin, Encrenaz, Fouchet, and
				DeWitt]{greathouse2016tracking}
				Greathouse, T.K.; Orton, G.S.; Cosentino, R.; Morales-Juberias, R.; Fletcher,
				L.N.; Giles, R.S.; Melin, H.; Encrenaz, T.A.; Fouchet, T.; DeWitt, C.N.
				\newblock Tracking Jupiter's Quasi-Quadrennial Oscillation and Mid-Latitude
				Zonal Waves with High Spectral Resolution Mid-Infrared Observations. In 
				\newblock  \emph{AAS/Division for Planetary Sciences Meeting Abstracts\# 48}; AAS: Washington, DC USA, {2016}; 
				Volume~48, p. 501.05. %MDPI: please add publisher and publisher's location (city, country)
				
				\bibitem[Orton and Yanamandra-Fisher(2005)]{orton2005saturn}
				Orton, G.; Yanamandra-Fisher, P.
				\newblock Saturn's temperature field from high-resolution middle-infrared
				imaging.
				\newblock {\em Science} {\bf 2005}, {\em 307},~696--698.

				
				\bibitem[Hue \em{et~al.}(2016)Hue, Greathouse, Cavali{\'e}, Dobrijevic, and
				Hersant]{hue20162d}
				Hue, V.; Greathouse, T.; Cavali{\'e}, T.; Dobrijevic, M.; Hersant, F.
				\newblock 2D photochemical modeling of Saturn’s stratosphere. Part II:
				Feedback between composition and temperature.
				\newblock {\em Icarus} {\bf 2016}, {\em 267},~334--343.

				\bibitem[Bardet \em{et~al.}(2022) Bardet, Spiga, Guerlet]{bardet2022joint} Bardet, D.; Spiga, A.; Guerlet, S. 
				\newblock Joint evolution of equatorial oscillation and interhemispheric circulation in Saturn’s stratosphere.
				\newblock {\em Nat. Astron.} {\bf 2022}, {\em 6(7)},~804--811.

				\bibitem[Fischer \em{et~al.}(2011)Fischer, Kurth,  Gurnett, Zarka,  Dyudina,  Ingersoll,Ewald, Porco,Wesley,  Go, Delcroix]{fischer2011giant}
				Fischer, G.; Kurth, W.S.; Gurnett, D.A.; Zarka, P.; Dyudina, U.A.; Ingersoll, A.P.; Ewald, S.P.; Porco, C.C.; Wesley, A.; Go, C.; Delcroix, M.
				\newblock A giant thunderstorm on Saturn.
				\newblock {\em Nature} {\bf 2011}, {\em 475(7354)},~75--77.

				
			\end{thebibliography}
	\end{document}